\documentclass[fleqn,usenatbib]{mnras}

\usepackage{newtxtext,newtxmath}
\usepackage{lscape}
\usepackage{pdflscape}

\usepackage[T1]{fontenc}
\DeclareRobustCommand{\VAN}[3]{#2}
\let\VANthebibliography\thebibliography
\def\thebibliography{\DeclareRobustCommand{\VAN}[3]{##3}\VANthebibliography}

\usepackage{graphicx}	
\usepackage{amsmath}	
\usepackage{booktabs,longtable}
\usepackage{xtab,afterpage}
\usepackage{adjustbox}
\extrafloats{200}
\usepackage{caption}
\usepackage{soul}
\usepackage{xcolor}

\usepackage{subcaption}
\usepackage{color}

\title[MPTA: Data Release and Modelling]{The MeerKAT Pulsar Timing Array: The $4.5$-year data release and the noise and stochastic signals of the millisecond pulsar population}

\author[Miles et al.]{
Matthew~T.~Miles$^{1,2}$\thanks{E-mail: \href{mailto:matthewmiles@swin.edu.au}{matthewmiles@swin.edu.au}},
Ryan~M.~Shannon$^{1,2}$,
Daniel~J.~Reardon$^{1,2}$,
Matthew~Bailes$^{1,2}$,
David~J.~Champion$^{3}$,\newauthor
Marisa~Geyer$^{4}$,
Pratyasha~Gitika$^{1,2}$,
Kathrin~Grunthal$^{3}$,
Michael~J.~Keith$^{5}$,
Michael~Kramer$^{3,5}$,\newauthor
Atharva~D.~Kulkarni$^{1,2}$,
Rowina~S.~Nathan$^{6,2}$,
Aditya~Parthasarathy$^{7,8,3}$,
Nataliya~K.~Porayko$^{9,3}$,\newauthor
Jaikhomba~Singha$^{4}$,
Gilles~Theureau$^{10,11}$,
Federico~Abbate$^{12,3}$,
Sarah~Buchner$^{13}$,
Andrew~D.~Cameron$^{1,2}$,\newauthor
Fernando~Camilo$^{13}$,
Beatrice~E.~Moreschi$^{9,12}$,
Golam~Shaifullah$^{9,12}$, 
Mohsen~Shamohammadi$^{1,2}$ \& \newauthor
Vivek~Venkatraman~Krishnan$^{3}$
\\
$^{1}$Centre for Astrophysics and Supercomputing, Swinburne University of Technology, PO Box 218, Hawthorn, VIC 3122, Australia\\
$^{2}$OzGrav: The ARC Centre of Excellence for Gravitational Wave Discovery \\
$^{3}$MPI f\"ur Radioastronomie, Auf dem H\"ugel 69, 53121 Bonn, Germany \\
$^{4}$High Energy Physics, Cosmology \& Astrophysics Theory (HEPCAT) Group, Department of Mathematics and Applied Mathematics, \\ University of Cape Town, Cape Town 7700, South Africa \\
$^{5}$Jodrell Bank Centre for Astrophysics, University of Manchester, Alan-Turing Building, Oxford Street, Manchester M13 9PL, UK \\
$^{6}$School of Physics and Astronomy, Monash University, Clayton VIC 3800, Australia \\
$^{7}$ASTRON, Netherlands Institute for Radio Astronomy, Oude Hoogeveensedijk 4, 7991 PD Dwingeloo, The Netherlands \\
$^{8}$Anton Pannekoek Institute for Astronomy, University of Amsterdam, Science Park 904, 1098 XH Amsterdam, The Netherlands \\
$^{9}$Dipartimento di Fisica ``G. Occhialini", Universit{\'a} degli Studi di Milano-Bicocca, Piazza della Scienza 3, I-20126 Milano, Italy\\
$^{10}$Laboratoire de Physique et Chimie de l'Environnement et de l'Espace  LPC2E UMR7328, Université d'Orléans, CNRS, CNES, OSUC,\\ Observatoire de Paris,  F-45071 Orléans, France \\
$^{11}$Laboratoire Univers et Théories, Observatoire de Paris, Université PSL, Université de Paris Cité,  CNRS, F-92190 Meudon, France \\
$^{12}$INAF -- Osservatorio Astronomico di Cagliari, Via della Scienza 5, I-09047 Selargius (CA), Italy \\
$^{13}$South African Radio Astronomy Observatory, 2 Fir Street, Black River Park, Observatory 7925, South Africa \\
}

\date{Accepted XXX. Received YYY; in original form ZZZ}

\pubyear{2024}

\begin{document}
\label{firstpage}
\pagerange{\pageref{firstpage}--\pageref{lastpage}}
\maketitle

\begin{abstract}
\newline
Pulsar timing arrays are ensembles of regularly observed millisecond pulsars timed to high precision. Each pulsar in an array could be affected by a suite of noise processes, most of which are astrophysically motivated. Analysing them carefully can be used to understand these physical processes. However, the primary purpose of these experiments is to detect signals that are common to all pulsars, in particular signals associated with a stochastic gravitational wave background. To detect this, it is paramount to appropriately characterise other signals that may otherwise impact array sensitivity or cause a spurious detection. Here we describe the second data release and first detailed noise analysis of the pulsars in the MeerKAT Pulsar Timing Array, comprising high-cadence and high-precision observations of $83$ millisecond pulsars over $4.5$ years. We use this analysis to search for a common signal in the data, finding a process with an amplitude of $\log_{10}\mathrm{A_{CURN}} = -14.25^{+0.21}_{-0.36}$ and spectral index $\gamma_\mathrm{CURN} = 3.60^{+1.31}_{-0.89}$. Fixing the spectral index at the value predicted for a background produced by the inspiral of binary supermassive black holes, we measure the amplitude to be $\log_{10}\mathrm{A_{CURN}} = -14.28^{+0.21}_{-0.21}$ at a significance expressed as a Bayes factor of $\ln(\mathcal{B}) = 4.46$. Under both assumptions, the amplitude that we recover is larger than those reported by other PTA experiments. We use the results of this analysis to forecast our sensitivity to a gravitational wave background possessing the spectral properties of the common signal we have measured.

\end{abstract}




\section{Introduction}
Pulsar timing arrays (PTAs) \citep{1990ApJ...361..300F} are regularly observed ensembles of millisecond pulsars (MSPs) that measure arrival times of the pulses emitted by pulsars over years to decades. MSPs are known to be particularly rotationally stable, allowing the times of arrival (ToAs) of their pulses to be predicted to precisions as small as tens of nanoseconds.

The predictability of MSP emission leads them to be ideal instruments to perform the principal goal of a PTA: to search for spatially and temporally correlated signals within their data set, with the aim of detecting and characterising gravitational waves in the nanohertz (nHz) frequency band. The dominant contributor of gravitational waves in this spectrum is likely to be the cosmological population of gravitationally radiating supermassive black hole binaries (SMBHBs) \citep{1995ApJ...446..543R, 2003ApJ...583..616J, 2003ApJ...590..691W, 2004ApJ...611..623S, 2011MNRAS.411.1467K, 2012A&A...545A.127R, 2017PhRvL.118r1102T} that emit gravitational waves at this frequency as they inspiral. However, there are alternate, exotic sources that may also contribute, including cosmic strings \citep{1976JPhA....9.1387K, PhysRevD.81.104028, 2012PhRvD..85l2003S, 2015MNRAS.453.2576L, 2018ApJ...859...47A}, cosmological phase transitions \citep{1980PhLB...91...99S, 2005PhyU...48.1235G}, and quantum fluctuations in the early universe \citep{2000gr.qc.....8027M, 2016PhRvX...6a1035L}. The most likely signal that PTAs are sensitive to is a stochastic gravitational wave background (SGWB), the incoherent superposition of gravitational-wave (GW) emission from many of these sources \citep{Hellings_Downs_1983}. 

One influence an SGWB will have on a PTA data set is in the emergence of a statically identical signal amongst the pulsars in the array. This signal is modelled as a red-noise process, and is often quantified in the Fourier domain as one that has a power-law power spectral density. When only the spectral characteristics of an SGWB are considered, this signal is often termed common uncorrelated\footnote{Here, uncorrelated refers to the spatial correlations that the process may also possess.} red noise (CURN). Assuming that the inspiral of SMBHBs is the source of the CURN, the power spectral density is
\begin{equation}
    S_\mathrm{GW}(f) = \frac{\mathrm{h}_{c}(f)^{2}}{12\pi^{2}f^{3}} = \frac{A^{2}_\mathrm{GW}}{12\pi^{2}} \left( \frac{f}{\mathrm{yr}^{-1}} \right)^{-13/3} \mathrm{yr}^{3},
\end{equation}
where $\mathrm{h}_{c}(f) \propto f^{-2/3}$ is the expected characteristic strain of the gravitational wave emission from a bound circular SMBHB when the only driver of the inspiral is gravitational radiation. This, in turn, equates to
\begin{equation}
    \mathrm{h}_{c} = A_\mathrm{GW} \left( \frac{f}{\mathrm{yr}^{-1}} \right)^{-2/3},
\end{equation}
where $\mathrm{A_{GW}}$ is the dimensionless gravitational wave amplitude at a frequency of $1$\,yr$^{-1}$, and $f$ is the fluctuation frequency to which the strain spectrum scales.

The detection of a CURN has been forecast to be an important initial step in the characterisation of an SGWB, but cannot in itself be treated as evidence for a detection. \citet{2021arXiv210712112G} and \citet{2022MNRAS.516..410Z} demonstrated that it is possible to spuriously detect CURN in PTA data sets, and its presence should be treated only as a potential indication of the presence of a common signal, rather than definitive evidence for one. 

In addition to a common spectrum process amongst the pulsars, the influence of a gravitational wave background is expected to be spatially correlated, arising from the quadrupolar signature of the local background on the Earth and depending on the angular separation of the pulsar pairs in an array. Under this assumption, the correlation between any two pulsars in an array (\textit{a} and \textit{b}) can be described by the overlap reduction function,
\begin{equation}
\label{eq: HD equation}
    \Gamma_{a,b}(\zeta) = \frac{1}{2} - \frac{1}{4}\left(\frac{1 - \cos{\zeta}}{2} \right) + \frac{3}{2}\left(\frac{1 - \cos{\zeta}}{2}\right)\ln\left({\frac{1 - \cos{\zeta}}{2}}\right),
\end{equation}
expressed in terms of their angular separation ($\zeta$). Commonly, this is referred to as the Hellings-Downs correlation function \citep{Hellings_Downs_1983}.

Searches for an SGWB have been performed on an individual basis by the European PTA \citep[EPTA;][]{2008AIPC..983..633J}, the Parkes PTA \citep[PPTA;][]{2008AIPC..983..584M}, the North American Nanohertz Observatory for Gravitational Waves \citep[NANOGrav;][]{2009arXiv0909.1058J}, the Chinese Pulsar Timing Array \citep[CPTA]{2023RAA....23g5024X}, and also in a joint effort through the International PTA \cite[IPTA;][]{2022MNRAS.510.4873A}. These searches have resulted in the detection of a CURN in each PTA data set \citep{NanoGravGWB, 2021arXiv210712112G, 2021MNRAS.508.4970C, 2022MNRAS.510.4873A, 2023RAA....23g5024X}, with an amplitude and spectral index that are in general agreement within reported uncertainties. While correlations with strong statistical significance ($3-4\sigma$) are emerging in the data sets, no collaboration has met a community defined protocol \citep{2023arXiv230404767A} required to claim a definitive detection \citep{2023ApJ...951L...8A, 2023arXiv230616214A, 2023ApJ...951L...6R, 2023RAA....23g5024X}. 

The influence of the background is thought to emerge in PTA data as a radio-frequency ($\nu$) independent (achromatic) time-correlated noise process. However, it is not the only astrophysical mechanism that can produce this. Often termed \textit{spin} or \textit{timing} noise, rotational instabilities in the pulsar can cause the arrival times of the pulsar emission to vary in a manner that is well described by a noise process such as this \citep{2010ApJ...725.1607S}. Millisecond pulsars are very stable rotators and have been described as nature's most precise clocks \citep{2018SSRv..214...30B}. However, spin noise inconsistent with the expected influence of the SGWB is observed in many MSPs and is common and strong in slow pulsars \citep{2019MNRAS.489.3810P}. While spin noise that is detected in slow pulsars is often of a far larger amplitude than that detected in MSPs, it is likely that they arise from the same or similar mechanisms \citep{2010ApJ...725.1607S}. In early data sets most MSPs did not show evidence for spin noise, and it was suggested that this was likely an observational bias as PTA data at these epochs were not precise enough to easily detect intrinsic noise processes \citep{2010ApJ...725.1607S}. Spin noise in MSPs has now been detected widely across multiple PTA data sets \citep{2023A&A...678A..49E, 2023ApJ...951L..10A, 2023ApJ...951L...7R}, even at short observational time-spans \citep{2023MNRAS.519.3976M}, demonstrating this reality.

While spin noise is considered intrinsic to the pulsar emission mechanism, the ionised interstellar medium (IISM) also causes time-correlated variations in pulsar arrival times. The variations induced by the IISM are radio-frequency dependent (chromatic), and mechanisms have been proposed that can scale the magnitude of these delays from $\nu^{-0.3}$ to $\nu^{-6.4}$ \citep{2010arXiv1010.3785C,2017MNRAS.464.2075S}. Of these various mechanisms, there are two which are by far the most prevalent. \textit{Dispersion measure} (DM) noise is a result of the stochastic variations in the column density of electrons along the line of sight to the pulsar, and scales to $\nu^{-2}$ \citep{2013MNRAS.429.2161K, 2015ApJ...801..130L}. Alongside spin noise, this process is thought to exist in all pulsar data sets to some extent. However, the detection of both of these processes is heavily dependent on the sensitivity of the pulsar data and the observational time span, often leading to the data not suggesting the presence of either process. The other principal contribution of the IISM is commonly termed \textit{Scattering} noise. This process is caused by inhomogeneities in the IISM, likely related to turbulence. Pulsar radiation is scattered off of these inhomogeneities, with the observed radiation being subject to multi-path propagation. The additional path length results in a delay in pulse arrival times \citep{1990ARA&A..28..561R, 1991Natur.354..121C, 2010arXiv1010.3785C}. The varying degree to which different radio frequencies will interact with the screen results in changes to the magnitude of the signal delay. This effect scales to $\sim \nu^{-4}$ \citep{1971ApJ...164..249L} if the inhomogeneities follow a Kolmogorov square law model. However the exact nature of the density inhomogeneities in the IISM allow for a range of possible scaling indices \citep{2016MNRAS.462.2587G}. Similarly to DM variations, as the pulsar-Earth line of sight changes, so too will the magnitude of the scattering variations.

Our local solar neighbourhood will also induce delays in the arrival times of pulsars in an array. As the line of sight between a pulsar and the Earth draws closer to the Sun, the pulse will be affected to a greater extent by the increase in the mean plasma density of the solar wind \citep{2021A&A...647A..84T}. This is a largely periodic effect and can be accounted for accordingly. Until recently it was commonplace for PTAs to assume a standard mean plasma density for all their pulsars; however, \citet{2023ApJ...951L...7R} demonstrated that for many pulsars in the PPTA this is not appropriate, especially for those found at an Ecliptic latitude close to $0^{\circ}$. Coupled with this, ignoring the stochastic variance of the plasma density may result in the emergence of dipolar spatial correlations in PTA data sets, motivating the need for a more precise model of these effects \citep{2022ApJ...929...39H}.

In addition to astrophysically motivated time-correlated noise processes, there also exists noise sources that are uncorrelated in time. These processes are often referred to as \textit{white} noise processes, named for their flat power spectral densities. White noise can be separated into EFAC, EQUAD, and ECORR \citep[e.g.][]{NanoGravGWB}. EFAC and EQUAD arise as a result of unaccounted-for systematic errors in the process of calculating the arrival times of the pulses, whereas ECORR is used to capture stochastic variations in the morphology and arrival times of individual pulses, a phenomenon referred to as \textit{jitter} \citep{2014MNRAS.443.1463S, 2019ApJ...872..193L, 2021MNRAS.502..407P}. 


As PTA experiments form using next-generation radio telescopes (e.g., The Deep Synoptic Array \cite[DSA2000;][]{2019BAAS...51g.255H}, Five-hundred-meter Aperture Spherical Telescope \cite[FAST;][]{2019SCPMA..6259502J}, Next Generation Very Large Array \cite[ngVLA;][]{2018ASPC..517....3M}, the MeerKAT radio telescope \citep{2016mks..confE...1J}, and the Square Kilometer Array \cite[SKA;][]{2009IEEEP..97.1482D}), understanding the best practices to correctly model these noise processes will have increased importance. While the added sensitivity from next-generation facilities will lead to ever greater constraints on the nature of an SGWB, they will also be sensitive to noise processes that are not currently obvious in PTA data sets that may impact results in the first years of an SGWB detection.

The MeerKAT Pulsar Timing Array \citep[MPTA;][]{2023MNRAS.519.3976M}, routinely observing $83$ MSPs to largely sub-microsecond precision, is the largest existing PTA experiment by number of pulsars observed. The MPTA makes use of the MeerKAT radio telescope, a 64-antenna interferometer, located in the Great Karoo region of South Africa. MeerKAT \citep{2016mks..confE...1J} is a precursor to the Square Kilometer Array Mid telescope \citep{2009IEEEP..97.1482D} and is actively demonstrating the performance of a next-generation radio telescope on a future SKA site. Notably, the MPTA observes $13$ pulsars with high DM $>100$ pc\,cm$^{-3}$, which will experience stronger propagation effects from the IISM \citep{2010arXiv1010.3785C}. Future, more sensitive, PTA experiments are likely to include more distant MSPs, which will also encounter such effects. By monitoring a subset of these pulsars now, the MPTA can assist future efforts in developing appropriate mitigation strategies.


In this paper, we present the preferred noise models for the MPTA based on the first four and a half years of observing. We show that processes that possess large chromatic variations can incorrectly be identified as achromatic, and comment on the inherent risk this poses in performing a gravitational wave analysis. We also include the first search for a common spectrum process in MPTA data, and provide examples of how noise misspecifications that are likely present in all PTA data sets can alter the inferred properties of a CURN. Through this, we describe a comprehensive process for noise analysis and modelling, towards the goal of detecting common signals in PTA data.

In Section \ref{Section: Observations}, we describe the data set we use for this work. In Section \ref{Section: Noise processes and models} we outline the different models that were considered for the pulsars in the MPTA data set. Section \ref{Section: PTA noise budget} describes the methodology we used for determining the appropriate models for the data. In Section \ref{Section: Results} we provide a detailed description of the preferred noise models for each pulsar in our data set, and the results of a search for a common spectrum process. In Section \ref{Section: Discussion} we discuss our results, and we conclude in Section \ref{Section: Conclusion}.

\section{Observations and data release}
\label{Section: Observations}


The data set used in this analysis is an extension of the first MPTA data release \cite[][]{2023MNRAS.519.3976M}.  Below we briefly summarise the data processing and differences between the two data releases. 

The MPTA is enabled by access to the MeerKAT radio telescope as a sub-theme of the MeerTime Large Survey project (LSP) \citep{2016mks..confE..11B,2020PASA...37...28B}, an LSP that has used MeerKAT, which is operated by the South African Radio Astronomy Observatory (SARAO). The data analysed in this work span February 2019 to August 2023\footnote{For PSR~J1713$+$0474 we restrict observations to before MJD 59319. After that date, the pulsar showed a large profile change \citep[e.g.][]{2021MNRAS.507L..57S}. If not accounted for, this introduces frequency-dependent biases in the pulse arrival time in excess of $50\,\mu$s.} (MJD $58526 - 60157$). Observations were obtained with the L-band receiver ($856$ - $1712$ MHz), and recorded with the Pulsar Timing User Supplied Equipment (PTUSE) backend recorders \citep{2020PASA...37...28B}. The integration times of the observations were tailored to individual pulsars in order to achieve a band averaged uncertainty of $1\,\mu$s, based on observations made as part of the MeerTime MSP census \cite[][]{2022PASA...39...27S}. An integration time of $256$\,s was chosen if this precision could be achieved in a shorter duration. This enabled a larger number of pulsars to be regularly observed with the MPTA time allocation, increasing the array sensitivity to a stochastic gravitational wave background \cite[][]{2013CQGra..30v4015S}.    

 The MPTA makes use of \textit{fold-mode} data products produced by the PTUSE machines. For each pulsar, these data are coherently dedispersed at a nominal dispersion measure, and folded at the topocentric period. The observations are written in \textsc{psrfits} \citep{2004PASA...21..302H} format, containing $8-$s sub-integrations of the pulsar observation at a phase resolution of $1024$ bins, with four polarisation products, with the early data being recorded with $928$ channels\footnote{Early observations were restricted to the inner $928$ channels of the $1024$ channels enabled by the MeerKAT CBF due to restrictions in ingest bandwidth of the MeerKAT correlator beamformer.} and the latter data with $1024$ frequency channels. Raw data from the MPTA are transferred to both the SARAO data archive and the MeerTime data archive and portal hosted on the OzStar supercomputer at Swinburne University of Technology.

Data stored at the MeerTime data portal are automatically processed using the MeerTime processing pipeline (\textsc{meerpipe}), which excises radio-frequency interference (RFI) via \textsc{meerguard}, a modified version of the \textsc{coastguard} RFI-excision algorithm \citep{2016MNRAS.458..868L}. 
 For observations with $1024$ channels, the outer $48$ MHz at the top and the bottom of the band were discarded to match the $928$ channel data and remove these less sensitive channels affected by bandpass roll-off that were not recorded in early MeerTime observations. 
For the purposes of this data release, we used observations that had been averaged to $32$ frequency channels across the bandwidth (unlike the first data release which had 16 channel subintegrations), fully averaged in time where the observation was less than $3000$ seconds, and converted to total intensity (Stokes I). 
We found that higher frequency resolution in this data release resulted in increased sensitivity to noise processes. 
Where necessary, observations that were longer than this  were split into integrations representative of the median observation length of the pulsar. The MPTA observes pulsars for a maximum of $2048$ seconds; however, the data set used in this analysis also included data collected by other projects within the MeerTime collaboration. Notably, the relativistic binary programme \citep{2021MNRAS.504.2094K} observes pulsars for longer integrations. If these observations were averaged completely, significant errors would be induced in the timing model of the pulsar, hence the limit employed on the maximum integration time for any arrival time calculation.

The core component of the data analysis and data release are the pulse arrival times and their uncertainties.
Using a Fourier domain Monte-Carlo algorithm (FDM) implemented in the \textsc{psrchive}\footnote{http://psrchive.sourceforge.net/} \texttt{pat} utility \citep{2004PASA...21..302H}, these were measured for $32$ sub-bands across the observing band using a portrait (a frequency-resolved timing template) developed using the \textsc{PulsePortraiture}\footnote{\url{github.com/pennucci/PulsePortraiture}} software \citep{2019ApJ...871...34P}. 
Updated portraits were created to match the $32$-channel resolution of this data set and correct for modest systematic drifts in the profiles used to produced DR1.
Of these arrival times, those were measured to have signal-to-noise ratio (S/N) of $<8$ were not included. These lower S/N observations are unlikely to add to the sensitivity of our searches for GWs and noise processes. Given our observing strategy aimed to achieve high precision arrival times, they also represent a small percentage ($\sim 10 \%$) reduction in the total number of arrival time measurements. The MPTA has previously shown results from a subset of $78$ of the pulsars it regularly observes \citep{2023MNRAS.519.3976M}. Here, we expand this sample and demonstrate our findings for the entire ensemble of the $83$ pulsars currently observed by the MPTA. 
In total, the data release comprises $245,907$ arrival-time measurements.  The median uncertainty for the (sub-banded) arrival time is 3.1\,$\mu$s. This equates to a  band averaged median arrival time of $3.1/\sqrt{32} \approx 0.5\,\mu$s. 

\begin{figure*}
    \includegraphics[width=\linewidth]{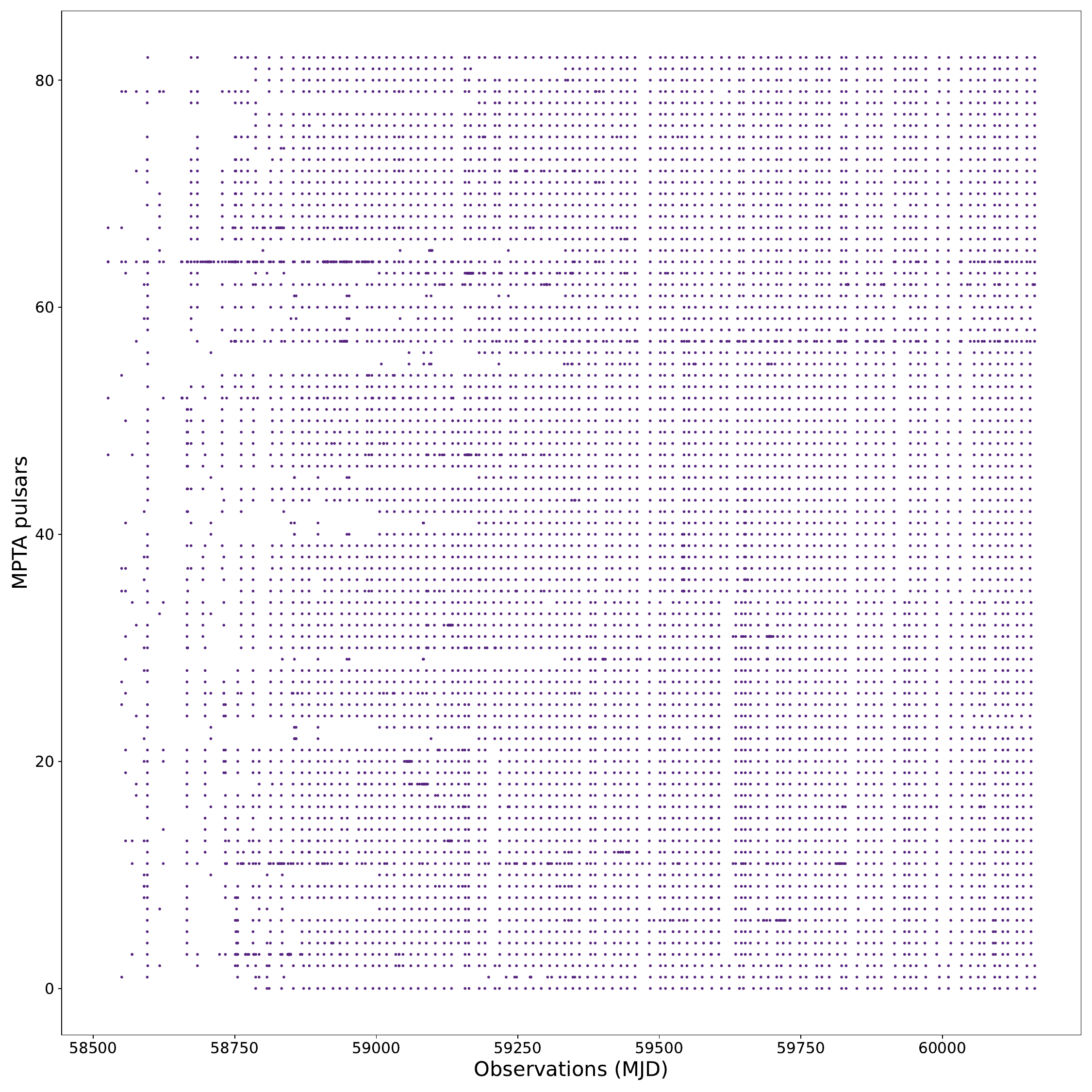}
    \caption{Observation epochs of the MPTA pulsars for this data release. Each time series shows the observations of the pulsars in this data  analysis in increase right ascension (i.e., PSR~J0030$+$0451 is presented at the bottom)}
    \label{fig: total_data}
\end{figure*}

In summary, our data release comprises derived pulse arrival times and uncertainties in \textsc{tempo2} compatible format with IPTA defined metadata \cite[][]{2010CQGra..27h4013H}, the pulse profiles and portraits used to derive the arrival times, and ephemerides that were used as the basis for the timing analysis we describe below. 
Pulsar ephemerides use the DE440 model of  the solar system for arrival time barycentric corrections, and the 2022 realisation of terrestrial time from the International Bureau of Weights and Measures (BIPM).
Compared to the first data release, we have removed three pulsars with poor timing precision: the probable mode changing MSP PSR~J1103$-$5403 \cite[][]{2023arXiv230402793N} which shows large excess white noise levels; the double neutron star system PSR~J1756$-$2251 \cite[][]{2004MNRAS.355..147F} which shows strong timing noise; and the black widow binary pulsar PSR~J1705$-$1903 \cite[][]{2019MNRAS.483.3673M} which has orbital phase dependent noise.
We have also added eight pulsars that were not included in the first data release:  
PSRs J0101$-$6422, J1231$-$1411, J1514$-$4946, J1804$-$2717, J1804$-$2858, J1843$-$1448, J1911$-$1114,  and J2236$-$5527. 
As a visual aid, we present the scope of the release in Figure \ref{fig: total_data}.

\section{Noise processes and models}
\label{Section: Noise processes and models}

To confidently detect and characterise any signal a PTA observes, complete models of the pulse arrival times for every pulsar are required. These contributions can be broadly separated into deterministic and stochastic components. The deterministic components are described by the fiducial timing model of the pulsar, referred to as the timing ephemeris, along with some radio-frequency dependent events that are less commonly detected. The parameters in the timing models describe the factors that are intrinsic to the pulsar and the IISM along the Earth-pulsar line of sight that can be modelled directly from the pulse arrival times. These include position, rotational characteristics, astrometry, dispersion, and binary parameters (where applicable). The deterministic events modelled in addition to the timing model describe both Gaussian-like and annually correlated features that are only rarely present in pulsar timing residuals. The stochastic components of pulsar arrival time deviations are described by noise processes. The combination of all noise processes in an array is referred to as the PTA noise budget and can be used to assess the sensitivity of the PTA to any common signal in the data.

\subsection{Deterministic timing model}
\label{Subsection: Deterministic Modelling}
The impact that the deterministic timing model has on the arrival time of the pulses can be separated into four primary components: model corrections to the inferred pulsar spin frequency and derivatives thereof($\Delta\mathrm{t}_f$), pulse arrival time corrections to the inertial rest frame of the solar system barycentre (SSB) ($\Delta \mathrm{t}_\mathrm{SSB}$), the dispersion of the pulse as it travels through the ionised interstellar medium ($\Delta \mathrm{t}_\mathrm{IISM}$), and arrival time variations from reflex motion about a companion star, if the pulsar is in a binary ($\Delta \mathrm{t}_\mathrm{B}$). The residual of the arrival time ($\mathrm{t}_\mathrm{res}$) from the deterministic model, in reference to the measured arrival time ($\mathrm{t}_\mathrm{ToA}$), is then
\begin{equation}
    \mathrm{t}_\mathrm{res} = \mathrm{t}_\mathrm{ToA} - \Delta\mathrm{t}_f - \Delta\mathrm{t}_\mathrm{SSB} - \Delta\mathrm{t}_\mathrm{IISM} -\Delta\mathrm{t}_\mathrm{B}.
\end{equation}

Pulsar timing software packages such as \textsc{TEMPO} \citep{2015ascl.soft09002N}, \textsc{TEMPO2} \citep{2006MNRAS.369..655H}, and \textsc{PINT} \citep{2021ApJ...911...45L} are commonly used to account for these effects. In this study we made use of \textsc{TEMPO2}, updating the timing ephemerides used in \citet{2023MNRAS.519.3976M} to model additional parameters that only became significant following the addition of two more years of data. Some of the pulsars in the MPTA are also timed by other collaborations, which possess much longer data spans. For these pulsars, some binary orbital parameters (particularly the parameters that have secular variations) were thought to be more accurate in other data sets due to the larger observational time span. In these cases, we made use of the binary parameter values published by the PPTA \cite[][]{2021arXiv210704609R}, and did not adjust them further. We considered the addition of parameters to the timing models through a rudimentary significance test, choosing to include a parameter where it was found to be significant at a level of $>3\sigma$. The astrometric contributions to $\Delta \mathrm{t}_\mathrm{SSB}$ are adjusted within the timing model by fitting the pulsar position, proper motion, and parallax. Further perturbations stemming from this term are corrected using a solar system ephemeris (SSE) supplied to the pulsar timing software that is assumed to be accurate. In this work, we have used the DE440 ephemeris published by NASA's Jet Propulsion Laboratory \citep{2021AJ....161..105P}. An incorrect SSE can result in dipolar spatial correlations through a PTA, an effect that could potentially impact spatially correlated processes in PTA data. To account for potential errors that may arise in this way, it has been common in past explorations of PTA data to use the \textsc{BayesEphem} SSE model \citep{2020ApJ...893..112V}, which acts to sample SSE parameters using a Bayesian approach. In this analysis, we have chosen not to make use of this as it is less likely to dominate pulsar-by-pulsar noise analyses or in searches for an uncorrelated common noise process \citep{2023ApJ...951L...6R}. Furthermore, the largest contributions to uncertainties are thought to arise from Jupiter, which has an orbital period of a factor of $\sim2.6$ greater than our current data sets.  

\subsection{White noise}
\label{Subsection: White Noise}
Temporally uncorrelated white noise processes are always present in radio-frequency PTA data and are fundamentally connected to the finite system temperature of the telescope receivers. White noise in excess of this is attributed to systematic errors in the estimation of arrival time uncertainties that emerge through the pulsar timing process. These are accounted for through three parameters: EFAC ($\mathrm{E_{F}}$), EQUAD ($\mathrm{E_{Q}}$), and ECORR ($\mathrm{E_{C}}$). Generally, these processes are strongly connected to observing systems, and it is common practice to search for their presence through each backend and receiver combination in use by the PTA. Depending on the complexity of an observing system, this can lead to dozens of white noise parameters that must be identified and measured. $\mathrm{E_{F}}$ is a scale factor applied directly to the arrival time uncertainties and accounts for unknown errors in the time-tagging algorithms used to determine the pulse arrival times. Usually, this is close to unity; deviations can be used as an indicator that there are issues with the PTA observing systems or template used in time tagging. There may exist mechanisms that can introduce errors on a system level that are not appropriately characterised by a scale factor such as $\mathrm{E_{F}}$. In these cases, $\mathrm{E_{Q}}$ is introduced as an error term added in quadrature to the arrival time uncertainties. The continued inclusion of $\mathrm{E_{Q}}$ in next-generation data sets could be attributed to underlying latent RFI that is not obvious enough in the data to be efficiently excised. If this was the case, it could be that the evidence for the inclusion of $\mathrm{E_{Q}}$ is more significant in pulsars possessing larger duty cycles or those that emit at a higher S/N, which could disguise the presence of RFI in a pulse. For the purpose of the MPTA noise determination, these processes were included via the convention described in \citet{2014MNRAS.437.3004L}:
\begin{equation}
    \sigma = \sqrt{\mathrm{E^{2}_{Q}} + \mathrm{E^{2}_{F}} \times \mathrm{\sigma^{2}_{ToA}}},
\end{equation}
where $\mathrm{\sigma_{ToA}}$ is the arrival time uncertainty reported by the chosen time-tagging algorithm.

While $\mathrm{E_{F}}$ and $\mathrm{E_{Q}}$ are required due to systematic uncertainties stemming from observations and arrival time calculations, $\mathrm{E_{C}}$ is motivated by a physical phenomenon known as \textit{pulse jitter} \citep{2014MNRAS.443.1463S,2021MNRAS.502..407P}. Pulse jitter is the term given to the observation that each individual pulse will vary stochastically in morphology and phase. By folding and averaging many pulses, a high S/N pulse that is representative of the probability distribution of the pulse energy through phase is created and subsequently used for timing. Due to the finite number of pulses that are averaged together, there will exist a  difference between the observed averaged pulse and the template used in the time-tagging algorithm. The $\mathrm{E_{C}}$ term can potentially account for this difference. Given that the arrival times calculated in each observational epoch are determined with the same series of pulses, $\mathrm{E_{C}}$ is assumed to have $100\%$ correlation across sub-banded observations collected in the same band or receiver, but no correlation between observing epochs. In addition to pulse jitter, $\mathrm{E_{C}}$ also appears to be capable of absorbing timing uncertainties introduced by a phenomenon known as mode-changing, where the pulsar emission strength and morphology varies intermittently. \citet{2022MNRAS.510.5908M} describes the discovery of multiple modes of emission in PSR~J1909$-$3744, demonstrating that the calculated value of jitter noise decreases when isolating only a single emission mode. This is also demonstrated to a greater extent in the millisecond pulsar PSR~J1103$-$5403 \citep{2023arXiv230402793N} which possesses strong evidence for mode-changing.

Although it is beneficial to search for temporally correlated processes in concert with the white processes, due to similarities in how they might emerge in the data, the latter are typically determined prior and fixed at their maximum likelihood values during the search for other processes. This reduces the number of parameters that are needed when searching for red-noise processes and thus saves a significant amount of computational effort. This methodology appears sound for $\mathrm{E_{F}}$ and $\mathrm{E_{Q}}$. However, there could be an observed degeneracy between $\mathrm{E_{C}}$ and red noise processes that possess a high fluctuation frequency. In addition, recent work in the MPTA has identified an apparent decorrelation of $\mathrm{E_{C}}$ through sub-banded observations collected in the same observing epoch \citep{2024MNRAS.528.3658K}. In principle, this appears to allow $\mathrm{E_{C}}$ to absorb power not only from red noise processes with high fluctuation frequencies but also from pulsars possessing evidence for frequency-dependent noise processes. If these noise terms are excluded from standard noise analyses, it follows that this would result in values of $\mathrm{E_{C}}$ that are routinely larger than expected, especially in cases where the pulsar emission is at a high S/N.

\subsection{Achromatic red noise}
\label{Subsection: Achromatic Red Noise}

Pulsars, particularly MSPs, are notable for the predictability of their emission. Nonetheless, rotational irregularities are likely present in all pulsars. Several mechanisms have been suggested for this phenomenon. Angular momentum exchange between the superfluid core and the crust of the neutron star can cause variations in the pulsar rotation \citep[e.g.][]{Alpar1986VortexCA, 1990MNRAS.246..364J, 2014MNRAS.437...21M}.  It is also thought that torques generated from the pulsar magnetosphere may also play a part in irregular rotation  \citep{1987ApJ...321..799C, 2010Sci...329..408L, 2010ApJ...725.1607S}. Alternate explanations stemming from the local environment of the pulsar, such as the presence of orbiting planets or asteroid belts, could also cause these irregularities \citep{2013ApJ...766....5S}. Despite the lack of a definitive cause, this phenomenon manifests in pulsar timing data as an achromatic stochastic wandering in pulse arrival times, which is correlated through time. In the frequency domain, this is described as a red-noise process, one which possesses larger amplitudes at lower fluctuation frequencies. Of all noise processes commonly present in pulsar timing data, achromatic red noise is arguably the most important to model correctly. This is due to the expected similarity of this signal to that from the influence of the SGWB, also expected to present in PTA data sets as an achromatic red noise process. 

All correlated noise has been modelled by the MPTA as stationary, stochastic processes in the Fourier domain defined by their power spectral densities. For an achromatic red noise process, this can be expressed as 
\begin{equation}
    \mathrm{P_{Red}}(f; \mathrm{A_{Red}, \gamma_\mathrm{Red}}) = \frac{\mathrm{A^{2}_{Red}}}{12\pi^{2}} \left (\frac{f}{f_\mathrm{c}} \right)^{-\gamma_\mathrm{Red}} \mathrm{yr}^{3},
\end{equation}
where $\mathrm{A_{Red}}$ is the amplitude of the signal, $\gamma_\mathrm{Red}$ is the associated spectral index, $f$ is the frequency range the signal is modelled over, and $f_\mathrm{c}$ is the characteristic reference frequency. For the purposes of this work, we have defined $f_\mathrm{c}$ to be $1 \mathrm{yr}^{-1}$. 


\subsection{Dispersion measure noise}
\label{Subsection: DM Noise}

Over time, variations in the column density of electrons are expected due to the turbulent nature of the IISM \citep{1991ApJ...382L..27P}. Functionally, this serves to alter the DM of the pulsar such that the nominal DM in the fiducial timing model does not fully correct for the dispersion of the free electrons. Pulsar timing residuals are sensitive to this effect, which results in a stochastic red-noise process similar to achromatic red noise, but where the magnitude of the signal delay is inversely proportional to the square of the radio frequency $\nu$ of the arrival time. As such the power spectral density is defined to be
\begin{equation}
    \mathrm{P_{DM}}(f; \mathrm{A_{DM}, \gamma_\mathrm{DM}}) = \frac{\mathrm{A^{2}_{DM}}}{12\pi^{2}} \left (\frac{f}{f_\mathrm{c}} \right)^{-\gamma_\mathrm{DM}} \left( \frac{\nu}{\nu_\mathrm{ref}} \right )^{-4}\mathrm{yr}^{3},
\end{equation}
where $\nu_\mathrm{ref}$ is the reference frequency for the process, fixed at $\nu_\mathrm{ref}=1400$ MHz for the MPTA. 

PTA experiments have chosen different methodologies for correcting DM variations. Some PTAs choose to model the DM noise process as a piecewise function, that operates to approximate the time realisation of the process \citep{2013MNRAS.429.2161K,2017ApJ...841..125J}. While this method offers advantages in computational efficiency, we have chosen to model DM variations using a Gaussian (stochastic noise) process. This decision was made as epoch-by-epoch methods of measuring DM can reduce data set sensitivities to an SGWB if the DM cannot be well constrained at individual epochs \citep{2013MNRAS.429.2161K,2015ApJ...801..130L}. 

\subsection{Scattering noise}
\label{Subsection: Chromatic Noise}

Dispersion measure variations are not the only frequency-dependent noise processes expected to be present in PTA data. Alternate chromatic noise processes can emerge due to the small-scale structures in the IISM between the pulsar and the Earth. These structures cause a multi-path propagation of pulses through the IISM, as they diffract pulsar radiation. The geometry of diffraction results in frequency dependence (chromaticity), which is often assumed to scale as $\nu^{-4}$ \citep{1971ApJ...164..249L}, the standard thin screen approximation where we assume there is an isotropic (Gaussian) distribution in the associated scattering angles. While this assumption likely holds true for delays stemming from a thin screen model, with electron density variations originating in Kolmogorov turbulence \citep{1977ARA&A..15..479R}, it may not necessarily reflect the chromaticity caused by a filled or more complex medium \citep{2016MNRAS.462.2587G,2016ApJ...817...16C}. Refractive effects further complicate the expected frequency dependence of multi-path propagation delays \citep{2017MNRAS.464.2075S}. To capture these potential variations, we have treated the degree of the chromaticity as a free parameter and modelled the scattering noise as a power spectral density of the form
\begin{equation}
    \mathrm{P_{Chrom}}(f; \mathrm{A_{Chrom}, \gamma_\mathrm{Chrom}, \beta}) = \frac{\mathrm{A^{2}_{Chrom}}}{12\pi^{2}} \left (\frac{f}{f_\mathrm{c}} \right)^{-\gamma_\mathrm{Chrom}} \left ( \frac{\nu}{\nu_\mathrm{ref}} \right )^{-2\beta} \mathrm{yr}^{3},
\end{equation}
where $\beta$ is the chromatic index of the noise process.

\subsection{Solar-wind models}
\label{Subsection: SW Noise}

As the Earth-pulsar line of sight changes in proximity to the Sun, the impact of the solar wind on the pulse arrival time varies. The solar wind has a similar dispersive effect on the pulse as dispersion measure noise. It is typically modelled using a spherically symmetric and time-independent model for the density, parameterized by the mean solar wind density at $1$ AU ($n_{\oplus}$). In \textsc{tempo2} this is set to a default value of $4 \ \mathrm{cm}^{-3}$ \citep{2006MNRAS.369..655H} and is often either fixed or ignored in PTA analyses, including in our first MPTA data release \citep{2023MNRAS.519.3976M}. This is of concern as improperly modelling solar wind in a PTA data set may induce dipolar spatial correlations in the PTA \citep{2016MNRAS.455.4339T}. 

The assessment by the PPTA on the variation of solar wind density as a function of ecliptic latitude \citep{2023ApJ...951L...7R} naturally implies that fixing the solar wind at a single value is not satisfactory. Recent work \citep{2022ApJ...929...39H, 2024MNRAS.528.3304N} further demonstrates that PTA data sets are sensitive to temporal variations in the solar wind, and \citet{2022ApJ...929...39H} present a model to account for stochastic variations plasma density, constraining subtle variations in the electron column density that would otherwise not be accounted for with a model assuming a constant solar wind. The variations in mean solar electron density are modelled to be a power law,
\begin{equation}
    \mathrm{P_{SW}}(f; \mathrm{A_{SW}, \gamma_\mathrm{SW}}) = \frac{\mathrm{A^{2}_{SW}}}{12\pi^{2}} \left (\frac{f}{f_\mathrm{c}} \right)^{-\gamma_\mathrm{SW}} \mathrm{cm^{-6}}\mathrm{yr},
\end{equation}
where the spectral index $\gamma_\mathrm{SW}$ is allowed to have a red or blue spectrum. The perturbations that are measured for each arrival time are done so after calculating the pulsar-Earth line of sight path through the solar system, taking into account the variations in column density as the proximity of the pulsar-Earth line of sight to the Sun changes over the course of a year.

The complexity of modelling the solar wind led us to employ three possible ways that it could be accounted for in the pulsar noise models. The $\mathrm{SW_{Full}}$ model describes where both the deterministic (the mean plasma density at $1$ AU) and the stochastic portion of the model were sampled for, the $\mathrm{SW_{Det}}$ model only sampled for the deterministic component, and the $\mathrm{SW_{Fixed}}$ model has a fixed mean plasma density of $4 \ \mathrm{cm}^{-3}$.

\subsection{Other deterministic models}
\label{Subsection: Other Deterministic Models}
The presence of discrete structures throughout the IISM can cause deviations from noise processes that are otherwise well described by the aforementioned power-law power spectral densities \citep[e.g][]{2015ApJ...808..113C}. In these cases, it may be more appropriate to model the deviations as a Gaussian deterministic waveform. To achieve this, we adopted the model described in \citet{2023ApJ...951L...7R}, defined to be
\begin{equation}
\label{Equation: Chrombump}
    t_\mathrm{Gauss}(t) = A_{g} \exp{\left (\frac{(t - t_{g,0})^{2}}{2\sigma_{g}^{2}} \right)} \times \left (\frac{\nu}{\nu_\mathrm{ref}} \right)^{-\beta_{g}},
\end{equation}
where $A_{g}$ is the amplitude of the waveform in the arrival times, $t_{g,0}$ is the epoch associated with the center of the Gaussian event, and $\sigma_{g}$ is the width of the event.
The motion of the Earth around the Sun can also induce variations that are more appropriate to model deterministically. These stem from the density gradient of the plasma between the Earth and the pulsar, and as such are well described as an annually varying function. To capture this, we describe the variations as a sinusoidal waveform as per \citet{2021MNRAS.502..478G}
\begin{equation}
\label{Equation: Chromannual}
    t_\mathrm{Annual}(t) = A_{s} \sin(2\pi t \times f_\mathrm{yr} + \phi) \times \left(\frac{\nu}{\nu_\mathrm{ref}}\right)^{-\beta_{s}},
\end{equation}
where $A_{s}$ is the amplitude of the sinusoid in seconds, $f_\mathrm{yr}$ is the frequency of a year, and $\phi$ is the dimensionless phase of the signal.

\subsection{Common uncorrelated red noise}
\label{Subsection: CURN}

An SGWB is usually expected to initially emerge as an achromatic red noise process common in both spectral index and amplitude through the ensemble of pulsars in a PTA. Of the PTAs that have searched for this signal, all have identified a signal consistent with what is expected of an SGWB \citep{NanoGravGWB, 2021arXiv210712112G, 2021MNRAS.508.4970C, 2022MNRAS.510.4873A}. The spectral properties of the common signal in each array are consistent with the expectation of a background formed from the incoherent superposition of GWs from inspiralling SMBHBs. However, when describing this process we have decided to instead refer to it as a common uncorrelated red noise (CURN), rather than a signal that is necessarily connected to the SGWB. \citet{2021ApJ...917L..19G} and \citet{2022MNRAS.516..410Z} have demonstrated that spurious detections of CURN can arise with strong support from the data where no common signal is present. Although, the occurrence of this appears to decrease as the intrinsic noise properties of the pulsars in the array are allowed to deviate further from commonality.

The CURN in each pulsar is modelled as an achromatic power spectral density in the frequency domain to be
\begin{equation}
    \mathrm{P_{CURN}}(f; \mathrm{A_{CURN}, \gamma_\mathrm{CURN}}) = \frac{\mathrm{A^{2}_{CURN}}}{12\pi^{2}} \left (\frac{f}{f_\mathrm{c}} \right)^{-\gamma_\mathrm{CURN}} \mathrm{yr}^{3}.
\end{equation}
Unlike the achromatic red noise process, the CURN is evaluated as a signal that is common among the pulsars in the array rather than intrinsic to them. The CURN model can be extended to include the overlap reduction function in Equation \ref{eq: HD equation}, where it then accounts for correlations between the pulsars in the array as a function of angular separation.

In this analysis, our focus is on identifying a shared signal within the data, while not exploring any spatially correlated signals. To assess the presence of this signal, we employed two distinct approaches. We first adopted a method to factorise the likelihood of each pulsar, evaluating the potential presence of a CURN while not requiring extensive computational resources \citep{2022PhRvD.105h4049T}. Following this, we assessed the full PTA likelihood in our search for a common signal. In both analyses, we included additional achromatic red noise processes into the preferred pulsar noise models (described below) where they were not already part of the fiducial noise model for the pulsar. This was implemented to minimize the risk of misspecifying the intrinsic pulsar noise as a potential shared signal at the expense of lowering our sensitivity to a CURN.

\section{A PTA noise budget}
\label{Section: PTA noise budget}
A PTA data set is inherently complex due to the number of noise processes that it can contain, especially as the data do not easily visually inform on the presence of many. Assuming that a process is not present in a data set without thorough investigation can lead to the incorrect characterisation of other processes, while adding all mentioned-above noise processes to describe the noise budget of each individual pulsar will unnecessary expand the parameter space of the problem. This could potentially adversely affect the search for spatial correlations in PTA data. For this reason, we have endeavoured to characterise the MPTA noise budget as comprehensively as possible by evaluating each pulsar for the presence of the noise processes described in Section \ref{Section: Noise processes and models}.

We constructed the noise model for each pulsar using Bayesian evidence comparisons to assess possible noise models, selecting the model possessing the highest probability given the arrival times and pulsar ephemeris. Following this, we used an Anderson-Darling statistic \citep{1a8d0b27-4e98-39f3-b2a1-d2944a87e07c} to test if the noise-reduced residuals had the expected Gaussian distribution. If the pulsar passed this test, and the reduced chi-squared ($\chi^{2}_\mathrm{red}$) of the noise-subtracted residuals was sufficiently close to unity\footnote{We defined this as $|\chi^{2}_\mathrm{red} - 1| < 0.1$}, we deemed the model acceptable. 

\subsection{Bayesian inference}
\label{Subsection: Bayesian Inference}

Our technique for selecting the most probable model for the data used Bayesian inference. The motivation behind using a Bayesian method for noise model selection is that it allows for direct comparisons between model classes, enabling the data to inform the preferred model. This is especially useful in PTA data sets as the presence of signals in the data is often difficult to characterise using other means.

To perform these comparisons, we used the \textsc{Enterprise} software package \citep{2019ascl.soft12015E} to model the different noise processes we considered. We used the \texttt{parallel-bilby} sampler \citep{pbilby_paper}, an extension of the \textsc{Bilby} \citep{Ashton+19} architecture to evaluate the posterior distribution through nested sampling. The \textsc{Bilby} architecture was integrated for PTA analyses using parts of the \textsc{Enterprise-warp}\footnote{https://github.com/bvgoncharov/enterprise\_warp} framework, used to pass the prior and likelihood information from \textsc{Enterprise} to \textsc{Bilby}. The decision to use \texttt{parallel-bilby} as our primary sampler was due to its efficiency in message passing interface (MPI) enabled sampling for high-dimensional models, as well as allowing for direct comparisons between the model evidences. Utilisation of \texttt{parallel-bilby} for PTA analyses was made possible by the efforts of \citet{2022MNRAS.517.1460S}, who have made their implementation publicly available: \url{https://github.com/anuradhaSamajdar/parallel_nested_sampling_pta}. During the assessment of a CURN in the data using the full PTA likelihood, we employed a Markov chain Monte Carlo sampling technique using the \texttt{PTMCMC} sampler \citep{justin_ellis_2017_1037579}, the standard sampling technique used in conjunction with \textsc{Enterprise}. 

The evidence can be calculated from the posterior distribution using
\begin{equation}
    \mathcal{Z} = \int\mathcal{L}(d|\theta)\pi(\theta) d\theta,
\end{equation}
for a likelihood function ($\mathcal{L}(d|\theta)$) and prior ($\pi(\theta)$), given the model parameters ($\theta$) and the data ($d$). This relates directly to the posterior distribution that the sampler constructs over the model parameters,
\begin{equation}
    \mathrm{p}(\theta | d) = \frac{\mathcal{L}(d | \theta)\pi(\theta)}{\mathcal{Z}}.
\end{equation}

The PTA likelihood function that is employed here can be described by the multivariate Gaussian distribution
\begin{equation}
\label{eq: PTA likelihood}
    \mathcal{L}(d | \theta) = \frac{\mathrm{exp}(-\frac{1}{2}\mathbf{\delta t}^{T}\mathbf{C}^{-1}\mathbf{\delta t})}{\sqrt{\mathrm{det}(2\pi \mathbf{C})}},
\end{equation}
where $\mathbf{\delta t}$ is a vector of timing residuals and $\mathbf{C}$ is the covariance matrix of the data \citep{2009MNRAS.395.1005V}.

To establish which was better suited to the data, the evidence for each model was directly compared to find a natural log of the Bayes factor
\begin{equation}
    \ln(\mathcal{B}) = \ln(\mathcal{Z}_\mathrm{A}) - \ln(\mathcal{Z}_\mathrm{B}),
\end{equation}
for any two models A and B with model parameters $\theta_\mathrm{A}$ and $\theta_\mathrm{B}$.

\subsection{Codified bayesian analysis}
\label{Subsection: CBA}

In PTA analyses, it is standard practice to analytically marginalise over the deterministic timing model parameters. This technique was also employed in this analysis. The red noise processes were modelled as Gaussian processes in the Fourier domain with a series of harmonically related sinusoids, with the fundamental frequency being the reciprocal of the observing span. By modelling the processes in this way, it is possible to marginalise over the amplitudes of individual Fourier components while searching for the amplitude and spectral index of the underlying power-law process. Due to the high observing cadence of the MPTA (approximately once every $14$ days for each pulsar), it was necessary to use a large number of Fourier components to model the correlated stochastic processes. The value was chosen such that the highest fluctuation frequency that the processes were modelled at was close to the nominal cadence of the MPTA. We thus chose $120$ components corresponding to $\sim 1/14$ days.

To characterise the noise in each pulsar, we first searched for white noise processes. These terms were searched for in conjunction with achromatic red noise and dispersion measure noise, to mitigate the potential of leakage of correlated noise in the data into the white noise parameters. The white noise term $\mathrm{E_{F}}$ is often close to unity and subsequently has little effect on the noise characterisation for most pulsars; as such it was included in all pulsar noise models. It is common for PTAs to assume the presence of $\mathrm{E_{Q}}$ and $\mathrm{E_{C}}$ for each pulsar, even where it is not clear if either or both are required. The $\mathrm{E_{C}}$ term is physically motivated, and where the pulsar is significantly bright it is thought to be needed to account for jitter noise. However, this is not always the case. Similarly, in sub-banded data, $\mathrm{E_{Q}}$ is not well motivated due to the presence of $\mathrm{E_{C}}$, unless the data were affected by RFI. In order to not increase the MPTA noise budget unnecessarily, these terms were only included where it was supported by their model evidence, or where the posterior of the parameter was clearly constrained.

Following this, all possible combinations of the time-correlated processes described in Section \ref{Section: Noise processes and models} were searched for, with the exception of a CURN and the two deterministic models described in Section \ref{Subsection: Other Deterministic Models}. During this process, the favoured white noise processes were held fixed at maximum {\em a-posteriori} (MAP) values. 

For each pulsar, we considered models that included up to four time-correlated processes. Given the similarity of the models described in Section \ref{Section: Noise processes and models}, these can be trivially misspecified even using sophisticated Bayesian selection techniques. To mitigate this, we required that more complex models with a greater number of processes must possess greater evidence than their simpler counterparts. In some cases, the evidence between alternate models was comparable within the uncertainty reported by the sampler. When this occurred, the joint posteriors of a model containing both processes were inspected to determine if one was clearly favoured over the other. If this was not evident, both processes were included in the model assigned to the pulsar. A different approach was taken when deciding upon the inclusion of the two additional deterministic models in Section \ref{Subsection: Other Deterministic Models} (described by Equations \ref{Equation: Chrombump} and \ref{Equation: Chromannual}). As these models are deterministic, and were not modelled in the Fourier domain, the risk of any misspecification with other models was thought to be minimal. As such, following the determination of the preferred model describing the pulsar data, it was assessed again by sampling the preferred model in addition to these deterministic processes. Following this, the most preferred model was sampled again in conjunction with the uncorrelated white noise terms, in the interest of reducing the covariance between the processes within the pulsar noise models.

Throughout our modelling, we included an additional noise process that was not taken to be representative of the true pulsar model, but one that we decided was required for any accurate attempt at describing the intrinsic pulsar noise. This was an additional achromatic red noise process, allowed to vary across the entire amplitude prior range, but with a spectral index fixed at $\gamma_\mathrm{Red}=13/3$. The motivation for the inclusion of this parameter was simple: in the search for a gravitational wave signal, which is the principal goal of a PTA, one would expect that in many pulsars both a common and intrinsic achromatic signal is present in the pulsar's timing residuals. However, modelling two identical signals in a single pulsar analysis would only result in extremely degenerate posterior distributions. To mitigate this, we instead fixed the spectral index of this process at the theoretical value expected of an SGWB, and sample it in conjunction with the models being assessed.


\label{Subsection: gaussianity}

To assess the suitability of the models as complete descriptions of the pulsar intrinsic noise processes, we tested the noise-reduced (whitened) and normalised residuals for indications of time-correlated processes remaining in the data. The models were first assessed by testing whether the noise-reduced normalised residuals represented a Gaussian distribution through an Anderson-Darling test. To achieve this, maximum-likelihood realisations of the noise processes were calculated and subtracted from the residuals using a modified version of the pulsar timing software \textsc{PINT}\footnote{The process of realising and subtracting the noise processes is trivially done using the \textsc{PINT} software, motivating its use. However, the pulsar timing models are still constructed using \textsc{Tempo2}.} \citep{2021ApJ...911...45L}, with values corresponding to the MAP values from the preferred noise model. As a final assessment of the quality of the noise model, the $\chi^{2}_\mathrm{red}$ was calculated using the whitened residuals. If a pulsar failed the Anderson-Darling test ($\mathrm{p} > 0.05$) or did not have appropriately whitened residuals ($|\chi^{2}_\mathrm{red} - 1| > 0.1$), it was taken as an indication that the pulsar noise processes or timing parameters were not well modelled. Where this was found to be the case, both were re-assessed by increasing the complexity of the noise model to include the next most favoured set of noise processes that built upon the initial selection.

\subsection{Search for common processes}

Following an assessment of Gaussianity, and any attempts at remodelling from this process, we searched for a common signal in the data. Both a full PTA likelihood analysis, following Equation \ref{eq: PTA likelihood}, and an analysis involving the factorisation of the individual pulsar likelihoods were performed. In the search for a common signal, all time-correlated noise processes identified in the MPTA were re-sampled simultaneously. In addition to this, achromatic red noise processes were included for pulsars, even if they did not have this term in their noise models, to mitigate any unidentified intrinsic pulsar noise being misspecified as a part of a shared signal between the pulsars. 

\section{Results}
\label{Section: Results}

\begin{figure*}
    \includegraphics[width=\linewidth]{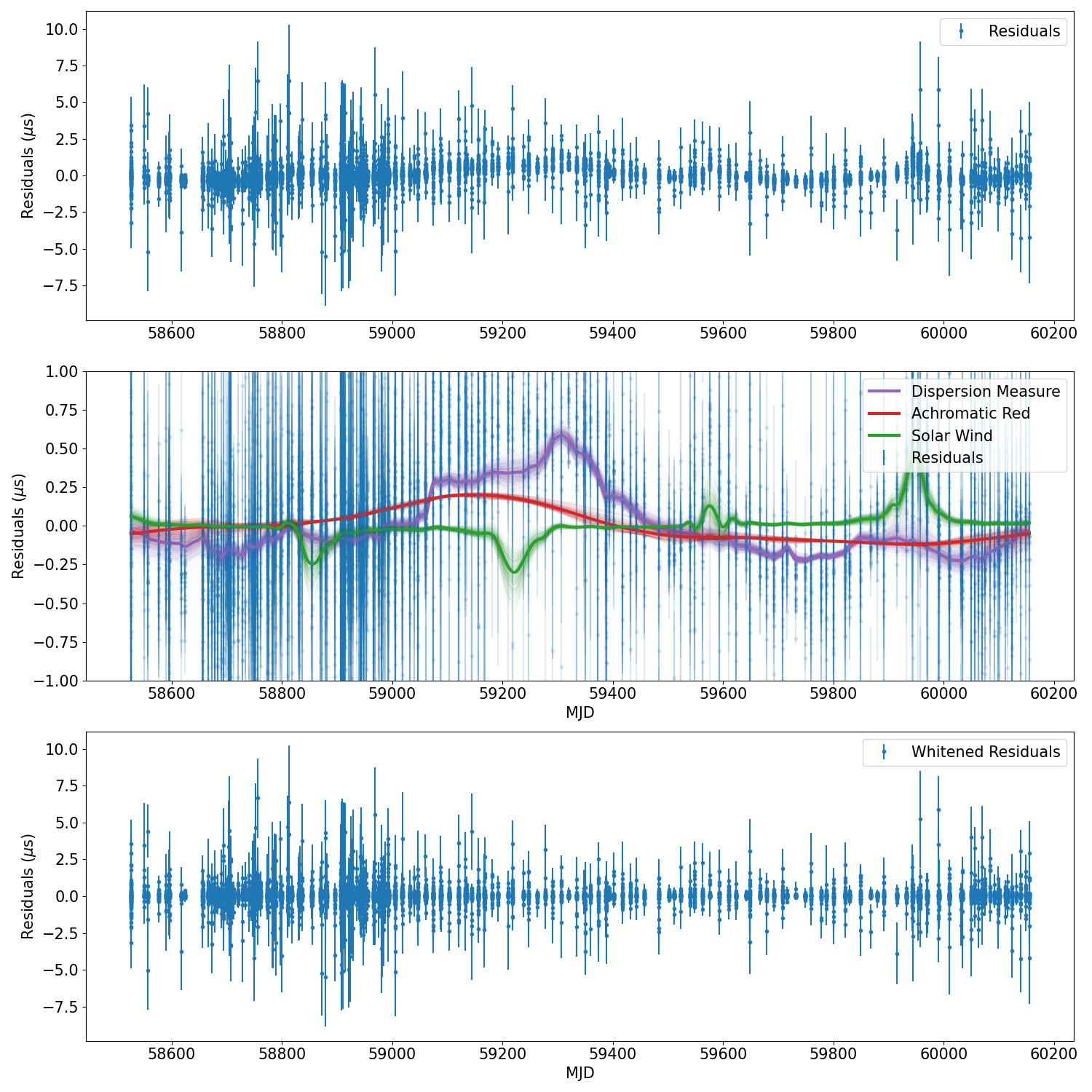}
    \caption{Timing residuals and noise process realisations of PSR~J1909$-$3744. (Top) The timing residuals (blue) of PSR~J1909$-$3744 with no removal of noise processes. (Middle) The realisations of dispersion measure noise (purple), achromatic red noise (red), and the impact of stochastic solar wind (green) overlaid on the residuals shown in the top panel (blue). We note that the reference frequency for the dispersion measure and solar wind realisations is $1400$ MHz,and the y-axis has been enlarged to better identify the sub-microsecond structures of the noise processes. The faint lines surrounding these realisations correspond to $1000$ random draws from the posterior distributions found in the analysis of the pulsar's noise properties, with the thicker line corresponding to the median values of these draws. (Bottom) The whitened residuals of the pulsar, calculated by removing the processes displayed in the middle panel at their maximum likelihood values.}
    \label{fig: 1909_noise_off}
\end{figure*}

The measured values of the apparent noise processes identified in the MPTA data are shown in Table \ref{Table: MPTA noise models} and Table \ref{Table: MPTA determinstic models}. In Figure \ref{fig: 1909_noise_off} we show the timing residuals before and after removing the time-realised noise processes, as well as the noise processes themselves, for the most precisely timed pulsar in the MPTA, PSR~J1909$-$3744. 

\subsection{White noise}

The process of calculating arrival times induces uncertainties in the measurements that are expected to exist in all pulsar timing data sets. To correct for these, white noise processes are commonly assigned to every observing backend and frequency band in use by the PTA. At the sensitivity to which we observe pulsars, we suspect this is not required. As such, we have assessed each pulsar for the requirement of both $\mathrm{E_{Q}}$ and $\mathrm{E_{C}}$ in our data. $\mathrm{E_{F}}$, which is expected to be close to unity when uncertainties on the TOAs are estimated reasonably, was included for each pulsar in the array.

As expected, the values of $\mathrm{E_{F}}$ detected in the MPTA are clustered about a mean of unity ($\overline{\mathrm{E_{F}}} = 1.02$) with a small standard deviation of $0.04$. We discuss any outlying pulsars in Section \ref{Section: Discussion}. We found that roughly a quarter of pulsars ($20/83$) show significant evidence supporting the presence of $\mathrm{E_{Q}}$, and slightly more ($29$/$83$) show evidence for the inclusion of $\mathrm{E_{C}}$. The prevalence of $\mathrm{E_{C}}$ over $\mathrm{E_{Q}}$ reflects the sensitivity of the MeerKAT telescope. This naturally leads to a large number of the pulsars observed by the MPTA that are \textit{jitter limited}, where intrinsic pulse-to-pulse morphology changes become the dominant source of arrival time uncertainty on short time scales. As addressed in Section \ref{Subsection: White Noise}, $\mathrm{E_{C}}$ adjusts for this margin of uncertainty alongside phenomena that manifests similarly in the data such as mode-changing, where the pulse energy distribution of the pulsar is multi-modal. 

While its inclusion is favoured in fewer MPTA pulsars, the distribution of $\mathrm{E_{Q}}$ observed in the MPTA had a mean of $\overline{\mathrm{E_{Q}}} = -6.39 \ \log_{10}(\mathrm{s})$, and a standard deviation of $0.48 \ \log_{10}(\mathrm{s})$. This distribution is similar to what is found for $\mathrm{E_{C}}$, which has a mean $\overline{\mathrm{E_{C}}} = -6.45 \ \log_{10}(\mathrm{s})$, and a standard deviation of $0.35 \ \log_{10}(\mathrm{s})$. The coincidence of the $\mathrm{E_{Q}}$ and $\mathrm{E_{C}}$ distributions suggests that they are modelling similar phenomena. It may be that the continued presence of $\mathrm{E_{Q}}$ in the MPTA sample is, in fact, adjusting for jitter where it is more suited as a global variable correlated across all arrival times, rather than in individual epochs. The small sample of $\mathrm{E_{Q}}$ values that are favoured for inclusion in the noise models leads the origin of this noise in the MPTA data set unclear. Further analysis of the distribution of jitter in the MPTA pulsars is currently underway (Gitika, et al., in prep). A previous assessment of data collected by the MPTA revealed that the largest $\mathrm{E_{C}}$ value is recorded for PSR~J1103$-$5403, a pulsar that is no longer included in MPTA analyses. This is not unexpected, as this pulsar possesses strong evidence for mode-changing behaviour, which is the reason it is now excluded from the data set. Nevertheless, it has been demonstrated that by accounting for this behaviour, the value of $\mathrm{E_{C}}$ can be decreased by a factor of $4.3$ \citep{2023arXiv230402793N}. 

\subsection{Achromatic red noise}

Understanding the distribution of achromatic red noise signals in a PTA is particularly important as an SGWB is often first expected to emerge as one, and the similarity between these processes can lead to the misidentification of one as the other. In the search for a common signal across the array, it is possible that the presence of achromatic noise in many pulsars can converge to a shared process where there is none present \citep{2022MNRAS.516..410Z, 2024ApJS..273...23V}. To mitigate this we have searched for additional achromatic red noise terms when assessing a common signal in the MPTA data, however, it is also useful to understand the intrinsic achromatic noise that is identified by our methodology.

Of the MPTA pulsars, 12 show significant evidence of an achromatic red noise process. The MAP amplitude distribution associated with these spans $-14.19$ to $-11.93$, with a spectral index range of $0.84$ to $3.47$. Some degree of intrinsic achromatic noise is expected in all pulsars, however, the values reported in this work consider only those processes affecting the arrival times to a sufficiently large degree such that they are included via the codified strategy described in Section \ref{Subsection: CBA}.

\subsection{Dispersion measure and scattering noise}

Every pulsar in the MPTA is expected to exhibit a certain level of noise caused by the interaction of radio pulsed radio emission with the IISM. Some of this process is modelled when fitting DM and its temporal derivatives as part of the deterministic timing model. However, the stochastic nature of the IISM can not be captured through this and may require additional modelling. Of the pulsars in the MPTA, $58$ display  DM or scattering variations that require stochastic models. Of these, dispersion measure noise is more prevalent in the MPTA. We note that there is a covariance between the power-law DM variations and dispersion due to the solar wind, which we describe below.

For $10$ pulsars, we observe scattering noise in the pulsar noise model where DM noise is not favoured. This may seem unusual as DM variations are expected to be present in all pulsars, while other chromatic noise processes are thought to be weaker. However, we note that some variations due to DM are accounted for in the deterministic timing model using first and second time derivatives of DM; there are no similar terms in the model to account for the effects of scattering noise. Due to this, the presence of scattering noise, where it strongly perturbs the arrival times, may present more obviously than the noise associated with DM, leading it to be favoured for inclusion in the pulsar noise model where DM noise is not.

\subsection{Solar wind: deterministic and stochastic}

\begin{figure*}
    \includegraphics[width=\linewidth]{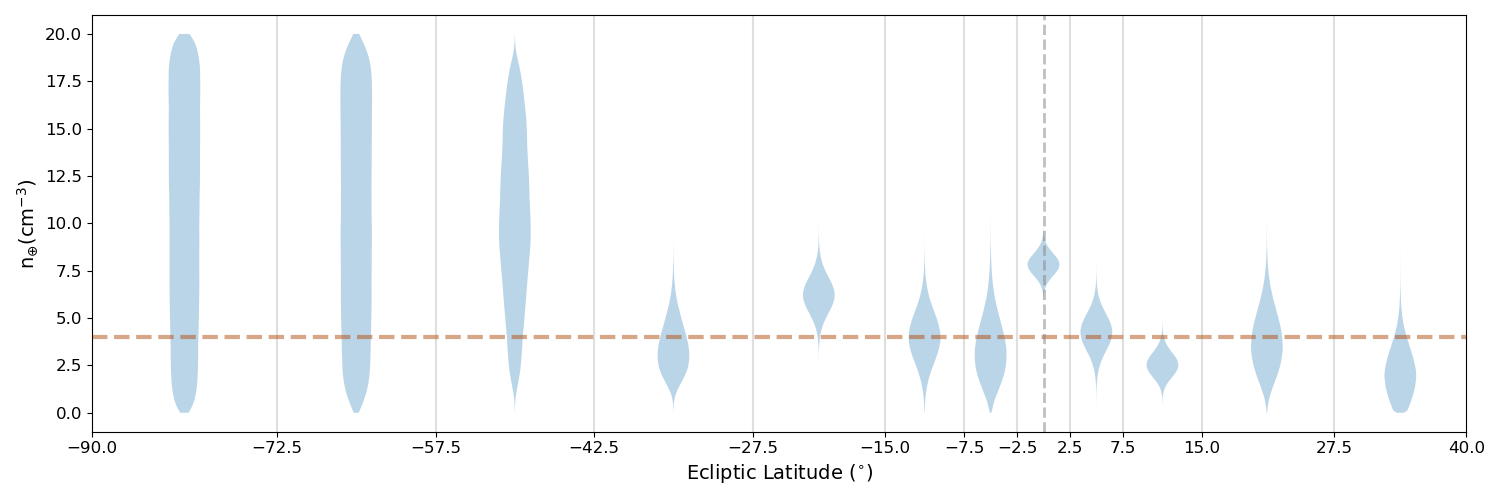}
    \caption[The distribution of mean solar wind density at 1AU ($n_{\oplus}$) for the MPTA as a function of ecliptic latitude.]{Posterior probability densities of $n_{\oplus}$ as a function of ecliptic latitude. The vertical lines separating the posteriors represent the bins of ecliptic latitude that were used to define the pulsars that were to be factorised. The pulsars approaching an ecliptic latitude of $0^{\circ}$ (dashed, vertical) show a clear increase in their derived mean solar wind densities, and are significantly different than the nominal value of $n_{\oplus}$ (brown dashed).}
    \label{fig: ELAT_vs_SW}
\end{figure*}

The majority ($58$) of the pulsars in the MPTA showed a preference for a value of the mean solar wind density at $1$ AU ($n_\oplus$) deviating from the nominal value of $4 \ \mathrm{cm}^{-3}$. This is not necessarily unexpected, as the Sun is in a different solar cycle to when this nominal value was chosen \citep{1998JGR...103.1969I}. Further, the sensitivity of the MeerKAT telescope and the relatively wide bandwidth of the L-band receiver likely make it more sensitive to chromatic processes that may not be as obvious in other data sets. In addition, observations with the MPTA began in proximity to the beginning of a new solar cycle (Solar Cycle $25$), in which case it is not unexpected that we observe an increased level of solar activity over our data span \citep{2020SoPh..295..163M}. Similar to \citet{2023ApJ...951L...7R}, we have included the distribution of $n_{\oplus}$ as a function of ecliptic latitude in Figure \ref{fig: ELAT_vs_SW} and find that the expected solar density is greater where the ecliptic latitude is low.

The stochastic component of the solar wind term is constrained in fewer pulsars than the deterministic counterpart. The degeneracy between the stochastic solar wind components, dispersion measure noise, and, to a lesser extent, scattering noise, can make it difficult to identify in many cases. Even so, the inclusion of this term is favoured in $25$ pulsars.

\subsection{Other deterministic models}

There were a set of $23$ pulsars that favour the inclusion of the additional deterministic models described in Section \ref{Subsection: Other Deterministic Models}. The parameter estimates constraining these processes are detailed in Table \ref{Table: MPTA determinstic models}. Of these, we observe that $15$ show support for a model accounting for a chromatic Gaussian event in their timing residuals, and another eight favour the inclusion of a deterministic waveform accounting for annual chromatic variations. No pulsars favour the inclusion of both processes. For two pulsars, PSR~J0610$-$2100 and PSR~J1902$-$5105, the values we report are taken from the CURN Bayesian analysis. We do this due to a marked increase in the precision constraint of the posterior during this step of the analysis.

The amplitude distribution of the chromatic Gaussian events ranges from $-7.68$ to $-5.30$ $\log_{10}(\mathrm{s)}$, with the upper limit corresponding to a deviation on the order of $\sim 5 \mu$s. The chromatic index constraint is far broader, ranging from $0.77$ to $8.95$. The annual chromatic variations have similar constraints in amplitude, ranging from $-8.96$ to $-5.48$. The smallest of these, corresponding to PSR~J0955$-$6150, possesses among the weakest constraints of the distribution, suggesting that it may be an artefact of chromatic time delays that are not as well suited to the strong DM process it possesses. The constraints on the chromatic index for these signals is not as varied as that observed in the chromatic Gaussian events, possessing MAP values between $0.91$ and $5.11$ with broad posterior distributions.

\subsection{A common uncorrelated red noise source}

Establishing fiducial noise models of the MPTA pulsars allowed us to explore the presence of noise processes common to the MPTA. In particular, we searched for an achromatic red noise process common to the pulsars as would be expected of a signal stemming from an SGWB, the aforementioned CURN. While only 12 of the pulsars in the MPTA display significant evidence for the inclusion of achromatic red noise into their fiducial noise model, this term is included in all pulsars when searching for a common spectrum process. The approach to model selection we have implemented will determine the most likely processes present in the data, but will miss sub-threshold terms. Given that the common spectrum process originating from an SGWB is thought to be achromatic, these additional noise terms are included in the model to be conservative and to reduce the risk of misidentifying sub-threshold intrinsic achromatic red noise as a common process instead. 

\begin{figure}
\centering
    \includegraphics[width=\columnwidth]{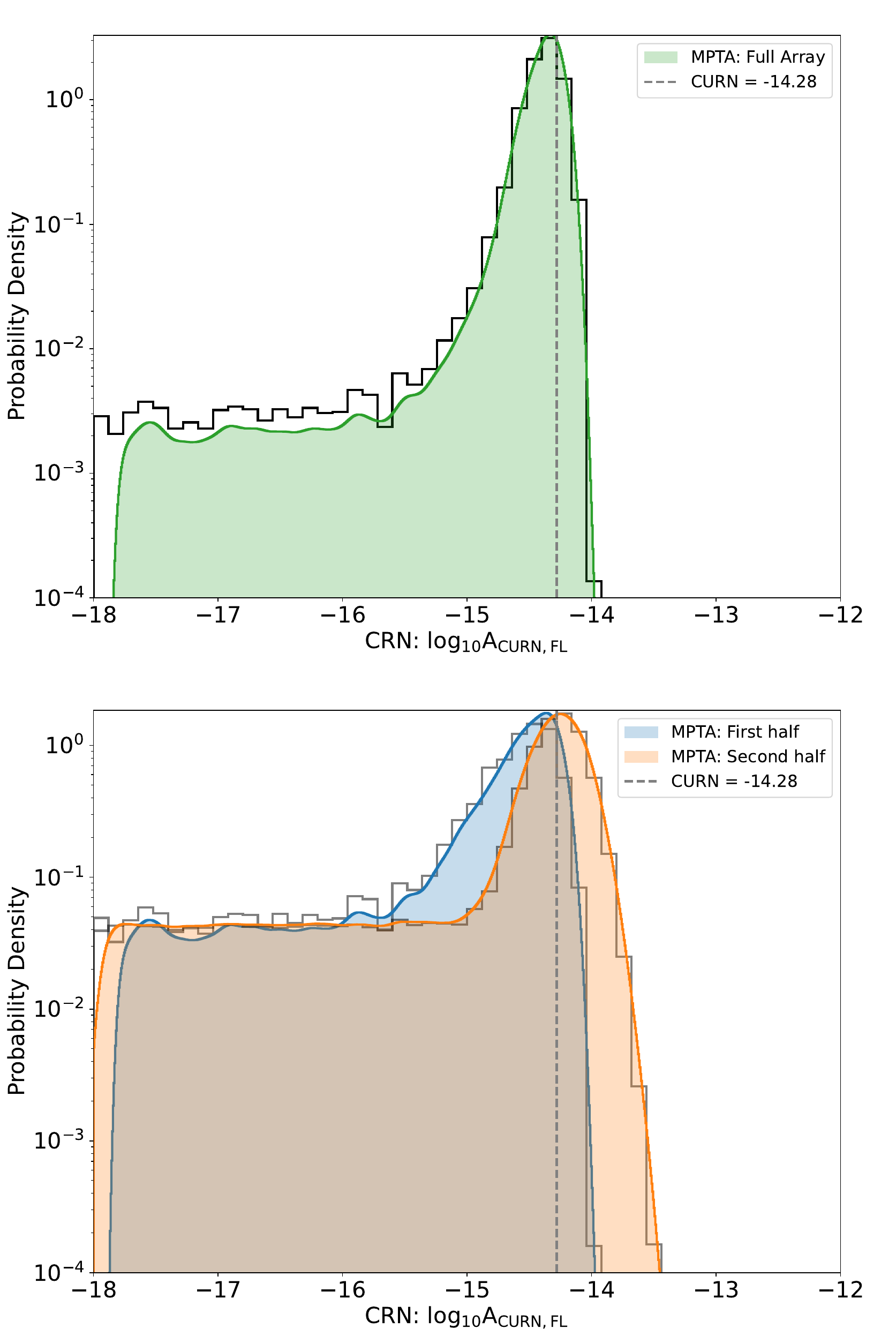}
    \caption{Factorised posterior product of the search for $\log_{10}\mathrm{A_{CURN}}$ at a fixed spectral index in individual MPTA pulsars. (Top) The probability density of the direct product of the full MPTA pulsar posteriors is provided (black line), with a kernel density estimate of the data also supplied (green shaded). To mitigate the chance of noise in the data influencing this result, we take the median and $1\sigma$ quantiles of the kernel density estimate as our reported value and use this same distribution to calculate the $\ln(\mathcal{B})$. (Bottom) The recovery of this signal in two halves of the MPTA, randomly split where no pulsar is in both halves. While the significance of the signal is lower in each individual half of the MPTA, the recovered amplitudes are consistent.}
    \label{fig: CRN_FL}
\end{figure}

\begin{figure*}
    \centering
    \includegraphics[width=\textwidth]{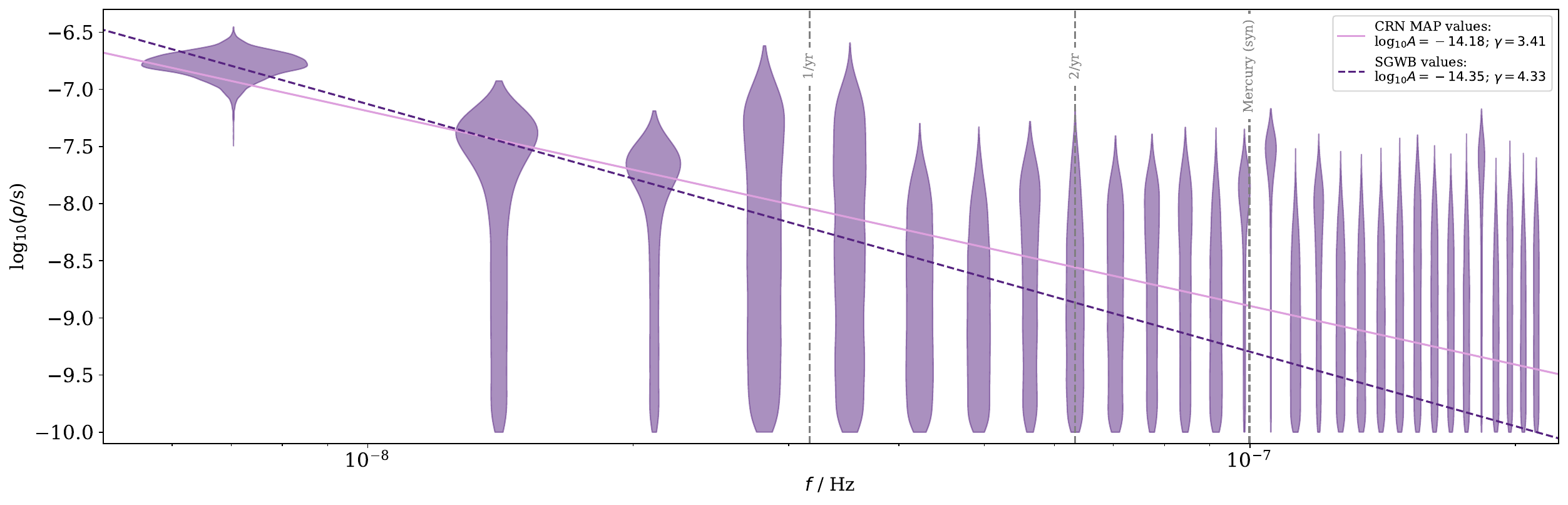}
    \caption{Free-spectrum measurement of common uncorrelated noise in the MPTA data. The amplitude of the common spectrum was sampled for $30$ frequencies ranging from 1/Tspan ($\sim 7.04$ nHz) to 30/Tspan ($\sim 211$ nHz). The violins show the posterior probability densities for each of the amplitudes sampled at these frequencies, of which only the first two are well constrained. The pink line overlaid on top of the spectrum represents the MAP parameter values recovered from the CURN Bayesian analysis, whereas the dashed purple line corresponds to the MAP parameter value taken from a small region of the posterior where $\gamma$ was close to $13/3$.}
    \label{fig: Full_PTA_free_spectrum}
\end{figure*}

\begin{figure}
    \includegraphics[width=\columnwidth]{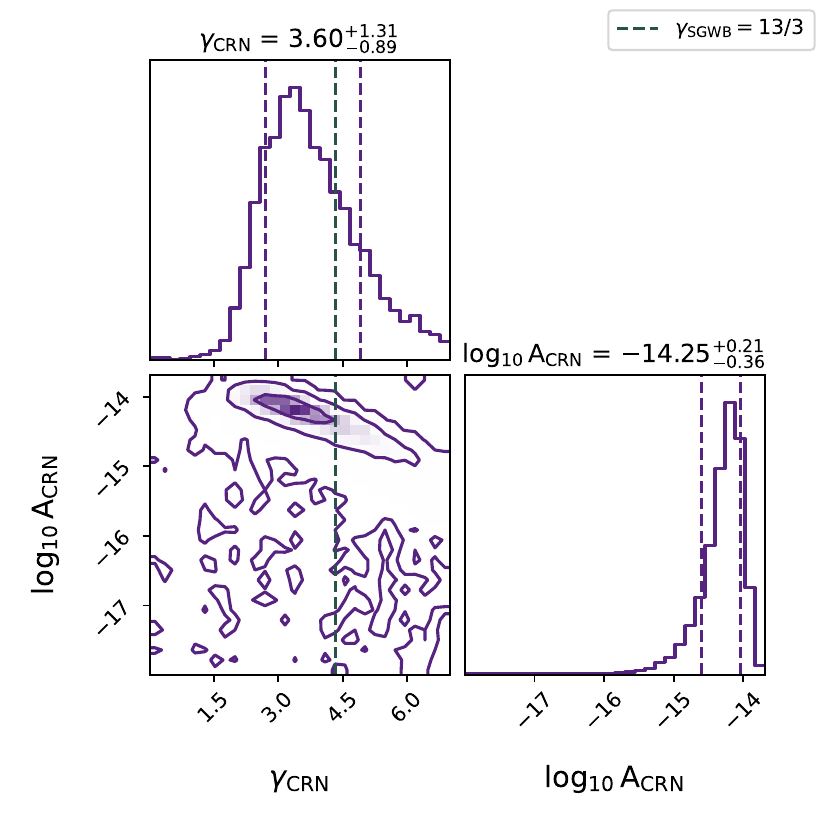}
    \caption{The two-dimensional marginal posterior distribution for the log-amplitude ($\log_{10}\mathrm{A_{CURN}}$) and spectral index ($\gamma_\mathrm{CURN}$) of the common uncorrelated signal identified in the MPTA data. The contours represent $1\sigma$, $2\sigma$ and $3\sigma$ confidence regions of the posteriors, and the values reported above each one-dimensional posterior are the median and corresponding $1\sigma$ values of the signal parameters. The spectral index of the process is consistent with a value representative of an SGWB, which we have overlaid for ease of comparison (green dashed line).}
    \label{fig: CRN_full_PTA}
\end{figure}

We found that there exists a common signal identifiable both through factorising the likelihood (Figure \ref{fig: CRN_FL}) of the MPTA pulsars and through a full PTA likelihood analysis (Figure \ref{fig: CRN_full_PTA}). Holding the spectral index fixed at $\gamma_\mathrm{CURN}=13/3$ during the factorised likelihood analysis, the common signal amplitude of the process is $\log_{10}\mathrm{A_{CURN}}=-14.28^{+0.21}_{-0.21}$. To check whether the presence of the signal is constrained to any particular set of pulsars, we also assess its presence by randomly splitting the array into two halves. We find that the signal remains present in both halves at a consistent amplitude, albeit to a lesser significance, which we show in the bottom panel of Figure \ref{fig: CRN_FL}. This amplitude is consistent with that found when we allowed the spectral index to vary during the full PTA likelihood analysis of $\log_{10}\mathrm{A_{CURN}}=-14.25^{+0.21}_{-0.21}$, with an associated constraint on the spectral index of $\gamma_\mathrm{CURN}=3.60^{+1.31}_{-0.89}$. To assess the spectral properties of the common noise we formed the free spectrum \citep{2013PhRvD..87j4021L}, in which the properties of a common process are measured at independent harmonically related sinusoids, shown in Figure \ref{fig: Full_PTA_free_spectrum}. It is apparent that the constraint on the spectral index is dominated by the first two frequency bins, of which the lowest frequency equates to approximately $1/\mathrm{T} \sim 7.04 \ \mathrm{nHz}$, with less power in higher frequency bins. 

To determine the significance of the detection of common red noise, we use the Savage-Dickey density ratio to calculate the Bayes factor. This was calculated for the factorised likelihood analysis by measuring the posterior probability distribution below a point where the prior range was clearly disfavoured ($p(\log_{10}\mathrm{A_{CURN, FL}} < -16.5)$), and taking the ratio of the probability and the prior density in that region ($\pi(\log_{10}\mathrm{A_{CURN, FL}} < -16.5)$), such that
\begin{equation}
    \mathcal{B}_\mathrm{CURN, FL} = \frac{\pi(\log_{10}\mathrm{A_{CURN, FL}} < -16.5)}{p(\log_{10}\mathrm{A_{CURN, FL}} < -16.5)}.
\end{equation}
Evaluating this by taking the average probability distribution below this region results in a Bayes factor of $\ln(\mathcal{B}) = 4.46$ in favour of a CURN. Assessing the results of the full PTA likelihood analysis in a similar fashion, but also allowing the spectral index of the process to vary, results in a Bayes factor of $\ln(\mathcal{B}) = 3.17$ in favour of CURN. While both results are significant, the Bayes factor when assessing the full PTA analysis is lower, likely stemming from a poorer constraint on the spectral index. This is not unexpected as, due to the short timescale of the MPTA data, a constrained posterior can only be achieved in two of the frequencies that we observe (Figure \ref{fig: Full_PTA_free_spectrum}).

\section{Discussion}
\label{Section: Discussion}

\subsection{Unusual white noise} 

While most pulsars show values of $\mathrm{E_{F}}$ consistent with (or close to) unity, one departs with a significance $>2\sigma$:  PSR~J0437$-$4715. This is the brightest MSP, and coupled with MeerKAT's relative sensitivity, the pulsar is expected to be entirely limited by jitter noise \citep{2021MNRAS.502..407P}. It may be possible that the larger value of $\mathrm{E_{F}}$ is adjusting the formal ToA uncertainties to account for aspects of jitter noise that are difficult to capture with a single $\mathrm{E_{C}}$ process (e.g. \citet{2024MNRAS.528.3658K}), or simply that the high signal-to-noise ratio observations of the pulsar are leading to underestimated errors during the timing process.

\subsection{Achromatic noise}


The spectral shape of spin or timing noise in slow pulsars has been observed to be $\gamma_\mathrm{Red} \sim 4-6$ \citep{2010arXiv1010.3785C}, a statistic that is at odds with what is observed by the MPTA. Rather, the distribution of this in the MPTA is better described as $\gamma_\mathrm{Red} \sim 1.5-3.5$ for ten of the twelve pulsars in the sample. Of the pulsars that do not fit this distribution, PSR~J1017$-$7156 and PSR~J2236$-$5527, both are shallower. In comparison to other PTA datasets, the achromatic noise that has been reported here agrees within uncertainties for all that have been reported by other PTAs, with the only exceptions being PSR~J1801$-$1417 and PSR~J2234$+$0944, as identified by the EPTA \citep{2023A&A...678A..49E} and NANOGrav \citep{2023ApJ...951L..10A}, respectively. 

We find evidence for the presence of a weak achromatic red noise in PSR~J1801$-$1417, in addition to a DM noise process of a similar amplitude and spectral index. The EPTA also reports the presence of a DM noise process, however the amplitude of this process is inconsistent with our measurements. The coincidence of the constraints on the achromatic red noise and DM noise in our data set may imply that the process is better described by only one of these. The pulsar has a moderate nominal DM ($57.26\, \mathrm{pc\, cm^{-3}}$), indicating that confusion between these processes is less likely. However, the short data span that the MPTA possesses naturally results in less accurate spectral characterisations. This could lead to an inability to discriminate between noise processes in some pulsar data sets, which may have occurred in this case.

For PSR~J2234$+$0944, we have found evidence for a strong ($\log_{10}\mathrm{A_{red}} = -12.83^{+0.15}_{-0.11}$) achromatic noise process where this has not been reported in other data sets. In the absence of other explanations, we propose this may be due to differing timing model ephemerides. As this is a binary (black widow) pulsar with a low mass non-degenerate companion, the time-correlated variations in the solution can induce noise-like structures in the timing residuals. It is possible that the differences between our solutions may have induced this noise in our data set, or perhaps below the noise in NANOGRAV data. However, PSR~J2234$+$0944 was previously observed by NANOGrav with the sensitive Arecibo telescope, making this less likely.

\subsection{Chromatic noise across the MSP population}

\begin{figure*}
\centering
\begin{subfigure}{0.5\linewidth}
  \centering
  \includegraphics[width=\linewidth]{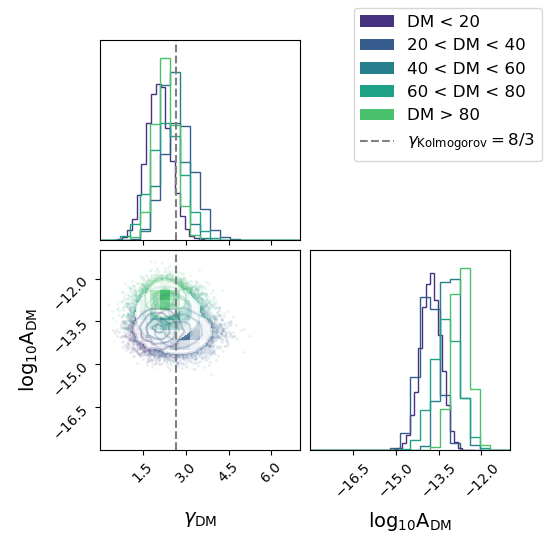}
  \caption{MPTA dispersion measure noise distribution.}
  \label{fig: DM_FL}
\end{subfigure}%
\begin{subfigure}{0.5\linewidth}
  \centering
  \includegraphics[width=\linewidth]{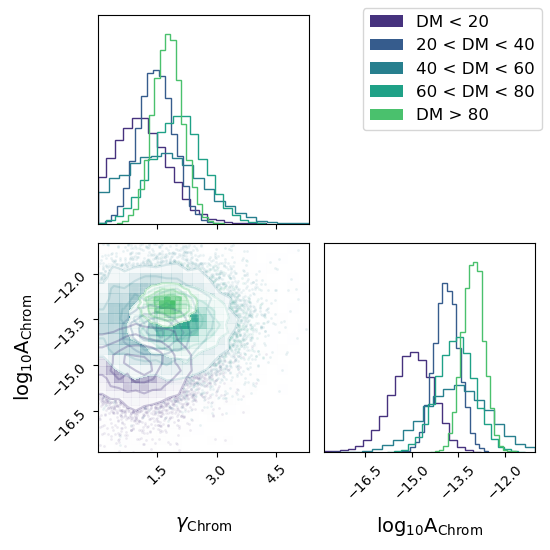}
  \caption{MPTA scattering noise distribution.}
  \label{fig: Chrom_FL}
\end{subfigure}
\caption[The distribution of dispersion measure and scattering noise in the MPTA.]{The distribution of noise processes originating from the IISM in the MPTA data, measured as the factorised likelihood of the processes through bins of characteristic DM. (a) The dispersion measure noise is well distributed about the expected Kolmogorov spectrum (grey, dashed) through all DM measurements, with clear growth in the amplitude of the stochastic process as the DM increases. (b) The scattering noise amplitude also appears to increase as a function of the DM, however, the constraints on the spectral index are much broader.}
\label{fig: Chrom_DM_FL}
\end{figure*}

The power spectral density of dispersion measure noise is nominally expected to follow a Kolmogorov spectrum ($\gamma_\mathrm{DM} \sim 8/3$) \citep{2013MNRAS.429.2161K} for DM variations arising from turbulence-driven density variations in the IISM. Within uncertainties, $27$ out of the $44$ pulsars that show evidence for dispersion measure noise overlap with this value. As a population, the distribution of this process in the MPTA is well constrained at this value, as shown in Figure \ref{fig: DM_FL}. In addition, there is a clear increase in the strength of the stochastic DM variations as a function of the nominal DM of the pulsar. This is not surprising as density variations are expected to be larger as longer paths (with larger DM) are explored in the IISM \citep{2016ApJ...817...16C}.

Most of the pulsars that are not consistent with $\gamma_\mathrm{DM} \sim 8/3$ show spectral indices shallower than that expected for Kolmogorov turbulence. For the majority of these, we noted a strong covariance between dispersion measure noise and other processes expected to vary at a high fluctuation frequency, namely the stochastic solar wind component and $\mathrm{E_{C}}$. Only four pulsars were found to have larger-than-expected spectral indices: PSRs J0613$-$0200, J1125$-$6014, J1721$-$2457, and J1804$-$2858. There is no clear covariance between the noise terms in these pulsars that could result in this, however, the IISM is inhomogeneous and deviations from the expected Kolmogorov turbulence are reasonable to observe in a large enough sample \citep{1990ARA&A..28..561R}.

Scattering noise is observed in $23$ of the MPTA pulsars. Of these, $13$ prefer chromatic indices that differ from $\beta = 4$, the value usually assumed for the scattering of radiation through the IISM. While the spectral indices of these processes do not appear to have any dependence on the measured DM of the pulsar, their amplitudes appear to strongly correlate in a similar manner to the stochastic DM process, as displayed in Figure \ref{fig: Chrom_FL}. The arrival time delays of PSR~J0437$-$4715 and PSR~J1643$-$1224 scale with frequency at $\beta > 6.4$, taking into account the corresponding posterior uncertainties. This is larger than expected, and likely indicates complicated scattering geometries in the IISM along the line of sight to the pulsar or could be related to refractive modulation of pulse broadening \citep[Reardon et al. 2024, in preparation]{2017MNRAS.464.2075S}. 

\begin{figure*}
\centering
\begin{subfigure}{0.5\linewidth}
  \centering
  \includegraphics[width=\linewidth]{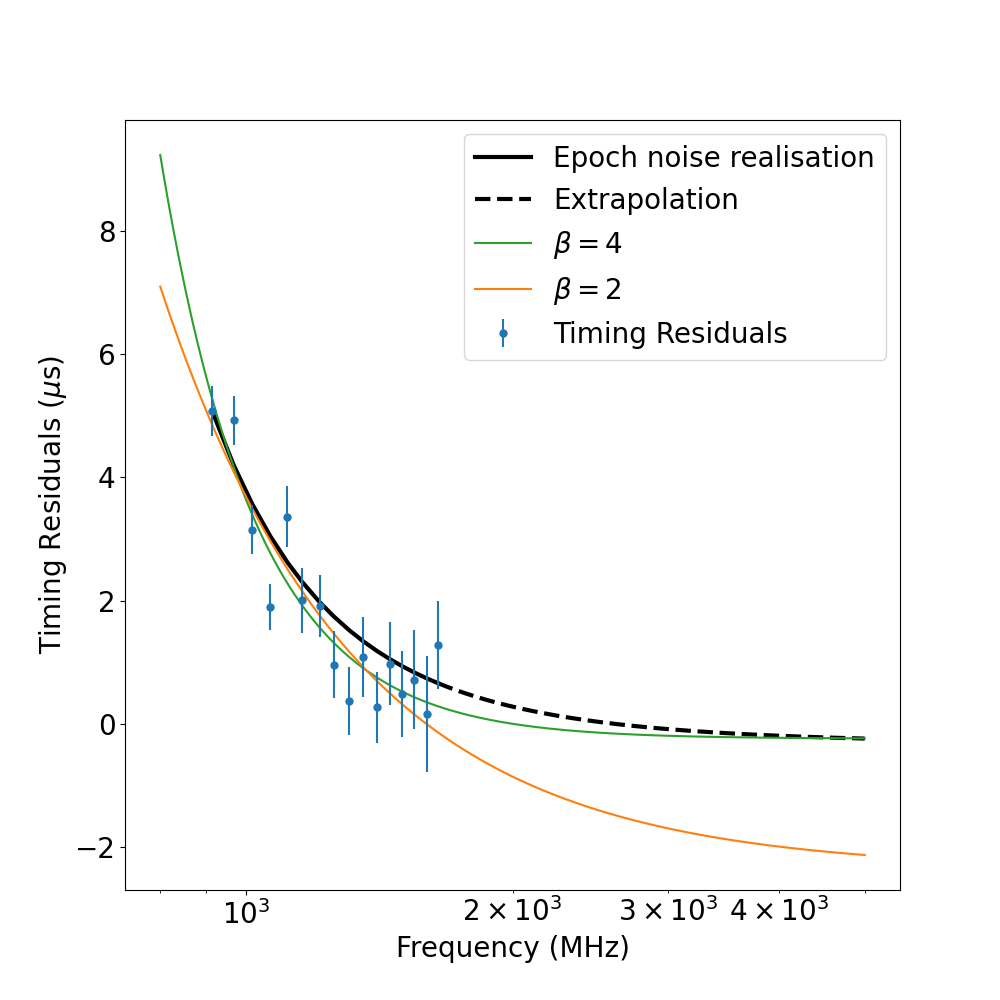}
  \captionsetup{width=.8\linewidth}
  \caption{Models fit to an observing epoch scattered primarily by a chromatic scattering process.}
  \label{fig: chrom_scatter}
\end{subfigure}%
\begin{subfigure}{0.5\linewidth}
  \centering
  \includegraphics[width=\linewidth]{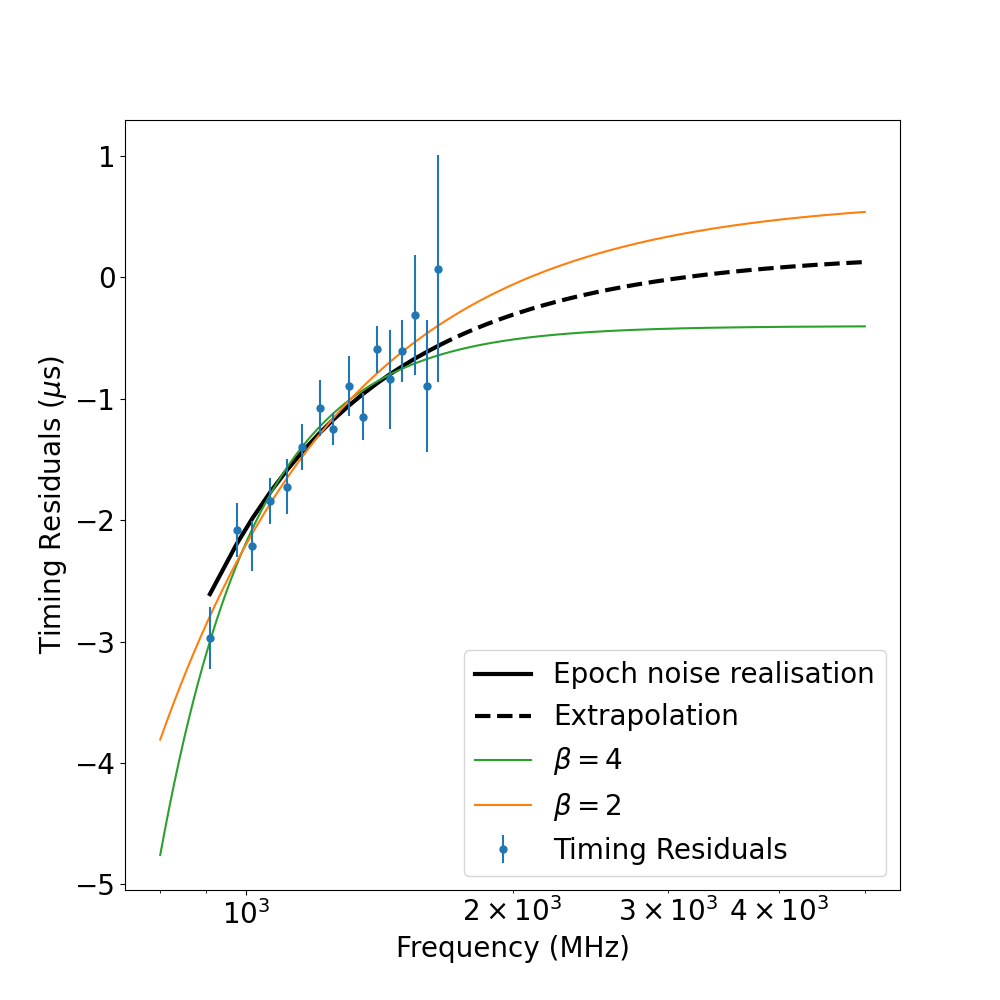}
  \captionsetup{width=.8\linewidth}
  \caption{Models fit to an observing epoch scattered primarily by a dispersion measure process.}
  \label{fig: DM_scatter}
\end{subfigure}
\caption{Comparison of deterministic and stochastic fits to noise processes observed in the MPTA for PSR~J1017$-$7156. Two power-law models with chromatic indices associated with scattering (green) and dispersion measure (orange) delays are overlaid on the timing residuals of two epochs of PSR~J1017$-$7156 (blue). Also included are \textsc{pint} realisation of the pulsar's noise processes (black) for these observing epochs, the parameters of which can be found in Table \ref{Table: MPTA noise models}. Each model has been extended through frequency to highlight how each process performs under an infinite frequency assumption. In panel (a) the $\beta=4$ model performs well under the infinite frequency assumption, implying this level of chromaticity is well-suited to model the scatter. In panel (b) the residuals associated with both power-law models do not approach $0\ \mu$s, implying both models are not well suited to account for this degree of arrival time scatter. However, the noise process that is realised by \textsc{pint} is able to capture this well, demonstrating the requirement for novel noise modelling techniques.}
\label{fig: scattered_arrival_times}
\end{figure*}

The effect of chromatic scattering as a function of frequency can be observed directly in the timing residuals. In Figure \ref{fig: scattered_arrival_times} we show two observing epochs of PSR~J1017$-$7156 alongside models of chromatic dispersion. Of the two epochs shown here, one is likely dominated by a scattering process (Figure \ref{fig: chrom_scatter}), and the other by a DM or solar wind process (Figure \ref{fig: DM_scatter}). To demonstrate the need for appropriate noise modelling of these processes, we extrapolate these processes to demonstrate their action as they approach infinite frequency. Both the power-law model associated with $\beta=4$ and the realised noise process for the epoch displayed in Figure \ref{fig: chrom_scatter} trend to $0\ \mu$s as they approach high frequencies, implying they are appropriate models of scatter broadening. Figure \ref{fig: DM_scatter} demonstrates that this is not always the case, revealing that the only model that trends towards $0\ \mu$s (as would be expected) is the realisation of the advanced noise model.

The dispersion measure noise we have observed in the MPTA is consistent in amplitude and spectral index for most pulsars that are also observed by the EPTA and the PPTA, the other PTAs that model chromatic variations as power-law Gaussian processes. However, there exist marginal differences between these realisations. For example, the PPTA report a different spectral index for DM noise for PSR~J1045$-$4509. They also report scattering noise and band noise in their data likely leading to this inconsistency. Given that the noise is not characterised in an identical fashion, across the same frequency range, with the same data products, at the same time, or over the same observing spans, these differences are not unexpected. In addition to these factors, the EPTA does not model the solar wind effects in their data in the same fashion as the MPTA. The strong covariance between the solar wind and the dispersion measure leads us to believe the differences between these processes for the pulsars we have in common are primarily due to our modelling techniques. An example of this is PSR~J1022$+$1001, in which we have identified a strong stochastic solar wind process ($n_{\oplus} = 10.63^{+1.38}_{-0.68}$, but is reported by the EPTA to possess a dispersion measure process with a shallow spectral index ($\gamma_\mathrm{DM} = 0.14$). Similar, albeit less significant, discrepancies are observed in comparison to the PPTA. The PPTA does not include stochastic variations in their solar wind models to the same extent as this analysis, and their observations are potentially more sensitive to achromatic red noise processes that are only obvious in longer data sets than the one used in this analysis. The combination of these factors is likely to influence the processes identified in the PPTA and the MPTA data sets.

\subsection{Unusual chromatic noise}

The noise processes identified in the MPTA are particularly complex. By using our codified model selection technique we have identified that almost all of the pulsars possess at least one chromatic noise process, some of which do not yet have satisfactory explanations. In particular, the noise analysis of PSRs J0437$-$4715 and J1643$-$1224 revealed that they prefer a chromatic indices of ${7.95}_{-0.67}^{+1.41}$ and ${8.83}_{-1.15}^{+1.96}$ respectively. These are unusual as the steepest predicted chromatic process has an index of $\beta=6.4$ \citep{2017MNRAS.464.2075S}. 

It is unclear if these processes are physical or related to artefacts or systematic errors. If the processes were physical, they would represent variations in the pulse arrival times at the lowest frequency of our observations on the order of $\sim 800$ times greater than that at the highest frequency. The PPTA, which observes Southern declination pulsars at far lower frequencies, would be ideally suited to assist in constraining this. One of the pulsars, PSR~J1643$-$1224, possesses a moderate DM of $62.4$ pc\,cm$^{-3}$, and is known to have unusual chromatic noise \citep{2017MNRAS.466.3706L}, which this measurement may lead insight into. However, PSR~J0437$-$4715 possesses the lowest DM in the array, leading us to consider the possibility that the processes are a consequence of the frequency-resolved portraits created to time them. Future work, including comparison and combination of the data sets with those obtained at other telescopes, is needed to conclusively determine the origin of the noise. 

\subsection{Impacts of noise misspecification}

The computational expense of PTA analyses often requires trade-offs between efficiency and completeness. One of the ways that some PTAs achieve this is to use analytic measurements of the IISM to account for dispersion measure, performed by using a piecewise model for dispersion measure variations (DMX) \citep{2013MNRAS.429.2161K}. The disadvantages of this in terms of sensitivity to an SGWB were briefly explored in Section \ref{Subsection: DM Noise}; however, there are additional flow-through effects that can occur from approximating a stochastic process in this way. 

\begin{figure}
    \includegraphics[width=\columnwidth]{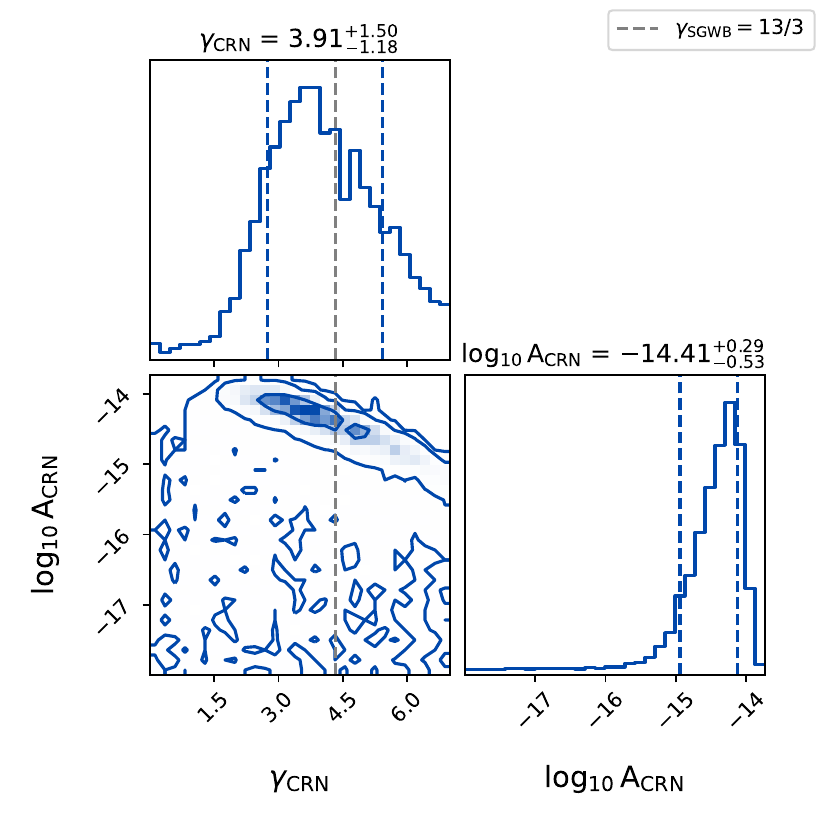}
    \caption[Two-dimensional marginal posterior distribution of the amplitude and spectral index of a common uncorrelated red noise process with intentionally misspecified intrinsic noise processes.]{The two-dimensional marginal posterior distribution for the log-amplitude ($\log_{10}\mathrm{A_{CURN}}$) and spectral index ($\gamma_\mathrm{CURN}$) of a common uncorrelated signal with intentionally misspecified pulsar intrinsic noise processes. The contours represent $1\sigma$, $2\sigma$ and $3\sigma$ confidence regions of the posteriors, and the values reported above each one-dimensional posterior are the median and corresponding $1\sigma$ values of the signal parameters. While the amplitude is constrained at an approximately similar value to that shown in Figure \ref{fig: CRN_full_PTA}, the posterior constraint is broader than that achieved by the detailed noise model.}
    \label{fig: Misspec_full_PTA}
\end{figure}

By analytically modelling the DM and scattering noise processes, the covariance between the chromatic and achromatic processes are not taken into account. Ultimately, this may cause residual noise in the data to be assigned to other processes. The effect of this has been observed in other PTAs \citep{2023ApJ...951L...8A, 2023ApJ...951L...6R}, and has also been observed in this analysis where deliberately misspecified noise processes, modelling only dispersion measure and achromatic red noise, are used to search for a CURN (Figure \ref{fig: Misspec_full_PTA}). 

As other PTA collaborations have noted, employing more detailed noise modelling has the effect of changing both the recovered amplitude and spectral index of a CURN. Even on the relatively short timescale that is available to the MPTA, we also note that this is the case. Properly determining the noise budget of the pulsars in an array also importantly improves the significance of the signal recovery. When we compared the MPTA detailed noise recovery to an example where the noise is deliberately misspecified (assuming only DM and achromatic red noise for each pulsar), we found the detailed models were able to recover the signal in a full PTA analysis at a significance of $\ln({\mathcal{B}}) = 3.17$, whereas the misspecified models could only recover it to a significance of $\ln({\mathcal{B}}) = 1.80$. If the CURN detected in PTA data sets is of an SGWB, then not only could the spectral properties of the background be incorrectly characterised through improper modelling, the significance to which it is detected may be strongly impacted, highlighting the importance of correctly characterising the noise processes in a PTA data set. We thus recommend approaches like the use of the codified Bayesian Analysis we have presented in Section \ref{Subsection: CBA} as a conservative and useful methodology for future noise analyses.

As a further demonstration of the importance of appropriately modelling noise processes, we analyse the sensitivity of a pulsar that was misspecified in a previous work by the MPTA, PSR~J1747$-$4036. Previously, this pulsar had been identified as showing achromatic red and dispersion measure noise \citep{2023MNRAS.519.3976M}. Following the noise analysis in this work, we have found that the pulsar also shows evidence for scattering noise, as well as a large value corresponding to the mean solar wind density at $1$ AU.
To illustrate the importance of the correct noise model, we assess the sensitivity of the pulsar to an SGWB under two scenarios, searching for an achromatic noise term with a characteristic SGWB spectral index for both models. We find an SGWB constrained at median and $1\sigma$ values of $\log_{10}\mathrm{A}=-12.58^{+0.45}_{-5.01}$ in the case of the misspecified model. In contrast, using the properly specified model, we recover a value of $\log_{10}\mathrm{A}=-14.03^{+1.59}_{-3.77}$. Directly comparing the preference of each model to the inclusion of an SGWB signal, we find a Bayes factor of $\mathcal{B}=36.8$ in favour of the misspecified model. If these models were used in a search for a common signal in the data, the larger value and relative support found using the misspecified model would influence the result.


\subsection{A common uncorrelated red noise process}

The common process identified in the MPTA data is consistent with predictions of an SGWB. The spectral index ($\gamma_\mathrm{CURN}=3.60^{+1.31}_{-0.89}$), while wide, is consistent at $1\sigma$ with the $\gamma=13/3$ spectral index expected of an SGWB from binary supermassive black holes inspiralling due to GW emission exclusively \citep{2001astro.ph..8028P}. Given the similarity in the datasets, the CURN in the MPTA data is likely of the same origin as other PTAs. However, the signal we have found has a larger amplitude. It is unclear if this is physical, or an artefact of the short time span of the MPTA resulting in difficulties resolving the spectral properties of the noise. 

In direct comparison to the results of other PTA experiments, the amplitude recovered by the MPTA is inconsistent within the reported uncertainties of the most recent findings \citep{2023ApJ...951L...8A, 2023arXiv230616214A, 2023ApJ...951L...6R, 2023RAA....23g5024X}. The degree of this inconsistency varies between different PTA data sets. Assuming a fixed spectral index, the EPTA recovers a signal possessing a log-amplitude of $-14.60^{+0.11}_{-0.14}$, the PPTA at $-14.69^{+0.05}_{-0.05}$, and NANOGrav at $-14.62^{+0.11}_{-0.12}$. In comparison with our own signal, recovered at an amplitude of $-14.28^{+0.21}_{-0.21}$, the most optimistic comparison we are able to make is to the EPTA result, culminating in a deviation of this signal from the results of other PTAs at a minimum of $1.35\sigma$. 

A recent analysis by the PPTA \citep{2023ApJ...951L...6R} has shown evidence of an apparent growth in the amplitude of the CURN in their data set, implying a non-stationarity in the common signal they detect. Additionally, there exists some evidence of this in analysis done by the EPTA when comparisons are performed between their datasets \citep{2023arXiv230616214A}. If this is physical, it would follow that the reported amplitude of the MPTA is further evidence of this growth, as our data set uses more recent data and has little overlap from those reported by most other PTA experiments. The CPTA undertook a search for an SGWB and CURN with an overlapping (but shorter) data set than ours. The amplitude and spectral index from this search are poorly constrained and are consistent with both our measurement and previous measurements of the CURN by other collaborations.




Assuming that the signal we have recovered is attributed to an SGWB, we can predict the MPTA sensitivity to angular correlations from an SGWB. We do this by using the \textsc{hasasia} \citep{2019JOSS....4.1775H} software package, which can be used to estimate the sensitivity of the MPTA as an SGWB, combining the sensitivities of each individual pulsar in the array. It calculates these over a gravitational-wave frequency range defined by the observation span of the PTA, marginalising over the individual pulsar timing models in conjunction with the noise properties of the pulsar. Doing this achieves an inverse-noise-weighted transmission function, from which the individual pulsar sensitivity can be calculated and subsequently combined. 

\begin{figure*}
    \includegraphics[width=\linewidth]{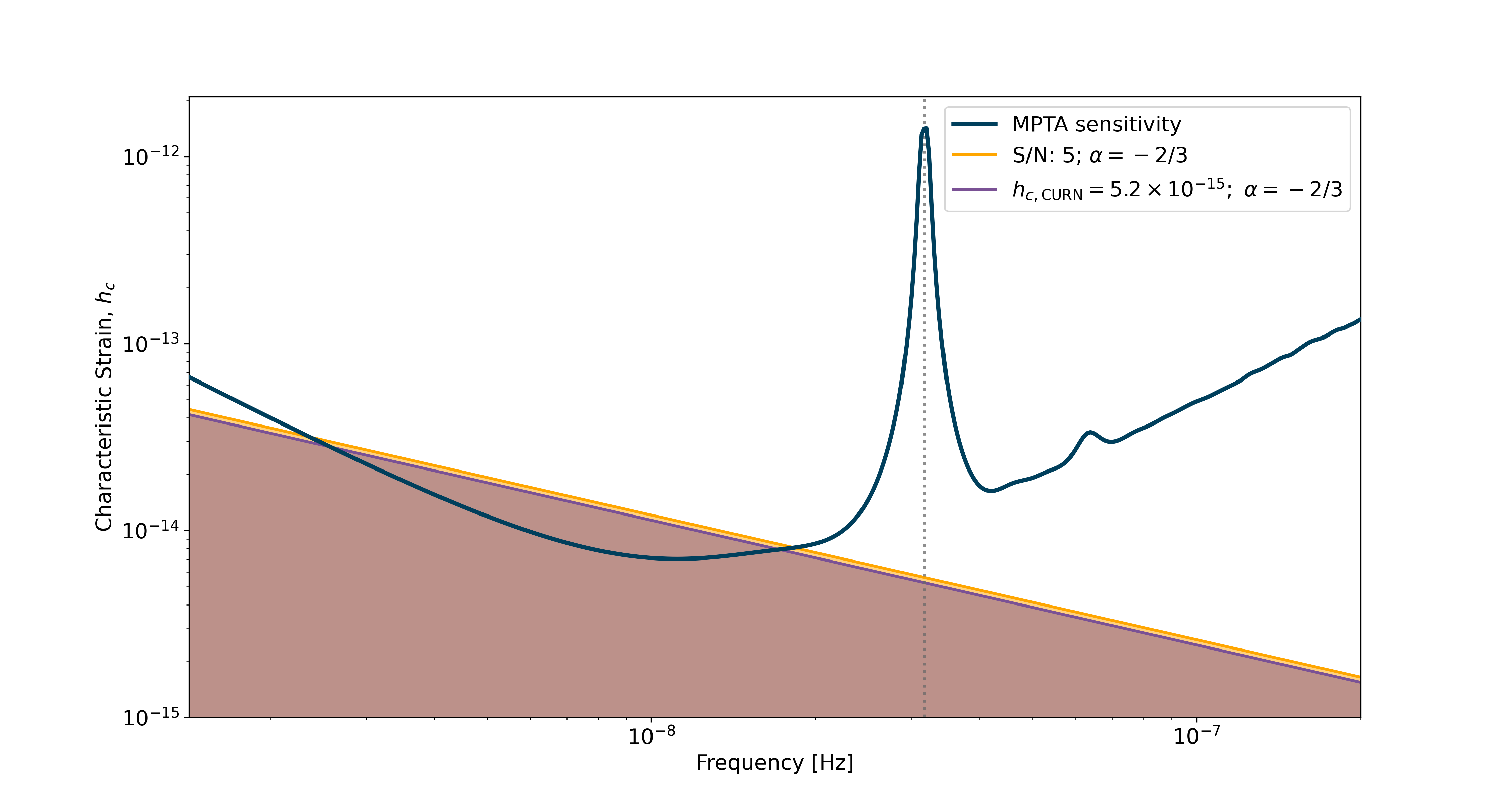}
    \caption[A prediction of the MPTA sensitivity to an SGWB.]{The sensitivity of the MPTA to an SGWB. By taking into account the noise models that have been determined in this analysis, we can estimate the sensitivity of the MPTA to an SGWB (dark blue). The orange-shaded region represents the amplitude that an SGWB would need to reach to achieve an optimal statistic S/N of $5$, whereas the purple-shaded region represents the amplitude that has been detected in the full PTA search for a CURN. The large peak in the sensitivity curve corresponds to the frequency associated with the Earth's orbit around the Sun (grey, dashed). From the estimation provided by \textsc{hasasia}, the amplitude that we have recovered in the full PTA analysis is predicted to be equivalent to an optimal statistic S/N of $\sim 4.5$, whereas the S/N of the result obtained whilst holding the spectral index fixed is even larger still.}
    \label{fig: MPTA_sensitivity}
\end{figure*}

The total sensitivity of the MPTA to an SGWB, as calculated by combining the sensitivities of the individual pulsars in the array, is displayed in Figure \ref{fig: MPTA_sensitivity}. The optimal statistic S/N of each pulsar pair can be combined by \textsc{hasasia} to offer a prediction of the significance of a detection at various amplitudes of an SGWB. Overplotted is the strain spectrum of an SGWB that would result in a optimal statistic S/N of $5$. This corresponds to a background with a characteristic strain amplitude of $ A_{\rm yr}  = 5.6 \times 10^{-15}$. We also show the strain spectrum of an SGWB that has an amplitude consistent with the CURN signal we have identified in this work. If an SGWB is responsible for the CURN, it should also be detected in spatial correlations at an optimal statistic S/N of $\sim 4.5$.

\section{Conclusion}
\label{Section: Conclusion}

We have presented a detailed noise analysis of the first $4.5$ years of MPTA observations. Through our codified Bayesian analysis, we found that the pulsars in our data set prefer noise processes that are not commonly considered or included as standard practice in PTA analyses. Additionally, a surprising number of pulsars disfavour white noise terms that until now have always been included in PTA analyses. Through the use of the preferred noise models, we present the first evidence for a common uncorrelated noise process in the MPTA data set. We have assessed its similarity to common processes identified in other PTAs, and found that while the spectral index is coincident with PTAs that employ detailed noise analyses, the amplitude of this process is larger than those found in other PTAs by at least $1.4\sigma$. While this is both exciting and unusual, the possibility remains that this could stem from the corruption of the signal by the intrinsic pulsar noise processes rather than as a characteristic of an SGWB. We provide an estimation of the MPTA sensitivity to an SGWB signal based on the noise budget determined in this work, from which we forecast the detection significance of the CURN recovered in this work if it is a signal of an SGWB. 

\section*{Acknowledgements}

The MeerKAT telescope is operated by the South African Radio Astronomy Observatory (SARAO), which is a facility of the National Research Foundation, an agency of the Department of Science and Innovation. SARAO acknowledges the ongoing advice and calibration of GPS systems by the National Metrology Institute of South Africa (NMISA) and the time space reference systems department department of the Paris Observatory.
MeerTime data is stored and processed on the OzStar and Ngarrgu Tindebeek supercomputers, operated by the Swinburne University of Technology.
This work was undertaken as part of the Australian Research Council Centre of Excellence for Gravitational Wave Discovery (CE170100004 and CE230100016).  RMS acknowledges support through ARC Future Fellowship FT190100155.
GT acknowledges financial support from “Programme National de Cosmologie and Galaxies” (PNCG), “Programme National Hautes Energies” (PNHE) and "Programme National Gravitation, Références, Astronomie, Métrologie" funded by CNRS/INSU-IN2P3-INP, CEA and CNES, France. We acknowledge financial support from Agence Nationale de la Recherche (ANR-18-CE31-0015), France.
MK acknowledges support by the MPG and the CAS-MPG Legacy Programme. Part of the PTUSE hardware was provided by MPIfR.
KG acknowledges continuing valuable support from the Max-Planck society.
KG acknowledges the support from the International Max Planck Research School (IMPRS) for Astronomy and Astrophysics at the Universities of Bonn and Cologne.
V.V.K acknowledges financial support from the European Research Council (ERC) starting grant ""COMPACT"" (Grant agreement number 101078094).
AP acknowledges financial support from the European Research Council (ERC) starting grant 'GIGA' (grant agreement number: 101116134) and through the NWO-I Veni fellowship. 
FA acknowledges that part of the research activities described in this paper were carried out with the contribution of the NextGenerationEU funds within the National Recovery and Resilience Plan (PNRR), Mission 4 - Education and Research, Component 2 - From Research to Business (M4C2), Investment Line 3.1 - Strengthening and creation of Research Infrastructures, Project IR0000034 – “STILES -Strengthening the Italian Leadership in ELT and SKA”.
Pulsar research at Jodrell Bank Centre for Astrophysics is supported by an STFC Consolidated Grant (ST/T000414/1; ST/X001229/1).
JS acknowledges funding from the South African Research Chairs Initiative of the Depart of Science and Technology and the National Research Foundation of South Africa. 
PG acknowledges support through SUT stipend SUPRA.
NKP is funded by the Deutsche Forschungsgemeinschaft (DFG, German Research
Foundation) – Projektnummer PO 2758/1–1, through the Walter–Benjamin programme.
Funding was provided for the PTUSE machines by the Max-Planck-Institut f\"{u}r Radioastronomie (MPlfR), also supported by the MPG-CAS LEGACY programme, Swinburne University of Technology, and the Australian SKA office.
This project utilises the MeerTime data portal, which is supported by Nick Swainston and the ADACS team.
We acknowledge and pay respects to the Elders and Traditional Owners of the land on which the Australian institutions stand, the Bunurong and Wurundjeri Peoples of the Kulin Nation.

\section*{Data Availability}

All data used in this work is available courtesy of AAO Data Central (\url{https://datacentral.org.au/}) at \url{https://doi.org/10.57891/j0vh-5g31}. 
The data provided includes sub-banded ToAs, the full data archives used to construct this data release, and the ephemerides that have been used to perform timing. Also included are the frequency resolved portraits used to calculate the ToAs used for this work.

The archives and portraits are in \textsc{psrfits} file format. The ephemerides are in a standard \texttt{ascii} text file format, and the arrival times are supplied as IFF data.

\onecolumn
\begin{landscape}
\setlength{\tabcolsep}{2.8pt}
{\renewcommand{\arraystretch}{1.75}%
\begin{longtable}[c]{@{}lcccccccccccccc@{}}
\captionsetup{width=0.9\linewidth}
\caption{The noise processes that are included for the MPTA pulsars. We report the MAP values and the $68\%$ confidence interval corresponding to the sampled posterior, where all included terms were sampled simultaneously for each pulsar. In some few cases, the MAP value has fallen outside of the the confidence interval we report. The parameters under the Uncorrelated Noise subheading are EFAC ($\mathrm{E_{F}}$), EQUAD $\mathrm{E_{Q}}$, and $\mathrm{E_{C}}$. Under the time-correlated noise subheading there is the amplitude ($\mathrm{log_{10}A}$) and spectral index ($\gamma$) of the achromatic red noise (Red), dispersion measure noise (DM), scattering noise (Chrom), and solar wind (SW). For pulsars with chromatic noise in their model, the chromatic index ($\beta$) is included, where $\beta$ is $4$ the pulsar favoured a model of chromatic noise with a fixed chromatic index. The amplitude of the fixed spectral index achromatic red noise process we search for alongside the others is included for reference as $\log_{10}\mathrm{A}_{13/3}$. $n_{\oplus}$, the deterministic value of the mean solar wind plasma density at 1 AU, is also presented for each model.}
\label{Table: MPTA noise models}\\
\toprule
\multicolumn{1}{l}{Pulsar} & \multicolumn{3}{c}{Uncorrelated Noise} & \multicolumn{10}{c}{Time-Correlated Noise} & Deterministic            \\* \cmidrule(lr){2-4} \cmidrule(lr){5-14} \cmidrule(l){15-15} 
 &
  $\mathrm{E_{F}}$ &
  $\mathrm{E_{Q}}$ &
  $\mathrm{E_{C}}$ &
  $\mathrm{log}_{10}\mathrm{A_{Red}}$ &
  $\gamma_{\textrm{Red}}$ &
  $\mathrm{log}_{10}\mathrm{A_{DM}}$ &
  $\gamma_{\textrm{DM}}$ &
  $\mathrm{log}_{10}\mathrm{A_{Chrom}}$ &
  $\gamma_{\textrm{Chrom}}$ &
  $\beta$ &
  $\mathrm{log}_{10}\textrm{A}_{\text{SW}}$ &
  $\gamma_{\textrm{SW}}$ &
  $\mathrm{log}_{10}\textrm{A}_{13/3}$ &
  $n_{\oplus} (\mathrm{cm}^{-3})$ \\* \midrule
\endfirsthead
\multicolumn{15}{c}%
{{\bfseries Table \thetable\ continued from previous page}} \\
\toprule
\multicolumn{1}{l}{Pulsar} & \multicolumn{3}{c}{Uncorrelated Noise} & \multicolumn{10}{c}{Time-Correlated Noise} & Deterministic            \\* \cmidrule(lr){2-4} \cmidrule(lr){5-14} \cmidrule(l){15-15} 
 &
  $\mathrm{E_{F}}$ &
  $\mathrm{E_{Q}}$ &
  $\mathrm{E_{C}}$ &
  $\mathrm{log}_{10}\mathrm{A_{Red}}$ &
  $\gamma_{\textrm{Red}}$ &
  $\mathrm{log}_{10}\mathrm{A_{DM}}$ &
  $\gamma_{\textrm{DM}}$ &
  $\mathrm{log}_{10}\mathrm{A_{Chrom}}$ &
  $\gamma_{\textrm{Chrom}}$ &
  $\beta$ &
  $\mathrm{log}_{10}\textrm{A}_{\text{SW}}$ &
  $\gamma_{\textrm{SW}}$ &
  $\mathrm{log}_{10}\textrm{A}_{13/3}$ &
  $n_{\oplus} (\mathrm{cm}^{-3})$ \\* \midrule
\endhead
J0030+0451 & ${1.03}_{-0.02}^{+0.01}$ & ${-6.45}_{-2.93}^{-0.17}$ & ${-6.59}_{-2.66}^{-0.10}$ & - & - & - & - & - & - & - & - & - & ${-16.49}_{-0.87}^{+1.74}$ & ${4.64}_{-1.33}^{+1.09}$ \\
J0101-6422 & ${0.99}_{-0.00}^{+0.03}$ & - & - & - & - & - & - & - & - & - & - & - & ${-16.68}_{-0.72}^{+2.01}$ & 4 \\
J0125-2327 & ${1.04}_{-0.02}^{+0.01}$ & ${-6.99}_{-1.84}^{+0.03}$ & ${-6.77}_{-0.10}^{+0.05}$ & - & - & ${-13.42}_{-0.30}^{+0.06}$ & ${2.72}_{-0.42}^{+1.45}$ & - & - & - & - & - & ${-14.96}_{-2.39}^{+0.36}$ & 4 \\
J0437-4715 & ${1.20}_{-0.01}^{+0.02}$ & - & ${-6.68}_{-0.04}^{+0.02}$ & - & - & ${-13.51}_{-0.06}^{+0.07}$ & ${1.11}_{-0.12}^{+0.19}$ & ${-15.55}_{-0.28}^{+0.24}$ & ${0.41}_{-0.27}^{+0.19}$ & ${7.95}_{-0.67}^{+1.41}$ & - & - & ${-15.86}_{-1.62}^{+0.70}$ & 4 \\
J0610-2100 & ${1.05}_{-0.02}^{+0.01}$ & - & - & - & - & ${-13.01}_{-0.08}^{+0.12}$ & ${1.91}_{-0.31}^{+0.55}$ & - & - & - & - & - & ${-16.81}_{-0.56}^{+2.39}$ & 4 \\
J0613-0200 & ${1.02}_{-0.02}^{+0.02}$ & ${-6.63}_{-0.54}^{+0.03}$ & - & - & - & ${-14.02}_{-0.50}^{+0.50}$ & ${3.73}_{-0.73}^{+2.52}$ & ${-14.32}_{-1.16}^{+0.39}$ & ${1.76}_{-0.42}^{+1.42}$ & ${6.11}_{-1.65}^{+4.29}$ & - & - & ${-15.45}_{-1.93}^{+0.77}$ & ${4.39}_{-1.77}^{+1.89}$ \\
J0614-3329 & ${0.95}_{-0.01}^{+0.02}$ & - & - & - & - & ${-13.46}_{-4.46}^{+0.06}$ & ${2.35}_{-0.90}^{+2.62}$ & - & - & - & ${-5.61}_{-2.72}^{+0.04}$ & ${1.88}_{-1.99}^{+0.92}$ & ${-16.36}_{-0.94}^{+2.15}$ & ${17.90}_{-9.13}^{+0.73}$ \\
J0636-3044 & ${1.04}_{-0.01}^{+0.02}$ & - & - & - & - & ${-13.72}_{-5.25}^{-0.63}$ & ${1.66}_{-0.75}^{+3.88}$ & - & - & - & - & - & ${-16.77}_{-0.55}^{+2.42}$ & 4 \\
J0711-6830 & ${1.03}_{-0.01}^{+0.02}$ & - & - & - & - & - & - & - & - & - & ${-5.77}_{-0.16}^{+0.13}$ & ${1.28}_{-0.18}^{+0.78}$ & ${-14.37}_{-2.85}^{+0.24}$ & ${10.69}_{-7.32}^{+6.21}$ \\
J0900-3144 & ${1.06}_{-0.02}^{+0.01}$ & - & ${-6.04}_{-2.85}^{+0.05}$ & ${-12.29}_{-3.92}^{+0.01}$ & ${2.28}_{-1.15}^{+1.86}$ & - & - & ${-13.28}_{-0.14}^{+0.21}$ & ${1.58}_{-0.19}^{+0.34}$ & ${5.25}_{-0.53}^{+0.99}$ & ${-6.16}_{-3.02}^{+0.46}$ & ${-0.20}_{-1.97}^{+2.46}$ & ${-12.65}_{-3.41}^{+0.19}$ & ${8.50}_{-5.58}^{+7.99}$ \\
J0931-1902 & ${0.94}_{-0.01}^{+0.02}$ & - & - & - & - & - & - & - & - & - & - & - & ${-14.60}_{-2.78}^{-0.20}$ & 4 \\
J0955-6150 & ${1.02}_{-0.01}^{+0.01}$ & ${-6.22}_{-2.75}^{-0.00}$ & - & - & - & ${-12.74}_{-0.08}^{+0.08}$ & ${2.19}_{-0.30}^{+0.34}$ & - & - & - & - & - & ${-14.41}_{-2.80}^{+0.39}$ & 4 \\
J1012-4235 & ${0.95}_{-0.01}^{+0.02}$ & ${-6.22}_{-2.47}^{+0.04}$ & - & - & - & ${-13.10}_{-0.98}^{+0.10}$ & ${3.41}_{-1.06}^{+2.17}$ & - & - & - & ${-5.55}_{-3.23}^{+0.10}$ & ${1.87}_{-3.17}^{+1.06}$ & ${-14.60}_{-2.65}^{+0.37}$ & ${5.51}_{-2.55}^{+11.15}$ \\
J1017-7156 & ${1.10}_{-0.01}^{+0.03}$ & ${-6.86}_{-0.11}^{+0.05}$ & ${-6.68}_{-1.42}^{+0.04}$ & ${-13.28}_{-4.57}^{+0.03}$ & ${1.31}_{-0.08}^{+4.56}$ & - & - & ${-13.42}_{-0.20}^{+0.26}$ & ${1.57}_{-0.30}^{+0.20}$ & ${3.85}_{-0.99}^{+0.43}$ & ${-5.27}_{-2.59}^{+0.04}$ & ${2.20}_{-0.88}^{+0.58}$ & ${-15.77}_{-1.37}^{+1.30}$ & ${8.70}_{-5.95}^{+6.27}$ \\
J1022+1001 & ${1.00}_{-0.01}^{+0.02}$ & - & ${-5.86}_{-0.03}^{+0.06}$ & - & - & ${-13.06}_{-0.19}^{+0.07}$ & ${0.93}_{-0.46}^{+0.18}$ & - & - & - & - & - & ${-14.37}_{-2.91}^{+0.11}$ & ${10.63}_{-0.68}^{+1.38}$ \\
J1024-0719 & ${1.03}_{-0.02}^{+0.01}$ & - & ${-6.86}_{-2.50}^{-0.16}$ & - & - & ${-14.19}_{-3.91}^{+0.19}$ & ${3.61}_{-1.65}^{+2.62}$ & - & - & - & - & - & ${-14.14}_{-2.90}^{+0.13}$ & 4 \\
J1036-8317 & ${1.00}_{-0.02}^{+0.01}$ & - & - & - & - & ${-13.57}_{-4.46}^{+0.19}$ & ${1.72}_{-0.49}^{+3.41}$ & - & - & - & ${-5.75}_{-3.27}^{+0.12}$ & ${2.19}_{-2.65}^{+1.04}$ & ${-13.56}_{-3.39}^{+0.18}$ & ${16.90}_{-11.16}^{+1.07}$ \\
J1045-4509 & ${1.01}_{-0.02}^{+0.02}$ & ${-6.18}_{-2.94}^{-0.01}$ & - & - & - & ${-12.33}_{-0.07}^{+0.08}$ & ${2.24}_{-0.13}^{+0.28}$ & - & - & - & - & - & ${-13.69}_{-3.42}^{-0.10}$ & 4 \\
J1101-6424 & ${0.94}_{-0.01}^{+0.03}$ & ${-5.73}_{-0.11}^{+0.03}$ & - & - & - & ${-12.70}_{-0.15}^{+0.05}$ & ${1.96}_{-0.22}^{+0.78}$ & - & - & - & - & - & ${-13.69}_{-3.32}^{+0.10}$ & 4 \\
J1125-5825 & ${0.93}_{-0.02}^{+0.01}$ & ${-6.18}_{-2.91}^{-0.04}$ & - & - & - & ${-12.74}_{-0.63}^{+0.06}$ & ${2.84}_{-0.30}^{+3.28}$ & - & - & - & ${-5.27}_{-3.38}^{+0.07}$ & ${1.56}_{-2.39}^{+0.57}$ & ${-15.68}_{-1.67}^{+1.22}$ & ${3.51}_{-0.88}^{+12.68}$ \\
J1125-6014 & ${0.97}_{-0.01}^{+0.02}$ & - & ${-6.77}_{-0.10}^{+0.04}$ & - & - & ${-13.19}_{-0.08}^{+0.10}$ & ${4.41}_{-0.65}^{+0.76}$ & - & - & - & - & - & ${-15.32}_{-2.08}^{+0.51}$ & ${17.50}_{-10.66}^{+0.58}$ \\
J1216-6410 & ${0.99}_{-0.01}^{+0.02}$ & - & - & - & - & ${-13.15}_{-0.10}^{+0.07}$ & ${2.46}_{-0.29}^{+0.57}$ & - & - & - & - & - & ${-14.05}_{-0.35}^{+0.30}$ & 4 \\
J1231-1411 & ${1.04}_{-0.02}^{+0.01}$ & - & - & - & - & - & - & - & - & - & - & - & ${-16.54}_{-0.82}^{+1.92}$ & ${7.13}_{-2.56}^{+2.52}$ \\
J1327-0755 & ${0.99}_{-0.04}^{+0.01}$ & - & - & - & - & - & - & - & - & - & ${-7.19}_{-1.63}^{+0.37}$ & ${-0.76}_{-2.29}^{+0.69}$ & ${-13.87}_{-2.06}^{+0.28}$ & ${8.50}_{-1.56}^{+1.76}$ \\
J1421-4409 & ${1.03}_{-0.01}^{+0.01}$ & - & - & - & - & ${-13.46}_{-4.61}^{+0.15}$ & ${2.42}_{-0.89}^{+3.10}$ & - & - & - & - & - & ${-17.81}_{+0.47}^{+3.58}$ & 4 \\
J1431-5740 & ${1.03}_{-0.01}^{+0.02}$ & - & - & ${-12.52}_{-5.99}^{-0.11}$ & ${2.63}_{-1.18}^{+2.97}$ & ${-12.20}_{-0.10}^{+0.08}$ & ${2.32}_{-0.25}^{+0.34}$ & ${-13.01}_{-0.32}^{+0.23}$ & ${2.08}_{-0.36}^{+0.54}$ & ${5.53}_{-0.70}^{+1.59}$ & - & - & ${-13.19}_{-3.44}^{+0.17}$ & ${16.50}_{-12.20}^{+0.86}$ \\
J1435-6100 & ${1.01}_{-0.01}^{+0.01}$ & - & - & - & - & ${-13.06}_{-2.73}^{+0.05}$ & ${1.15}_{-0.25}^{+1.31}$ & - & - & - & ${-5.44}_{-3.79}^{+0.02}$ & ${1.16}_{-2.74}^{+1.06}$ & ${-14.73}_{-2.44}^{+0.47}$ & ${3.50}_{-1.66}^{+9.60}$ \\
J1446-4701 & ${1.05}_{-0.01}^{+0.02}$ & - & ${-6.63}_{-0.20}^{+0.06}$ & - & - & - & - & - & - & - & - & - & ${-15.73}_{-1.73}^{+0.76}$ & ${1.07}_{-0.52}^{+3.66}$ \\
J1455-3330 & ${1.00}_{-0.01}^{+0.02}$ & - & - & ${-13.51}_{-4.70}^{+0.02}$ & ${2.39}_{-0.98}^{+3.57}$ & - & - & - & - & - & - & - & ${-13.83}_{-2.26}^{+0.28}$ & ${9.03}_{-3.06}^{+1.75}$ \\
J1514-4946 & ${0.97}_{-0.03}^{+0.02}$ & - & - & - & - & - & - & - & - & - & - & - & ${-17.63}_{+0.21}^{+3.23}$ & 4 \\
J1525-5545 & ${0.99}_{-0.01}^{+0.01}$ & - & ${-5.82}_{-0.06}^{+0.04}$ & - & - & ${-12.38}_{-6.12}^{+0.04}$ & ${2.35}_{-0.55}^{+2.88}$ & ${-12.97}_{-0.12}^{+0.18}$ & ${1.20}_{-0.14}^{+0.19}$ & ${5.49}_{-0.56}^{+0.56}$ & ${-4.72}_{-2.84}^{+0.06}$ & ${1.63}_{-2.86}^{+0.58}$ & ${-13.42}_{-3.76}^{-0.23}$ &${17.09}_{-13.02}^{+0.04}$ \\
J1543-5149 & ${1.05}_{-0.02}^{+0.01}$ & - & - & - & - & ${-14.25}_{-4.68}^{+0.08}$ & ${1.99}_{-0.78}^{+3.76}$ & - & - & - & - & - & ${-13.96}_{-3.26}^{+0.07}$ & ${1.50}_{-0.33}^{+8.33}$ \\
J1545-4550 & ${1.03}_{-0.01}^{+0.01}$ & - & ${-6.72}_{-0.26}^{+0.03}$ & - & - & ${-13.58}_{-3.59}^{+0.14}$ & ${1.40}_{-0.65}^{+4.50}$ & ${-13.51}_{-0.50}^{+0.06}$ & ${1.64}_{-0.48}^{+1.02}$ & 4 & - & - & ${-14.23}_{-2.88}^{+0.10}$ & ${6.89}_{-3.50}^{+6.53}$ \\
J1547-5709 & ${0.99}_{-0.01}^{+0.02}$ & - & - & - & - & ${-13.11}_{-3.97}^{+0.11}$ & ${0.87}_{-0.38}^{+2.86}$ & - & - & - & - & - & ${-13.01}_{-0.27}^{+0.18}$ & 4 \\
J1600-3053 & ${1.01}_{-0.01}^{+0.03}$ & ${-6.81}_{-0.30}^{+0.06}$ & - & - & - & ${-13.10}_{-0.06}^{+0.10}$ & ${1.81}_{-0.21}^{+0.27}$ & ${-13.51}_{-0.19}^{+0.08}$ & ${1.57}_{-0.44}^{+0.30}$ & 4 & - & - & ${-13.51}_{-2.53}^{+0.20}$ & ${2.72}_{-0.99}^{+1.42}$ \\
J1603-7202 & ${1.06}_{-0.02}^{+0.01}$ & - & ${-6.09}_{-0.06}^{+0.03}$ & - & - & ${-13.35}_{-5.30}^{+0.03}$ & ${1.10}_{-0.22}^{+4.52}$ & ${-13.60}_{-0.52}^{+0.07}$ & ${0.95}_{-0.31}^{+0.62}$ & 4 & - & - & ${-14.96}_{-2.33}^{+0.34}$ & ${9.90}_{-6.60}^{+6.56}$ \\
J1614-2230 & ${1.02}_{-0.01}^{+0.02}$ & - & - & - & - & ${-13.15}_{-0.09}^{+0.08}$ & ${2.32}_{-0.21}^{+0.55}$ & - & - & - & ${-6.55}_{-0.80}^{+0.18}$ & ${0.24}_{-1.52}^{+0.21}$ & ${-17.26}_{-0.08}^{+2.96}$ & ${8.13}_{-1.20}^{+1.71}$ \\
J1629-6902 & ${1.04}_{-0.02}^{+0.01}$ & ${-6.54}_{-2.63}^{-0.03}$ & ${-6.68}_{-0.69}^{+0.07}$ & - & - & ${-14.87}_{-3.11}^{+0.62}$ & ${5.83}_{-4.08}^{+0.31}$ & - & - & - & - & - & ${-14.32}_{-2.83}^{+0.11}$ & 4 \\
J1643-1224 & ${0.97}_{-0.04}^{+0.02}$ & ${-6.13}_{-0.05}^{+0.04}$ & ${-6.31}_{-0.10}^{+0.06}$ & - & - & ${-12.74}_{-0.11}^{+0.26}$ & ${1.97}_{-0.41}^{+0.59}$ & ${-13.87}_{-0.41}^{+0.30}$ & ${2.38}_{-0.20}^{+0.58}$ & ${8.83}_{-1.15}^{+1.96}$ & ${-8.31}_{-1.12}^{+1.31}$ & ${-1.96}_{-0.09}^{+4.80}$ & ${-12.88}_{-0.35}^{+0.20}$ & ${1.52}_{-0.62}^{+3.44}$ \\
J1652-4838 & ${0.94}_{-0.02}^{+0.03}$ & ${-5.95}_{-0.09}^{+0.05}$ & ${-6.13}_{-2.23}^{+0.02}$ & ${-12.61}_{-0.66}^{+0.09}$ & ${1.51}_{-0.16}^{+2.03}$ & - & - & ${-12.70}_{-0.12}^{+0.10}$ & ${1.26}_{-0.19}^{+0.27}$ & ${2.98}_{-0.17}^{+0.57}$ & ${-8.92}_{-0.61}^{+1.79}$ & ${-0.68}_{-1.92}^{+2.89}$ & ${-13.10}_{-4.05}^{-0.04}$ & ${2.41}_{-0.32}^{+11.07}$ \\
J1653-2054 & ${1.01}_{-0.01}^{+0.02}$ & - & - & - & - & ${-12.47}_{-0.08}^{+0.08}$ & ${1.75}_{-0.24}^{+0.23}$ & - & - & - & ${-6.27}_{-2.51}^{+0.18}$ & ${2.36}_{-5.00}^{+0.33}$ & ${-15.55}_{-1.74}^{+1.45}$ & ${7.47}_{-1.93}^{+4.49}$ \\
J1658-5324 & ${0.99}_{-0.02}^{+0.02}$ & - & - & - & - & - & - & - & - & - & ${-6.10}_{-2.75}^{+0.11}$ & ${2.36}_{-3.03}^{+0.92}$ & ${-14.01}_{-3.14}^{+0.06}$ & ${1.50}_{-0.58}^{+6.18}$ \\
J1708-3506 & ${1.03}_{-0.04}^{+0.01}$ & ${-5.82}_{-3.21}^{-0.06}$ & - & - & - & ${-12.74}_{-0.23}^{+0.18}$ & ${1.22}_{-0.36}^{+0.53}$ & ${-13.60}_{-1.34}^{+0.33}$ & ${5.30}_{-3.20}^{+1.09}$ & 4 & - & - & ${-13.37}_{-3.11}^{+0.25}$ & 4 \\
J1713+0747 & ${1.07}_{-0.02}^{+0.02}$ & - & ${-6.86}_{-0.09}^{+0.05}$ & - & - & - & - & ${-14.69}_{-3.36}^{+0.18}$ & ${0.60}_{-0.26}^{+3.66}$ & 4 & - & - & ${-16.59}_{-0.57}^{+2.55}$ & 4 \\
J1719-1438 & ${1.03}_{-0.02}^{+0.01}$ & - & - & - & - & ${-13.28}_{-5.02}^{+0.02}$ & ${2.35}_{-0.58}^{+2.88}$ & - & - & - & - & - & ${-13.24}_{-2.32}^{+0.16}$ & 4 \\
J1721-2457 & ${1.07}_{-0.02}^{+0.01}$ & - & - & - & - & ${-12.88}_{-0.42}^{+0.13}$ & ${4.32}_{-1.47}^{+1.52}$ & - & - & - & - & - & ${-12.92}_{-0.76}^{+0.24}$ & ${9.88}_{-2.28}^{+1.67}$ \\
J1730-2304 & ${1.02}_{-0.01}^{+0.01}$ & - & ${-6.40}_{-0.22}^{+0.07}$ & - & - & ${-13.19}_{-0.17}^{+0.06}$ & ${1.29}_{-0.17}^{+0.99}$ & - & - & - & ${-7.92}_{-1.34}^{+1.14}$ & ${-1.61}_{-1.06}^{+3.93}$ & ${-15.50}_{-1.81}^{+0.81}$ & ${6.26}_{-1.24}^{+1.96}$ \\
J1732-5049 & ${1.04}_{-0.01}^{+0.01}$ & - & - & - & - & - & - & - & - & - & ${-5.72}_{-0.30}^{+0.14}$ & ${2.04}_{-0.83}^{+0.87}$ & ${-15.73}_{-1.52}^{+1.30}$ & ${14.50}_{-9.34}^{+2.34}$ \\
J1737-0811 & ${1.00}_{-0.01}^{+0.01}$ & - & - & - & - & ${-12.88}_{-0.16}^{+0.18}$ & ${1.68}_{-0.41}^{+0.54}$ & - & - & - & - & - & ${-14.05}_{-3.17}^{+0.19}$ & ${6.10}_{-3.58}^{+5.65}$ \\
J1744-1134 & ${1.03}_{-0.01}^{+0.02}$ & ${-7.04}_{-0.04}^{+0.05}$ & ${-6.59}_{-0.06}^{+0.03}$ & - & - & - & - & - & - & - & ${-6.43}_{-0.37}^{+0.15}$ & ${0.91}_{-0.81}^{+0.66}$ & ${-16.18}_{-1.24}^{+1.18}$ & ${3.73}_{-1.10}^{+1.09}$ \\
J1747-4036 & ${1.04}_{-0.02}^{+0.01}$ & ${-6.31}_{-1.57}^{+0.10}$ & - & ${-12.65}_{-5.14}^{+0.08}$ & ${2.48}_{-0.98}^{+1.49}$ & ${-12.83}_{-0.27}^{+0.21}$ & ${1.41}_{-0.74}^{+0.29}$ & ${-12.97}_{-0.20}^{+0.17}$ & ${2.61}_{-0.40}^{+1.86}$ & ${4.05}_{-0.51}^{+1.08}$ & - & - & ${-12.88}_{-3.78}^{+0.11}$ & ${18.17}_{-5.58}^{+0.68}$ \\
J1751-2857 & ${1.00}_{-0.01}^{+0.02}$ & - & - & - & - & ${-12.79}_{-0.09}^{+0.12}$ & ${2.31}_{-0.49}^{+0.93}$ & - & - & - & ${-8.04}_{-1.03}^{+1.24}$ & ${-2.32}_{-0.69}^{+3.91}$ & ${-13.15}_{-0.22}^{+0.22}$ & ${2.00}_{-0.87}^{+4.80}$ \\
J1757-5322 & ${1.03}_{-0.01}^{+0.02}$ & - & ${-6.49}_{-2.09}^{+0.03}$ & - & - & - & - & ${-13.69}_{-3.58}^{+0.04}$ & ${1.58}_{-0.60}^{+2.51}$ & 4 & - & - & ${-15.91}_{-1.41}^{+1.43}$ & ${7.69}_{-4.06}^{+7.16}$ \\
J1801-1417 & ${1.01}_{-0.01}^{+0.02}$ & - & - & ${-13.02}_{-5.64}^{+0.01}$ & ${3.26}_{-1.71}^{+2.44}$ & ${-13.21}_{-5.28}^{+0.01}$ & ${2.35}_{-0.88}^{+3.13}$ & - & - & - & - & - & ${-13.06}_{-3.00}^{+0.08}$ & ${3.80}_{-1.46}^{+2.76}$ \\
J1802-2124 & ${1.05}_{-0.02}^{+0.01}$ & - & ${-6.00}_{-0.08}^{+0.03}$ & - & - & ${-12.47}_{-0.09}^{+0.11}$ & ${2.72}_{-0.31}^{+0.70}$ & ${-13.28}_{-0.20}^{+0.14}$ & ${1.68}_{-0.17}^{+0.28}$ & ${6.52}_{-0.64}^{+0.79}$ & - & - & ${-16.45}_{-0.80}^{+2.26}$ & ${6.34}_{-1.99}^{+1.39}$ \\
J1804-2717 & ${1.01}_{-0.02}^{+0.02}$ & - & ${-6.09}_{-0.10}^{+0.08}$ & - & - & - & - & - & - & - & - & - & ${-13.51}_{-3.63}^{-0.02}$ & ${4.32}_{-2.10}^{+2.03}$ \\
J1804-2858 & ${1.07}_{-0.02}^{+0.01}$ & - & - & ${-11.93}_{-0.09}^{+0.19}$ & ${2.13}_{-0.23}^{+0.48}$ & ${-11.47}_{-0.08}^{+0.14}$ & ${4.02}_{-0.51}^{+0.72}$ & ${-12.56}_{-0.18}^{+0.25}$ & ${1.87}_{-0.29}^{+0.40}$ & ${6.26}_{-0.62}^{+0.84}$ & - & - & ${-15.68}_{-1.38}^{+2.55}$ & ${5.30}_{-2.49}^{+9.26}$ \\
J1811-2405 & ${1.04}_{-0.02}^{+0.01}$ & ${-6.73}_{-2.31}^{+0.07}$ & - & - & - & ${-13.10}_{-0.08}^{+0.07}$ & ${2.40}_{-0.28}^{+0.35}$ & - & - & - & ${-8.37}_{-0.57}^{+1.20}$ & ${-2.21}_{-1.00}^{+1.85}$ & ${-17.81}_{+0.44}^{+3.11}$ & ${7.56}_{-0.64}^{+1.08}$ \\
J1825-0319 & ${1.02}_{-0.01}^{+0.01}$ & - & - & - & - & - & - & ${-12.42}_{-0.10}^{+0.09}$ & ${2.04}_{-0.22}^{+0.30}$ & ${2.53}_{-0.18}^{+0.37}$ & ${-8.59}_{-0.87}^{+2.24}$ & ${1.72}_{-3.22}^{+1.38}$ & ${-17.58}_{+0.42}^{+3.89}$ & ${2.10}_{-0.76}^{+9.25}$ \\
J1832-0836 & ${0.98}_{-0.01}^{+0.02}$ & - & - & - & - & ${-12.83}_{-0.06}^{+0.10}$ & ${2.55}_{-0.27}^{+0.39}$ & - & - & - & - & - & ${-14.87}_{-2.46}^{+0.29}$ & 4 \\
J1843-1113 & ${1.01}_{-0.01}^{+0.01}$ & - & - & - & - & ${-12.97}_{-0.05}^{+0.11}$ & ${2.60}_{-0.25}^{+0.52}$ & - & - & - & - & - & ${-15.91}_{-1.42}^{+1.30}$ & ${1.17}_{-0.81}^{+1.22}$ \\
J1843-1448 & ${0.96}_{-0.02}^{+0.06}$ & ${-5.18}_{-0.13}^{+0.06}$ & - & - & - & ${-13.08}_{-4.38}^{+0.19}$ & ${3.33}_{-1.34}^{+2.55}$ & - & - & - & - & - & ${-14.55}_{-2.56}^{+0.76}$ & ${5.50}_{-2.44}^{+8.34}$ \\
J1902-5105 & ${1.06}_{-0.01}^{+0.01}$ & - & - & - & - & ${-13.33}_{-5.07}^{-0.18}$ & ${3.72}_{-2.07}^{+2.05}$ & ${-13.51}_{-0.08}^{+0.09}$ & ${1.23}_{-0.19}^{+0.26}$ & 4 & - & - & ${-13.65}_{-2.42}^{+0.14}$ & 4 \\
J1903-7051 & ${1.04}_{-0.01}^{+0.01}$ & - & ${-6.95}_{-1.46}^{+0.10}$ & - & - & ${-13.56}_{-0.82}^{+0.03}$ & ${2.64}_{-0.38}^{+3.08}$ & - & - & - & - & - & ${-14.64}_{-2.65}^{+0.18}$ & ${7.31}_{-4.15}^{+7.82}$ \\
J1909-3744 & ${1.04}_{-0.02}^{+0.00}$ & ${-7.17}_{-0.03}^{-0.00}$ & ${-7.17}_{-0.06}^{+0.02}$ & - & - & ${-13.60}_{-0.07}^{+0.07}$ & ${2.04}_{-0.18}^{+0.28}$ & - & - & - & ${-6.43}_{-0.19}^{+0.10}$ & ${1.39}_{-0.42}^{+0.21}$ & ${-14.28}_{-0.21}^{+0.17}$ & ${4.96}_{-1.24}^{+0.86}$ \\
J1911-1114 & ${1.02}_{-0.02}^{+0.02}$ & - & - & - & - & - & - & ${-13.87}_{-0.34}^{+0.54}$ & ${2.74}_{-0.46}^{+1.55}$ & ${4.89}_{-1.67}^{+2.30}$ & - & - & ${-14.64}_{-2.44}^{+1.29}$ & ${5.78}_{-1.74}^{+2.86}$ \\
J1918-0642 & ${1.02}_{-0.01}^{+0.01}$ & - & ${-6.54}_{-0.06}^{+0.08}$ & - & - & - & - & - & - & - & - & - & ${-16.04}_{-1.31}^{+1.35}$ & ${2.52}_{-1.38}^{+2.12}$ \\
J1933-6211 & ${1.05}_{-0.01}^{+0.02}$ & - & ${-6.59}_{-0.38}^{+0.08}$ & - & - & ${-13.66}_{-1.40}^{+0.06}$ & ${1.57}_{-0.15}^{+4.31}$ & - & - & - & - & - & ${-14.10}_{-3.21}^{-0.27}$ & 4 \\
J1946-5403 & ${0.97}_{-0.02}^{+0.02}$ & - & - & - & - & - & - & - & - & - & - & - & ${-14.46}_{-2.76}^{+0.05}$ & 4 \\
J2010-1323 & ${1.03}_{-0.01}^{+0.02}$ & - & - & - & - & - & - & - & - & - & - & - & ${-14.14}_{-2.47}^{+0.15}$ & ${3.35}_{-0.90}^{+0.52}$ \\
J2039-3616 & ${1.06}_{-0.03}^{+0.01}$ & - & - & - & - & - & - & - & - & - & - & - & ${-16.04}_{-1.37}^{+1.16}$ & 4 \\
J2124-3358 & ${1.10}_{-0.01}^{+0.02}$ & - & ${-6.63}_{-1.19}^{+0.10}$ & - & - & - & - & - & - & - & ${-6.48}_{-2.41}^{+0.09}$ & ${0.19}_{-2.26}^{+2.30}$ & ${-16.72}_{-0.71}^{+1.98}$ & ${7.78}_{-2.04}^{+3.49}$ \\
J2129-5721 & ${1.03}_{-0.01}^{+0.02}$ & - & - & - & - & - & - & ${-14.01}_{-0.31}^{+0.06}$ & ${1.01}_{-0.45}^{+0.62}$ & 4 & - & - & ${-13.83}_{-0.21}^{+0.16}$ & ${1.28}_{-0.47}^{+4.87}$ \\
J2145-0750 & ${1.04}_{-0.01}^{+0.01}$ & - & ${-6.09}_{-0.03}^{+0.04}$ & - & - & - & - & - & - & - & ${-6.55}_{-1.05}^{+0.22}$ & ${0.70}_{-2.18}^{+0.62}$ & ${-14.01}_{-3.13}^{+0.07}$ & ${5.37}_{-1.04}^{+1.75}$ \\
J2150-0326 & ${1.03}_{-0.01}^{+0.02}$ & - & - & - & - & - & - & ${-13.51}_{-0.35}^{+0.07}$ & ${0.92}_{-0.69}^{+0.18}$ & 4 & - & - & ${-13.65}_{-3.36}^{+0.01}$ & ${0.88}_{-0.30}^{+2.49}$ \\
J2222-0137 & ${1.06}_{-0.01}^{+0.01}$ & - & ${-6.04}_{-0.03}^{+0.05}$ & - & - & - & - & ${-13.96}_{-1.24}^{+0.13}$ & ${2.42}_{-0.74}^{+2.77}$ & 4 & - & - & ${-17.76}_{+0.39}^{+3.36}$ & ${1.52}_{-0.65}^{+1.50}$ \\
J2229+2643 & ${1.07}_{-0.01}^{+0.02}$ & - & - & ${-14.19}_{-4.68}^{+0.09}$ & ${3.15}_{-2.01}^{+2.54}$ & ${-14.23}_{-4.40}^{+0.15}$ & ${2.69}_{-1.48}^{+3.13}$ & - & - & - & - & - & ${-13.87}_{-3.15}^{+0.03}$ & 4 \\
J2234+0944 & ${1.02}_{-0.02}^{+0.02}$ & - & - & ${-12.83}_{-0.11}^{+0.15}$ & ${1.99}_{-0.27}^{+0.67}$ & - & - & - & - & - & ${-6.16}_{-2.96}^{+0.14}$ & ${2.44}_{-2.40}^{+0.87}$ & ${-17.63}_{+0.42}^{+3.71}$ & ${8.85}_{-4.01}^{+3.60}$ \\
J2236-5527 & ${1.01}_{-0.01}^{+0.03}$ & - & ${-6.13}_{-0.54}^{+0.08}$ & ${-13.30}_{-5.50}^{+0.08}$ & ${0.84}_{-0.21}^{+4.61}$ & - & - & - & - & - & - & - & ${-15.45}_{-1.82}^{+1.33}$ & 4 \\
J2241-5236 & ${1.05}_{-0.01}^{+0.01}$ & - & - & - & - & - & - & - & - & - & ${-6.16}_{-0.10}^{+0.06}$ & ${1.81}_{-0.30}^{+0.18}$ & ${-14.82}_{-1.57}^{+0.28}$ & ${5.86}_{-2.32}^{+1.59}$ \\
J2317+1439 & ${1.00}_{-0.01}^{+0.02}$ & - & - & ${-13.51}_{-5.21}^{+0.04}$ & ${3.47}_{-1.96}^{+2.43}$ & ${-14.25}_{-4.55}^{+0.05}$ & ${3.28}_{-2.02}^{+2.39}$ & - & - & - & - & - & ${-13.33}_{-2.23}^{+0.10}$ & 4 \\
J2322+2057 & ${1.02}_{-0.02}^{+0.01}$ & ${-6.64}_{-2.61}^{+0.04}$ & - & - & - & - & - & - & - & - & - & - & ${-14.10}_{-2.78}^{+0.23}$ & 4 \\
J2322-2650 & ${0.95}_{-0.01}^{+0.02}$ & - & - & - & - & - & - & - & - & - & - & - & ${-17.63}_{+0.38}^{+3.39}$ & ${9.27}_{-3.51}^{+6.49}$ \\

\end{longtable}}
\end{landscape}
\twocolumn

\onecolumn
\setlength{\tabcolsep}{3.5pt}
{\renewcommand{\arraystretch}{1.75}%
\begin{longtable}[c]{@{}lcccccccc@{}}
\captionsetup{width=0.9\linewidth}
\caption{The deterministic noise processes that are included for the MPTA pulsars. We report the MAP values and the $68\%$ confidence interval corresponding to the sampled posterior. In some few cases, the MAP value has fallen outside of the the confidence interval we report. The parameters under the Chromatic Gaussian Event subheading are the log of the amplitude in $\log_{10}(\mathrm{s})$ ($\mathrm{log}_{10}\mathrm{A_{g}}$), the chromatic index of the event ($\beta_{\textrm{g}}$), the arrival time the event is centered on in MJD ($t_{g,0}$), the width or duration of the event in MJD ($\sigma_{g}$), and the corresponding sign of the delay ($\mathrm{Sign\, [+/-]}$). Under the Annual Chromatic Variations subheading we present the log of the amplitude in $\log_{10}(\mathrm{s})$ ($\mathrm{log}_{10}\mathrm{A_{s}}$), the chromatic index of the annual variation ($\beta_\textrm{s}$), and the dimensionless phase of the waveform ($\phi$). Where the pulsar name is displayed in bold, the parameter values we report are taken from the CURN Bayesian analysis due to a marked increase in the precision constraint of the posterior during this analysis.}
\label{Table: MPTA determinstic models}\\
\toprule
\multicolumn{1}{l}{Pulsar} & \multicolumn{5}{c}{Chromatic Gaussian Event} & \multicolumn{3}{c}{Annual Chromatic Variations}           \\* \cmidrule(lr){2-6} \cmidrule(lr){7-9}
 &
  $\mathrm{log}_{10}\mathrm{A_{g}}$ &
  $\beta_{\textrm{g}}$ &
  $t_{g,0}$ &
  $\sigma_{g}$ &
  $\mathrm{Sign\, [+/-]}$ &
  $\mathrm{log}_{10}\mathrm{A_{s}}$ &
  $\beta_\textrm{s}$ &
  $\phi$  \\* \midrule
\endfirsthead
\multicolumn{9}{c}%
{{\bfseries Table \thetable\ continued from previous page}} \\
\toprule
\multicolumn{1}{l}{Pulsar} & \multicolumn{5}{c}{Chromatic Gaussian Event} & \multicolumn{3}{c}{Annual Chromatic Variations}           \\* \cmidrule(lr){2-6} \cmidrule(lr){7-9}
 &
  $\mathrm{log}_{10}\mathrm{A_{g}}$ &
  $\beta_{\textrm{g}}$ &
  $t_{g,0}$ &
  $\sigma_{g}$ &
  $\mathrm{Sign\, [+/-]}$ &
  $\mathrm{log}_{10}\mathrm{A_{s}}$ &
  $\beta_\textrm{s}$ &
  $\phi$  \\* \midrule
\endhead
\textbf{J0610-2100} & ${-5.68}_{-2.78}^{+0.06}$ & ${1.47}_{-0.59}^{+4.83}$ & ${58872.64}_{-1.52}^{+611.04}$ & ${13.99}_{-2.86}^{+806.21}$ & $+$ &  &  &  \\
J0613-0200 &  &  &  &  &  & ${-7.28}_{-9.34}^{+0.22}$ & ${4.16}_{-2.00}^{+6.81}$ & ${2.55}_{-0.78}^{+1.98}$ \\
J0614-3329 &  &  &  &  &  & ${-7.33}_{-9.87}^{-0.21}$ & ${5.11}_{-2.81}^{+6.17}$ & ${3.86}_{-2.34}^{+1.02}$ \\
J0955-6150 &  &  &  &  &  & ${-8.96}_{-8.70}^{+0.24}$ & ${4.61}_{-2.56}^{+7.06}$ & ${4.62}_{-3.57}^{+0.65}$ \\
J1017-7156 & ${-7.68}_{-0.63}^{+0.88}$ & ${8.95}_{-3.32}^{+1.85}$ & ${59381.10}_{-302.77}^{+385.29}$ & ${1244.40}_{-407.19}^{+224.27}$ & $+$ &  &  &  \\
J1022+1001 & ${-6.68}_{-0.75}^{+2.12}$ & ${6.06}_{-5.24}^{+3.05}$ & ${60056.84}_{-422.39}^{+12.00}$ & ${1031.43}_{-397.25}^{+346.74}$ & $-$ &  &  &  \\
J1024-0719 & ${-6.31}_{-2.88}^{+0.53}$ & ${2.09}_{-0.77}^{+7.18}$ & ${59383.87}_{-415.60}^{+617.48}$ & ${263.66}_{-26.74}^{+1053.21}$ & $+$ &  &  &  \\
J1045-4509 &  &  &  &  &  & ${-6.63}_{-10.38}^{-0.05}$ & ${4.29}_{-1.83}^{+6.48}$ & ${6.06}_{-5.45}^{-0.17}$ \\
J1125-6014 & ${-6.74}_{-1.54}^{+1.23}$ & ${4.05}_{-2.66}^{+4.20}$ & ${58829.32}_{-147.74}^{+553.79}$ & ${623.77}_{-212.71}^{+640.51}$ & $+$ &  &  &  \\
J1231-1411 &  &  &  &  &  & ${-7.13}_{-7.96}^{+0.31}$ & ${4.11}_{-2.01}^{+6.68}$ & ${4.56}_{-1.88}^{+0.55}$ \\
J1421-4409 & ${-6.27}_{-2.42}^{+0.26}$ & ${4.13}_{-1.12}^{+5.29}$ & ${59633.39}_{-535.56}^{+251.43}$ & ${278.90}_{-122.43}^{+736.57}$ & $-$ &  &  &  \\
J1600-3053 & ${-6.13}_{-1.01}^{+0.55}$ & ${4.17}_{-1.02}^{+2.54}$ & ${58738.82}_{-116.58}^{+222.06}$ & ${937.13}_{-258.68}^{+382.94}$ & $+$ &  &  &  \\
J1643-1224 &  &  &  &  &  & ${-5.48}_{-2.10}^{+0.50}$ & ${0.91}_{-0.46}^{+3.95}$ & ${3.17}_{-0.17}^{+0.27}$ \\
J1652-4838 & ${-6.90}_{-1.41}^{+0.92}$ & ${6.02}_{-3.47}^{+5.20}$ & ${58962.00}_{-185.10}^{+510.37}$ & ${1349.36}_{-876.74}^{+21.37}$ & $+$ &  &  &  \\
J1721-2457 & ${-6.69}_{-2.29}^{+0.59}$ & ${6.93}_{-4.55}^{+4.66}$ & ${60008.81}_{-699.10}^{+55.30}$ & ${661.02}_{-440.62}^{+361.53}$ & $+$ &  &  &  \\
J1737-0811 & ${-5.00}_{-2.09}^{+0.03}$ & ${0.77}_{-0.32}^{+2.68}$ & ${58682.16}_{-4.18}^{+330.24}$ & ${15.93}_{-2.07}^{+445.53}$ & $+$ &  &  &  \\
J1747-4036 & ${-4.46}_{-0.04}^{+0.34}$ & ${1.03}_{-0.71}^{+0.56}$ & ${59149.10}_{-251.23}^{+394.24}$ & ${1320.09}_{-460.96}^{+89.80}$ & $-$ &  &  &  \\
J1804-2858 &  &  &  &  &  & ${-5.78}_{-11.37}^{+0.18}$ & ${4.84}_{-2.10}^{+6.20}$ & ${1.41}_{-0.40}^{+3.51}$ \\
J1832-0836 & ${-5.64}_{-0.18}^{+0.55}$ & ${3.21}_{-0.74}^{+1.00}$ & ${58729.18}_{-80.10}^{+356.38}$ & ${1204.83}_{-388.03}^{+230.77}$ & $+$ &  &  &  \\
\textbf{J1902-5105} & ${-5.82}_{-3.46}^{+0.06}$ & ${1.52}_{-0.68}^{+7.13}$ & ${59517.64}_{-535.04}^{+367.60}$ & ${185.53}_{-280.04}^{+757.79}$ & $+$ &  &  &  \\
J1918-0642 & ${-6.40}_{-0.60}^{+0.11}$ & ${3.83}_{-0.67}^{+3.18}$ & ${59829.56}_{-19.90}^{+31.30}$ & ${108.56}_{-20.43}^{+36.17}$ & $+$ &  &  &  \\
J2129-5721 &  &  &  &  &  & ${-6.68}_{-9.40}^{+0.13}$ & ${2.35}_{-1.13}^{+7.53}$ & ${4.56}_{-2.71}^{+0.53}$ \\
J2150-0326 & ${-5.30}_{-3.91}^{+0.25}$ & ${2.03}_{-0.86}^{+6.19}$ & ${59438.40}_{-392.92}^{+370.07}$ & ${285.09}_{-23.88}^{+843.29}$ & $+$ &  &  &  \\

\end{longtable}}

\twocolumn



\bibliographystyle{mnras}
\bibliography{ref} 

\begin{thebibliography}{}
\makeatletter
\relax
\def\mn@urlcharsother{\let\do\@makeother \do\$\do\&\do\#\do\^\do\_\do\%\do\~}
\def\mn@doi{\begingroup\mn@urlcharsother \@ifnextchar [ {\mn@doi@} {\mn@doi@[]}}
\def\mn@doi@[#1]#2{\def\@tempa{#1}\ifx\@tempa\@empty \href {http://dx.doi.org/#2} {doi:#2}\else \href {http://dx.doi.org/#2} {#1}\fi \endgroup}
\def\mn@eprint#1#2{\mn@eprint@#1:#2::\@nil}
\def\mn@eprint@arXiv#1{\href {http://arxiv.org/abs/#1} {{\tt arXiv:#1}}}
\def\mn@eprint@dblp#1{\href {http://dblp.uni-trier.de/rec/bibtex/#1.xml} {dblp:#1}}
\def\mn@eprint@#1:#2:#3:#4\@nil{\def\@tempa {#1}\def\@tempb {#2}\def\@tempc {#3}\ifx \@tempc \@empty \let \@tempc \@tempb \let \@tempb \@tempa \fi \ifx \@tempb \@empty \def\@tempb {arXiv}\fi \@ifundefined {mn@eprint@\@tempb}{\@tempb:\@tempc}{\expandafter \expandafter \csname mn@eprint@\@tempb\endcsname \expandafter{\@tempc}}}

\bibitem[\protect\citeauthoryear{{Agazie} et~al.,}{{Agazie} et~al.}{2023a}]{2023ApJ...951L...8A}
{Agazie} G.,  et~al., 2023a, \mn@doi [\apjl] {10.3847/2041-8213/acdac6}, \href {https://ui.adsabs.harvard.edu/abs/2023ApJ...951L...8A} {951, L8}

\bibitem[\protect\citeauthoryear{{Agazie} et~al.,}{{Agazie} et~al.}{2023b}]{2023ApJ...951L..10A}
{Agazie} G.,  et~al., 2023b, \mn@doi [\apjl] {10.3847/2041-8213/acda88}, \href {https://ui.adsabs.harvard.edu/abs/2023ApJ...951L..10A} {951, L10}

\bibitem[\protect\citeauthoryear{{Allen}, {Dhurandhar}, {Gupta}, {McLaughlin}, {Natarajan}, {Shannon}, {Thrane}  \& {Vecchio}}{{Allen} et~al.}{2023}]{2023arXiv230404767A}
{Allen} B.,  {Dhurandhar} S.,  {Gupta} Y.,  {McLaughlin} M.,  {Natarajan} P.,  {Shannon} R.~M.,  {Thrane} E.,   {Vecchio} A.,  2023, \mn@doi [arXiv e-prints] {10.48550/arXiv.2304.04767}, \href {https://ui.adsabs.harvard.edu/abs/2023arXiv230404767A} {p. arXiv:2304.04767}

\bibitem[\protect\citeauthoryear{Alpar, Nandkumar  \& Pines}{Alpar et~al.}{1986}]{Alpar1986VortexCA}
Alpar M.~A.,  Nandkumar R.,   Pines D.~S.,  1986, The Astrophysical Journal, 311, 197

\bibitem[\protect\citeauthoryear{{Antoniadis} et~al.,}{{Antoniadis} et~al.}{2022}]{2022MNRAS.510.4873A}
{Antoniadis} J.,  et~al., 2022, \mn@doi [\mnras] {10.1093/mnras/stab3418}, \href {https://ui.adsabs.harvard.edu/abs/2022MNRAS.510.4873A} {510, 4873}

\bibitem[\protect\citeauthoryear{{Antoniadis} et~al.,}{{Antoniadis} et~al.}{2023}]{2023arXiv230616214A}
{Antoniadis} J.,  et~al., 2023, \mn@doi [arXiv e-prints] {10.48550/arXiv.2306.16214}, \href {https://ui.adsabs.harvard.edu/abs/2023arXiv230616214A} {p. arXiv:2306.16214}

\bibitem[\protect\citeauthoryear{{Arzoumanian} et~al.,}{{Arzoumanian} et~al.}{2018}]{2018ApJ...859...47A}
{Arzoumanian} Z.,  et~al., 2018, \mn@doi [\apj] {10.3847/1538-4357/aabd3b}, \href {https://ui.adsabs.harvard.edu/abs/2018ApJ...859...47A} {859, 47}

\bibitem[\protect\citeauthoryear{{Arzoumanian} et~al.,}{{Arzoumanian} et~al.}{2020}]{NanoGravGWB}
{Arzoumanian} Z.,  et~al., 2020, \mn@doi [\apjl] {10.3847/2041-8213/abd401}, \href {https://ui.adsabs.harvard.edu/abs/2020ApJ...905L..34A} {905, L34}

\bibitem[\protect\citeauthoryear{{Ashton} et~al.,}{{Ashton} et~al.}{2019}]{Ashton+19}
{Ashton} G.,  et~al., 2019, \mn@doi [\apjs] {10.3847/1538-4365/ab06fc}, \href {https://ui.adsabs.harvard.edu/abs/2019ApJS..241...27A} {241, 27}

\bibitem[\protect\citeauthoryear{{Bailes} et~al.,}{{Bailes} et~al.}{2016}]{2016mks..confE..11B}
{Bailes} M.,  et~al., 2016, in MeerKAT Science: On the Pathway to the SKA. p.~11 (\mn@eprint {arXiv} {1803.07424})

\bibitem[\protect\citeauthoryear{{Bailes} et~al.,}{{Bailes} et~al.}{2020}]{2020PASA...37...28B}
{Bailes} M.,  et~al., 2020, \mn@doi [\pasa] {10.1017/pasa.2020.19}, \href {https://ui.adsabs.harvard.edu/abs/2020PASA...37...28B} {37, e028}

\bibitem[\protect\citeauthoryear{{Becker}, {Kramer}  \& {Sesana}}{{Becker} et~al.}{2018}]{2018SSRv..214...30B}
{Becker} W.,  {Kramer} M.,   {Sesana} A.,  2018, \mn@doi [\ssr] {10.1007/s11214-017-0459-0}, \href {https://ui.adsabs.harvard.edu/abs/2018SSRv..214...30B} {214, 30}

\bibitem[\protect\citeauthoryear{{Chen} et~al.,}{{Chen} et~al.}{2021}]{2021MNRAS.508.4970C}
{Chen} S.,  et~al., 2021, \mn@doi [\mnras] {10.1093/mnras/stab2833}, \href {https://ui.adsabs.harvard.edu/abs/2021MNRAS.508.4970C} {508, 4970}

\bibitem[\protect\citeauthoryear{{Cheng}}{{Cheng}}{1987}]{1987ApJ...321..799C}
{Cheng} K.~S.,  1987, \mn@doi [\apj] {10.1086/165672}, \href {https://ui.adsabs.harvard.edu/abs/1987ApJ...321..799C} {321, 799}

\bibitem[\protect\citeauthoryear{{Coles} et~al.,}{{Coles} et~al.}{2015}]{2015ApJ...808..113C}
{Coles} W.~A.,  et~al., 2015, \mn@doi [\apj] {10.1088/0004-637X/808/2/113}, \href {https://ui.adsabs.harvard.edu/abs/2015ApJ...808..113C} {808, 113}

\bibitem[\protect\citeauthoryear{{Cordes} \& {Shannon}}{{Cordes} \& {Shannon}}{2010}]{2010arXiv1010.3785C}
{Cordes} J.~M.,  {Shannon} R.~M.,  2010, arXiv e-prints, \href {https://ui.adsabs.harvard.edu/abs/2010arXiv1010.3785C} {p. arXiv:1010.3785}

\bibitem[\protect\citeauthoryear{{Cordes}, {Weisberg}, {Frail}, {Spangler}  \& {Ryan}}{{Cordes} et~al.}{1991}]{1991Natur.354..121C}
{Cordes} J.~M.,  {Weisberg} J.~M.,  {Frail} D.~A.,  {Spangler} S.~R.,   {Ryan} M.,  1991, \mn@doi [\nat] {10.1038/354121a0}, \href {https://ui.adsabs.harvard.edu/abs/1991Natur.354..121C} {354, 121}

\bibitem[\protect\citeauthoryear{{Cordes}, {Shannon}  \& {Stinebring}}{{Cordes} et~al.}{2016}]{2016ApJ...817...16C}
{Cordes} J.~M.,  {Shannon} R.~M.,   {Stinebring} D.~R.,  2016, \mn@doi [\apj] {10.3847/0004-637X/817/1/16}, \href {https://ui.adsabs.harvard.edu/abs/2016ApJ...817...16C} {817, 16}

\bibitem[\protect\citeauthoryear{{Dewdney}, {Hall}, {Schilizzi}  \& {Lazio}}{{Dewdney} et~al.}{2009}]{2009IEEEP..97.1482D}
{Dewdney} P.~E.,  {Hall} P.~J.,  {Schilizzi} R.~T.,   {Lazio} T.~J.~L.~W.,  2009, \mn@doi [IEEE Proceedings] {10.1109/JPROC.2009.2021005}, \href {https://ui.adsabs.harvard.edu/abs/2009IEEEP..97.1482D} {97, 1482}

\bibitem[\protect\citeauthoryear{{EPTA Collaboration} et~al.,}{{EPTA Collaboration} et~al.}{2023}]{2023A&A...678A..49E}
{EPTA Collaboration} et~al., 2023, \mn@doi [\aap] {10.1051/0004-6361/202346842}, \href {https://ui.adsabs.harvard.edu/abs/2023A&A...678A..49E} {678, A49}

\bibitem[\protect\citeauthoryear{Ellis \& van Haasteren}{Ellis \& van Haasteren}{2017}]{justin_ellis_2017_1037579}
Ellis J.,  van Haasteren R.,  2017, jellis18/PTMCMCSampler: Official Release, \mn@doi{10.5281/zenodo.1037579}, \url {https://doi.org/10.5281/zenodo.1037579}

\bibitem[\protect\citeauthoryear{{Ellis}, {Vallisneri}, {Taylor}  \& {Baker}}{{Ellis} et~al.}{2019}]{2019ascl.soft12015E}
{Ellis} J.~A.,  {Vallisneri} M.,  {Taylor} S.~R.,   {Baker} P.~T.,  2019, {ENTERPRISE: Enhanced Numerical Toolbox Enabling a Robust PulsaR Inference SuitE}, Astrophysics Source Code Library, record ascl:1912.015 (\mn@eprint {ascl} {1912.015})

\bibitem[\protect\citeauthoryear{{Faulkner} et~al.,}{{Faulkner} et~al.}{2004}]{2004MNRAS.355..147F}
{Faulkner} A.~J.,  et~al., 2004, \mn@doi [\mnras] {10.1111/j.1365-2966.2004.08310.x}, \href {https://ui.adsabs.harvard.edu/abs/2004MNRAS.355..147F} {355, 147}

\bibitem[\protect\citeauthoryear{{Foster} \& {Backer}}{{Foster} \& {Backer}}{1990}]{1990ApJ...361..300F}
{Foster} R.~S.,  {Backer} D.~C.,  1990, \mn@doi [\apj] {10.1086/169195}, \href {https://ui.adsabs.harvard.edu/abs/1990ApJ...361..300F} {361, 300}

\bibitem[\protect\citeauthoryear{{Geyer} \& {Karastergiou}}{{Geyer} \& {Karastergiou}}{2016}]{2016MNRAS.462.2587G}
{Geyer} M.,  {Karastergiou} A.,  2016, \mn@doi [\mnras] {10.1093/mnras/stw1724}, \href {https://ui.adsabs.harvard.edu/abs/2016MNRAS.462.2587G} {462, 2587}

\bibitem[\protect\citeauthoryear{{Goncharov} et~al.,}{{Goncharov} et~al.}{2021a}]{2021MNRAS.502..478G}
{Goncharov} B.,  et~al., 2021a, \mn@doi [\mnras] {10.1093/mnras/staa3411}, \href {https://ui.adsabs.harvard.edu/abs/2021MNRAS.502..478G} {502, 478}

\bibitem[\protect\citeauthoryear{{Goncharov} et~al.,}{{Goncharov} et~al.}{2021b}]{2021arXiv210712112G}
{Goncharov} B.,  et~al., 2021b, \mn@doi [\apjl] {10.3847/2041-8213/ac17f4}, \href {https://ui.adsabs.harvard.edu/abs/2021ApJ...917L..19G} {917, L19}

\bibitem[\protect\citeauthoryear{{Goncharov} et~al.,}{{Goncharov} et~al.}{2021c}]{2021ApJ...917L..19G}
{Goncharov} B.,  et~al., 2021c, \mn@doi [\apjl] {10.3847/2041-8213/ac17f4}, \href {https://ui.adsabs.harvard.edu/abs/2021ApJ...917L..19G} {917, L19}

\bibitem[\protect\citeauthoryear{{Grishchuk}}{{Grishchuk}}{2005}]{2005PhyU...48.1235G}
{Grishchuk} L.~P.,  2005, \mn@doi [Physics Uspekhi] {10.1070/PU2005v048n12ABEH005795}, \href {https://ui.adsabs.harvard.edu/abs/2005PhyU...48.1235G} {48, 1235}

\bibitem[\protect\citeauthoryear{{Hallinan} et~al.,}{{Hallinan} et~al.}{2019}]{2019BAAS...51g.255H}
{Hallinan} G.,  et~al., 2019, in Bulletin of the American Astronomical Society. p.~255 (\mn@eprint {arXiv} {1907.07648}), \mn@doi{10.48550/arXiv.1907.07648}

\bibitem[\protect\citeauthoryear{{Hazboun}, {Romano}  \& {Smith}}{{Hazboun} et~al.}{2019}]{2019JOSS....4.1775H}
{Hazboun} J.,  {Romano} J.,   {Smith} T.,  2019, \mn@doi [The Journal of Open Source Software] {10.21105/joss.01775}, \href {https://ui.adsabs.harvard.edu/abs/2019JOSS....4.1775H} {4, 1775}

\bibitem[\protect\citeauthoryear{{Hazboun} et~al.,}{{Hazboun} et~al.}{2022}]{2022ApJ...929...39H}
{Hazboun} J.~S.,  et~al., 2022, \mn@doi [\apj] {10.3847/1538-4357/ac5829}, \href {https://ui.adsabs.harvard.edu/abs/2022ApJ...929...39H} {929, 39}

\bibitem[\protect\citeauthoryear{{Hellings} \& {Downs}}{{Hellings} \& {Downs}}{1983}]{Hellings_Downs_1983}
{Hellings} R.~W.,  {Downs} G.~S.,  1983, \mn@doi [\apjl] {10.1086/183954}, \href {https://ui.adsabs.harvard.edu/abs/1983ApJ...265L..39H} {265, L39}

\bibitem[\protect\citeauthoryear{{Hobbs}, {Edwards}  \& {Manchester}}{{Hobbs} et~al.}{2006}]{2006MNRAS.369..655H}
{Hobbs} G.~B.,  {Edwards} R.~T.,   {Manchester} R.~N.,  2006, \mn@doi [\mnras] {10.1111/j.1365-2966.2006.10302.x}, \href {https://ui.adsabs.harvard.edu/abs/2006MNRAS.369..655H} {369, 655}

\bibitem[\protect\citeauthoryear{{Hobbs} et~al.,}{{Hobbs} et~al.}{2010}]{2010CQGra..27h4013H}
{Hobbs} G.,  et~al., 2010, \mn@doi [Classical and Quantum Gravity] {10.1088/0264-9381/27/8/084013}, \href {https://ui.adsabs.harvard.edu/abs/2010CQGra..27h4013H} {27, 084013}

\bibitem[\protect\citeauthoryear{{Hotan}, {van Straten}  \& {Manchester}}{{Hotan} et~al.}{2004}]{2004PASA...21..302H}
{Hotan} A.~W.,  {van Straten} W.,   {Manchester} R.~N.,  2004, \mn@doi [\pasa] {10.1071/AS04022}, \href {https://ui.adsabs.harvard.edu/abs/2004PASA...21..302H} {21, 302}

\bibitem[\protect\citeauthoryear{{Issautier}, {Meyer-Vernet}, {Moncuquet}  \& {Hoang}}{{Issautier} et~al.}{1998}]{1998JGR...103.1969I}
{Issautier} K.,  {Meyer-Vernet} N.,  {Moncuquet} M.,   {Hoang} S.,  1998, \mn@doi [\jgr] {10.1029/97JA02661}, \href {https://ui.adsabs.harvard.edu/abs/1998JGR...103.1969I} {103, 1969}

\bibitem[\protect\citeauthoryear{{Jaffe} \& {Backer}}{{Jaffe} \& {Backer}}{2003}]{2003ApJ...583..616J}
{Jaffe} A.~H.,  {Backer} D.~C.,  2003, \mn@doi [\apj] {10.1086/345443}, \href {https://ui.adsabs.harvard.edu/abs/2003ApJ...583..616J} {583, 616}

\bibitem[\protect\citeauthoryear{{Janssen}, {Stappers}, {Kramer}, {Purver}, {Jessner}  \& {Cognard}}{{Janssen} et~al.}{2008}]{2008AIPC..983..633J}
{Janssen} G.~H.,  {Stappers} B.~W.,  {Kramer} M.,  {Purver} M.,  {Jessner} A.,   {Cognard} I.,  2008, in {Bassa} C.,  {Wang} Z.,  {Cumming} A.,   {Kaspi} V.~M.,  eds,  American Institute of Physics Conference Series Vol. 983, 40 Years of Pulsars: Millisecond Pulsars, Magnetars and More. pp 633--635, \mn@doi{10.1063/1.2900317}

\bibitem[\protect\citeauthoryear{{Jenet} et~al.,}{{Jenet} et~al.}{2009}]{2009arXiv0909.1058J}
{Jenet} F.,  et~al., 2009, arXiv e-prints, \href {https://ui.adsabs.harvard.edu/abs/2009arXiv0909.1058J} {p. arXiv:0909.1058}

\bibitem[\protect\citeauthoryear{{Jiang} et~al.,}{{Jiang} et~al.}{2019}]{2019SCPMA..6259502J}
{Jiang} P.,  et~al., 2019, \mn@doi [Science China Physics, Mechanics, and Astronomy] {10.1007/s11433-018-9376-1}, \href {https://ui.adsabs.harvard.edu/abs/2019SCPMA..6259502J} {62, 959502}

\bibitem[\protect\citeauthoryear{{Jonas} \& {MeerKAT Team}}{{Jonas} \& {MeerKAT Team}}{2016}]{2016mks..confE...1J}
{Jonas} J.,  {MeerKAT Team} 2016, in MeerKAT Science: On the Pathway to the SKA. p.~1, \mn@doi{10.22323/1.277.0001}

\bibitem[\protect\citeauthoryear{{Jones}}{{Jones}}{1990}]{1990MNRAS.246..364J}
{Jones} P.~B.,  1990, \mnras, \href {https://ui.adsabs.harvard.edu/abs/1990MNRAS.246..364J} {246, 364}

\bibitem[\protect\citeauthoryear{{Jones} et~al.,}{{Jones} et~al.}{2017}]{2017ApJ...841..125J}
{Jones} M.~L.,  et~al., 2017, \mn@doi [\apj] {10.3847/1538-4357/aa73df}, \href {https://ui.adsabs.harvard.edu/abs/2017ApJ...841..125J} {841, 125}

\bibitem[\protect\citeauthoryear{{Keith} et~al.,}{{Keith} et~al.}{2013}]{2013MNRAS.429.2161K}
{Keith} M.~J.,  et~al., 2013, \mn@doi [\mnras] {10.1093/mnras/sts486}, \href {https://ui.adsabs.harvard.edu/abs/2013MNRAS.429.2161K} {429, 2161}

\bibitem[\protect\citeauthoryear{{Kibble}}{{Kibble}}{1976}]{1976JPhA....9.1387K}
{Kibble} T.~W.~B.,  1976, \mn@doi [Journal of Physics A Mathematical General] {10.1088/0305-4470/9/8/029}, \href {https://ui.adsabs.harvard.edu/abs/1976JPhA....9.1387K} {9, 1387}

\bibitem[\protect\citeauthoryear{{Kocsis} \& {Sesana}}{{Kocsis} \& {Sesana}}{2011}]{2011MNRAS.411.1467K}
{Kocsis} B.,  {Sesana} A.,  2011, \mn@doi [\mnras] {10.1111/j.1365-2966.2010.17782.x}, \href {https://ui.adsabs.harvard.edu/abs/2011MNRAS.411.1467K} {411, 1467}

\bibitem[\protect\citeauthoryear{{Kramer} et~al.,}{{Kramer} et~al.}{2021}]{2021MNRAS.504.2094K}
{Kramer} M.,  et~al., 2021, \mn@doi [\mnras] {10.1093/mnras/stab375}, \href {https://ui.adsabs.harvard.edu/abs/2021MNRAS.504.2094K} {504, 2094}

\bibitem[\protect\citeauthoryear{{Kulkarni}, {Shannon}, {Reardon}, {Miles}, {Bailes}  \& {Shamohammadi}}{{Kulkarni} et~al.}{2024}]{2024MNRAS.528.3658K}
{Kulkarni} A.~D.,  {Shannon} R.~M.,  {Reardon} D.~J.,  {Miles} M.~T.,  {Bailes} M.,   {Shamohammadi} M.,  2024, \mn@doi [\mnras] {10.1093/mnras/stae041}, \href {https://ui.adsabs.harvard.edu/abs/2024MNRAS.528.3658K} {528, 3658}

\bibitem[\protect\citeauthoryear{{Lam}, {Cordes}, {Chatterjee}  \& {Dolch}}{{Lam} et~al.}{2015}]{2015ApJ...801..130L}
{Lam} M.~T.,  {Cordes} J.~M.,  {Chatterjee} S.,   {Dolch} T.,  2015, \mn@doi [\apj] {10.1088/0004-637X/801/2/130}, \href {https://ui.adsabs.harvard.edu/abs/2015ApJ...801..130L} {801, 130}

\bibitem[\protect\citeauthoryear{{Lam} et~al.,}{{Lam} et~al.}{2019}]{2019ApJ...872..193L}
{Lam} M.~T.,  et~al., 2019, \mn@doi [\apj] {10.3847/1538-4357/ab01cd}, \href {https://ui.adsabs.harvard.edu/abs/2019ApJ...872..193L} {872, 193}

\bibitem[\protect\citeauthoryear{{Lang}}{{Lang}}{1971}]{1971ApJ...164..249L}
{Lang} K.~R.,  1971, \mn@doi [\apj] {10.1086/150836}, \href {https://ui.adsabs.harvard.edu/abs/1971ApJ...164..249L} {164, 249}

\bibitem[\protect\citeauthoryear{{Lasky} et~al.,}{{Lasky} et~al.}{2016}]{2016PhRvX...6a1035L}
{Lasky} P.~D.,  et~al., 2016, \mn@doi [Physical Review X] {10.1103/PhysRevX.6.011035}, \href {https://ui.adsabs.harvard.edu/abs/2016PhRvX...6a1035L} {6, 011035}

\bibitem[\protect\citeauthoryear{{Lazarus}, {Karuppusamy}, {Graikou}, {Caballero}, {Champion}, {Lee}, {Verbiest}  \& {Kramer}}{{Lazarus} et~al.}{2016}]{2016MNRAS.458..868L}
{Lazarus} P.,  {Karuppusamy} R.,  {Graikou} E.,  {Caballero} R.~N.,  {Champion} D.~J.,  {Lee} K.~J.,  {Verbiest} J.~P.~W.,   {Kramer} M.,  2016, \mn@doi [\mnras] {10.1093/mnras/stw189}, \href {https://ui.adsabs.harvard.edu/abs/2016MNRAS.458..868L} {458, 868}

\bibitem[\protect\citeauthoryear{{Lentati}, {Alexander}, {Hobson}, {Taylor}, {Gair}, {Balan}  \& {van Haasteren}}{{Lentati} et~al.}{2013}]{2013PhRvD..87j4021L}
{Lentati} L.,  {Alexander} P.,  {Hobson} M.~P.,  {Taylor} S.,  {Gair} J.,  {Balan} S.~T.,   {van Haasteren} R.,  2013, \mn@doi [\prd] {10.1103/PhysRevD.87.104021}, \href {https://ui.adsabs.harvard.edu/abs/2013PhRvD..87j4021L} {87, 104021}

\bibitem[\protect\citeauthoryear{{Lentati}, {Alexander}, {Hobson}, {Feroz}, {van Haasteren}, {Lee}  \& {Shannon}}{{Lentati} et~al.}{2014}]{2014MNRAS.437.3004L}
{Lentati} L.,  {Alexander} P.,  {Hobson} M.~P.,  {Feroz} F.,  {van Haasteren} R.,  {Lee} K.~J.,   {Shannon} R.~M.,  2014, \mn@doi [\mnras] {10.1093/mnras/stt2122}, \href {https://ui.adsabs.harvard.edu/abs/2014MNRAS.437.3004L} {437, 3004}

\bibitem[\protect\citeauthoryear{{Lentati} et~al.,}{{Lentati} et~al.}{2015}]{2015MNRAS.453.2576L}
{Lentati} L.,  et~al., 2015, \mn@doi [\mnras] {10.1093/mnras/stv1538}, \href {https://ui.adsabs.harvard.edu/abs/2015MNRAS.453.2576L} {453, 2576}

\bibitem[\protect\citeauthoryear{{Lentati} et~al.,}{{Lentati} et~al.}{2017}]{2017MNRAS.466.3706L}
{Lentati} L.,  et~al., 2017, \mn@doi [\mnras] {10.1093/mnras/stw3359}, \href {https://ui.adsabs.harvard.edu/abs/2017MNRAS.466.3706L} {466, 3706}

\bibitem[\protect\citeauthoryear{{Luo} et~al.,}{{Luo} et~al.}{2021}]{2021ApJ...911...45L}
{Luo} J.,  et~al., 2021, \mn@doi [\apj] {10.3847/1538-4357/abe62f}, \href {https://ui.adsabs.harvard.edu/abs/2021ApJ...911...45L} {911, 45}

\bibitem[\protect\citeauthoryear{{Lyne}, {Hobbs}, {Kramer}, {Stairs}  \& {Stappers}}{{Lyne} et~al.}{2010}]{2010Sci...329..408L}
{Lyne} A.,  {Hobbs} G.,  {Kramer} M.,  {Stairs} I.,   {Stappers} B.,  2010, \mn@doi [Science] {10.1126/science.1186683}, \href {https://ui.adsabs.harvard.edu/abs/2010Sci...329..408L} {329, 408}

\bibitem[\protect\citeauthoryear{{Maggiore}}{{Maggiore}}{2000}]{2000gr.qc.....8027M}
{Maggiore} M.,  2000, \mn@doi [arXiv e-prints] {10.48550/arXiv.gr-qc/0008027}, \href {https://ui.adsabs.harvard.edu/abs/2000gr.qc.....8027M} {pp gr--qc/0008027}

\bibitem[\protect\citeauthoryear{{Manchester}}{{Manchester}}{2008}]{2008AIPC..983..584M}
{Manchester} R.~N.,  2008, in {Bassa} C.,  {Wang} Z.,  {Cumming} A.,   {Kaspi} V.~M.,  eds,  American Institute of Physics Conference Series Vol. 983, 40 Years of Pulsars: Millisecond Pulsars, Magnetars and More. pp 584--592 (\mn@eprint {arXiv} {0710.5026}), \mn@doi{10.1063/1.2900303}

\bibitem[\protect\citeauthoryear{{McIntosh}, {Chapman}, {Leamon}, {Egeland}  \& {Watkins}}{{McIntosh} et~al.}{2020}]{2020SoPh..295..163M}
{McIntosh} S.~W.,  {Chapman} S.,  {Leamon} R.~J.,  {Egeland} R.,   {Watkins} N.~W.,  2020, \mn@doi [\solphys] {10.1007/s11207-020-01723-y}, \href {https://ui.adsabs.harvard.edu/abs/2020SoPh..295..163M} {295, 163}

\bibitem[\protect\citeauthoryear{{Melatos} \& {Link}}{{Melatos} \& {Link}}{2014}]{2014MNRAS.437...21M}
{Melatos} A.,  {Link} B.,  2014, \mn@doi [\mnras] {10.1093/mnras/stt1828}, \href {https://ui.adsabs.harvard.edu/abs/2014MNRAS.437...21M} {437, 21}

\bibitem[\protect\citeauthoryear{{Miles}, {Shannon}, {Bailes}, {Reardon}, {Buchner}, {Middleton}  \& {Spiewak}}{{Miles} et~al.}{2022}]{2022MNRAS.510.5908M}
{Miles} M.~T.,  {Shannon} R.~M.,  {Bailes} M.,  {Reardon} D.~J.,  {Buchner} S.,  {Middleton} H.,   {Spiewak} R.,  2022, \mn@doi [\mnras] {10.1093/mnras/stab3549}, \href {https://ui.adsabs.harvard.edu/abs/2022MNRAS.510.5908M} {510, 5908}

\bibitem[\protect\citeauthoryear{{Miles} et~al.,}{{Miles} et~al.}{2023}]{2023MNRAS.519.3976M}
{Miles} M.~T.,  et~al., 2023, \mn@doi [\mnras] {10.1093/mnras/stac3644}, \href {https://ui.adsabs.harvard.edu/abs/2023MNRAS.519.3976M} {519, 3976}

\bibitem[\protect\citeauthoryear{{Morello} et~al.,}{{Morello} et~al.}{2019}]{2019MNRAS.483.3673M}
{Morello} V.,  et~al., 2019, \mn@doi [\mnras] {10.1093/mnras/sty3328}, \href {https://ui.adsabs.harvard.edu/abs/2019MNRAS.483.3673M} {483, 3673}

\bibitem[\protect\citeauthoryear{{Murphy} et~al.,}{{Murphy} et~al.}{2018}]{2018ASPC..517....3M}
{Murphy} E.~J.,  et~al., 2018, in {Murphy} E.,  ed.,  Astronomical Society of the Pacific Conference Series Vol. 517, Science with a Next Generation Very Large Array. p.~3 (\mn@eprint {arXiv} {1810.07524}), \mn@doi{10.48550/arXiv.1810.07524}

\bibitem[\protect\citeauthoryear{{Nathan}, {Miles}, {Ashton}, {Lasky}, {Thrane}, {Reardon}, {Shannon}  \& {Cameron}}{{Nathan} et~al.}{2023}]{2023arXiv230402793N}
{Nathan} R.~S.,  {Miles} M.~T.,  {Ashton} G.,  {Lasky} P.~D.,  {Thrane} E.,  {Reardon} D.~J.,  {Shannon} R.~M.,   {Cameron} A.~D.,  2023, \mn@doi [arXiv e-prints] {10.48550/arXiv.2304.02793}, \href {https://ui.adsabs.harvard.edu/abs/2023arXiv230402793N} {p. arXiv:2304.02793}

\bibitem[\protect\citeauthoryear{{Nice} et~al.,}{{Nice} et~al.}{2015}]{2015ascl.soft09002N}
{Nice} D.,  et~al., 2015, {Tempo: Pulsar timing data analysis}, Astrophysics Source Code Library, record ascl:1509.002

\bibitem[\protect\citeauthoryear{{Ni{\c{t}}u} et~al.,}{{Ni{\c{t}}u} et~al.}{2024}]{2024MNRAS.528.3304N}
{Ni{\c{t}}u} I.~C.,  et~al., 2024, \mn@doi [\mnras] {10.1093/mnras/stae220}, \href {https://ui.adsabs.harvard.edu/abs/2024MNRAS.528.3304N} {528, 3304}

\bibitem[\protect\citeauthoryear{\"Olmez, Mandic  \& Siemens}{\"Olmez et~al.}{2010}]{PhysRevD.81.104028}
\"Olmez S.,  Mandic V.,   Siemens X.,  2010, \mn@doi [Phys. Rev. D] {10.1103/PhysRevD.81.104028}, 81, 104028

\bibitem[\protect\citeauthoryear{{Park}, {Folkner}, {Williams}  \& {Boggs}}{{Park} et~al.}{2021}]{2021AJ....161..105P}
{Park} R.~S.,  {Folkner} W.~M.,  {Williams} J.~G.,   {Boggs} D.~H.,  2021, \mn@doi [\aj] {10.3847/1538-3881/abd414}, \href {https://ui.adsabs.harvard.edu/abs/2021AJ....161..105P} {161, 105}

\bibitem[\protect\citeauthoryear{{Parthasarathy} et~al.,}{{Parthasarathy} et~al.}{2019}]{2019MNRAS.489.3810P}
{Parthasarathy} A.,  et~al., 2019, \mn@doi [\mnras] {10.1093/mnras/stz2383}, \href {https://ui.adsabs.harvard.edu/abs/2019MNRAS.489.3810P} {489, 3810}

\bibitem[\protect\citeauthoryear{{Parthasarathy} et~al.,}{{Parthasarathy} et~al.}{2021}]{2021MNRAS.502..407P}
{Parthasarathy} A.,  et~al., 2021, \mn@doi [\mnras] {10.1093/mnras/stab037}, \href {https://ui.adsabs.harvard.edu/abs/2021MNRAS.502..407P} {502, 407}

\bibitem[\protect\citeauthoryear{{Pennucci}}{{Pennucci}}{2019}]{2019ApJ...871...34P}
{Pennucci} T.~T.,  2019, \mn@doi [\apj] {10.3847/1538-4357/aaf6ef}, \href {https://ui.adsabs.harvard.edu/abs/2019ApJ...871...34P} {871, 34}

\bibitem[\protect\citeauthoryear{{Phillips} \& {Wolszczan}}{{Phillips} \& {Wolszczan}}{1991}]{1991ApJ...382L..27P}
{Phillips} J.~A.,  {Wolszczan} A.,  1991, \mn@doi [\apjl] {10.1086/186206}, \href {https://ui.adsabs.harvard.edu/abs/1991ApJ...382L..27P} {382, L27}

\bibitem[\protect\citeauthoryear{{Phinney}}{{Phinney}}{2001}]{2001astro.ph..8028P}
{Phinney} E.~S.,  2001, \mn@doi [arXiv e-prints] {10.48550/arXiv.astro-ph/0108028}, \href {https://ui.adsabs.harvard.edu/abs/2001astro.ph..8028P} {pp astro--ph/0108028}

\bibitem[\protect\citeauthoryear{{Rajagopal} \& {Romani}}{{Rajagopal} \& {Romani}}{1995}]{1995ApJ...446..543R}
{Rajagopal} M.,  {Romani} R.~W.,  1995, \mn@doi [\apj] {10.1086/175813}, \href {https://ui.adsabs.harvard.edu/abs/1995ApJ...446..543R} {446, 543}

\bibitem[\protect\citeauthoryear{{Reardon} et~al.,}{{Reardon} et~al.}{2021}]{2021arXiv210704609R}
{Reardon} D.~J.,  et~al., 2021, arXiv e-prints, \href {https://ui.adsabs.harvard.edu/abs/2021arXiv210704609R} {p. arXiv:2107.04609}

\bibitem[\protect\citeauthoryear{{Reardon} et~al.,}{{Reardon} et~al.}{2023a}]{2023ApJ...951L...6R}
{Reardon} D.~J.,  et~al., 2023a, \mn@doi [\apjl] {10.3847/2041-8213/acdd02}, \href {https://ui.adsabs.harvard.edu/abs/2023ApJ...951L...6R} {951, L6}

\bibitem[\protect\citeauthoryear{{Reardon} et~al.,}{{Reardon} et~al.}{2023b}]{2023ApJ...951L...7R}
{Reardon} D.~J.,  et~al., 2023b, \mn@doi [\apjl] {10.3847/2041-8213/acdd03}, \href {https://ui.adsabs.harvard.edu/abs/2023ApJ...951L...7R} {951, L7}

\bibitem[\protect\citeauthoryear{{Rickett}}{{Rickett}}{1977}]{1977ARA&A..15..479R}
{Rickett} B.~J.,  1977, \mn@doi [\araa] {10.1146/annurev.aa.15.090177.002403}, \href {https://ui.adsabs.harvard.edu/abs/1977ARA&A..15..479R} {15, 479}

\bibitem[\protect\citeauthoryear{{Rickett}}{{Rickett}}{1990}]{1990ARA&A..28..561R}
{Rickett} B.~J.,  1990, \mn@doi [\araa] {10.1146/annurev.aa.28.090190.003021}, \href {https://ui.adsabs.harvard.edu/abs/1990ARA&A..28..561R} {28, 561}

\bibitem[\protect\citeauthoryear{{Roedig}, {Sesana}, {Dotti}, {Cuadra}, {Amaro-Seoane}  \& {Haardt}}{{Roedig} et~al.}{2012}]{2012A&A...545A.127R}
{Roedig} C.,  {Sesana} A.,  {Dotti} M.,  {Cuadra} J.,  {Amaro-Seoane} P.,   {Haardt} F.,  2012, \mn@doi [\aap] {10.1051/0004-6361/201219986}, \href {https://ui.adsabs.harvard.edu/abs/2012A&A...545A.127R} {545, A127}

\bibitem[\protect\citeauthoryear{{Samajdar} et~al.,}{{Samajdar} et~al.}{2022}]{2022MNRAS.517.1460S}
{Samajdar} A.,  et~al., 2022, \mn@doi [\mnras] {10.1093/mnras/stac2810}, \href {https://ui.adsabs.harvard.edu/abs/2022MNRAS.517.1460S} {517, 1460}

\bibitem[\protect\citeauthoryear{{Sanidas}, {Battye}  \& {Stappers}}{{Sanidas} et~al.}{2012}]{2012PhRvD..85l2003S}
{Sanidas} S.~A.,  {Battye} R.~A.,   {Stappers} B.~W.,  2012, \mn@doi [\prd] {10.1103/PhysRevD.85.122003}, \href {https://ui.adsabs.harvard.edu/abs/2012PhRvD..85l2003S} {85, 122003}

\bibitem[\protect\citeauthoryear{{Sesana}, {Haardt}, {Madau}  \& {Volonteri}}{{Sesana} et~al.}{2004}]{2004ApJ...611..623S}
{Sesana} A.,  {Haardt} F.,  {Madau} P.,   {Volonteri} M.,  2004, \mn@doi [\apj] {10.1086/422185}, \href {https://ui.adsabs.harvard.edu/abs/2004ApJ...611..623S} {611, 623}

\bibitem[\protect\citeauthoryear{{Shannon} \& {Cordes}}{{Shannon} \& {Cordes}}{2010}]{2010ApJ...725.1607S}
{Shannon} R.~M.,  {Cordes} J.~M.,  2010, \mn@doi [\apj] {10.1088/0004-637X/725/2/1607}, \href {https://ui.adsabs.harvard.edu/abs/2010ApJ...725.1607S} {725, 1607}

\bibitem[\protect\citeauthoryear{{Shannon} \& {Cordes}}{{Shannon} \& {Cordes}}{2017}]{2017MNRAS.464.2075S}
{Shannon} R.~M.,  {Cordes} J.~M.,  2017, \mn@doi [\mnras] {10.1093/mnras/stw2449}, \href {https://ui.adsabs.harvard.edu/abs/2017MNRAS.464.2075S} {464, 2075}

\bibitem[\protect\citeauthoryear{{Shannon} et~al.,}{{Shannon} et~al.}{2013}]{2013ApJ...766....5S}
{Shannon} R.~M.,  et~al., 2013, \mn@doi [\apj] {10.1088/0004-637X/766/1/5}, \href {https://ui.adsabs.harvard.edu/abs/2013ApJ...766....5S} {766, 5}

\bibitem[\protect\citeauthoryear{{Shannon} et~al.,}{{Shannon} et~al.}{2014}]{2014MNRAS.443.1463S}
{Shannon} R.~M.,  et~al., 2014, \mn@doi [\mnras] {10.1093/mnras/stu1213}, \href {https://ui.adsabs.harvard.edu/abs/2014MNRAS.443.1463S} {443, 1463}

\bibitem[\protect\citeauthoryear{{Siemens}, {Ellis}, {Jenet}  \& {Romano}}{{Siemens} et~al.}{2013}]{2013CQGra..30v4015S}
{Siemens} X.,  {Ellis} J.,  {Jenet} F.,   {Romano} J.~D.,  2013, \mn@doi [Classical and Quantum Gravity] {10.1088/0264-9381/30/22/224015}, \href {https://ui.adsabs.harvard.edu/abs/2013CQGra..30v4015S} {30, 224015}

\bibitem[\protect\citeauthoryear{{Singha} et~al.,}{{Singha} et~al.}{2021}]{2021MNRAS.507L..57S}
{Singha} J.,  et~al., 2021, \mn@doi [\mnras] {10.1093/mnrasl/slab098}, \href {https://ui.adsabs.harvard.edu/abs/2021MNRAS.507L..57S} {507, L57}

\bibitem[\protect\citeauthoryear{{Smith}, {Ashton}, {Vajpeyi}  \& {Talbot}}{{Smith} et~al.}{2020}]{pbilby_paper}
{Smith} R. J.~E.,  {Ashton} G.,  {Vajpeyi} A.,   {Talbot} C.,  2020, \mn@doi [\mnras] {10.1093/mnras/staa2483}, \href {https://ui.adsabs.harvard.edu/abs/2020MNRAS.498.4492S} {498, 4492}

\bibitem[\protect\citeauthoryear{{Spiewak} et~al.,}{{Spiewak} et~al.}{2022}]{2022PASA...39...27S}
{Spiewak} R.,  et~al., 2022, \mn@doi [\pasa] {10.1017/pasa.2022.19}, \href {https://ui.adsabs.harvard.edu/abs/2022PASA...39...27S} {39, e027}

\bibitem[\protect\citeauthoryear{{Starobinsky}}{{Starobinsky}}{1980}]{1980PhLB...91...99S}
{Starobinsky} A.~A.,  1980, \mn@doi [Physics Letters B] {10.1016/0370-2693(80)90670-X}, \href {https://ui.adsabs.harvard.edu/abs/1980PhLB...91...99S} {91, 99}

\bibitem[\protect\citeauthoryear{Stephens}{Stephens}{1974}]{1a8d0b27-4e98-39f3-b2a1-d2944a87e07c}
Stephens M.~A.,  1974, Journal of the American Statistical Association, 69, 730

\bibitem[\protect\citeauthoryear{{Taylor}, {Simon}  \& {Sampson}}{{Taylor} et~al.}{2017}]{2017PhRvL.118r1102T}
{Taylor} S.~R.,  {Simon} J.,   {Sampson} L.,  2017, \mn@doi [\prl] {10.1103/PhysRevLett.118.181102}, \href {https://ui.adsabs.harvard.edu/abs/2017PhRvL.118r1102T} {118, 181102}

\bibitem[\protect\citeauthoryear{{Taylor}, {Simon}, {Schult}, {Pol}  \& {Lamb}}{{Taylor} et~al.}{2022}]{2022PhRvD.105h4049T}
{Taylor} S.~R.,  {Simon} J.,  {Schult} L.,  {Pol} N.,   {Lamb} W.~G.,  2022, \mn@doi [\prd] {10.1103/PhysRevD.105.084049}, \href {https://ui.adsabs.harvard.edu/abs/2022PhRvD.105h4049T} {105, 084049}

\bibitem[\protect\citeauthoryear{{Tiburzi} et~al.,}{{Tiburzi} et~al.}{2016}]{2016MNRAS.455.4339T}
{Tiburzi} C.,  et~al., 2016, \mn@doi [\mnras] {10.1093/mnras/stv2143}, \href {https://ui.adsabs.harvard.edu/abs/2016MNRAS.455.4339T} {455, 4339}

\bibitem[\protect\citeauthoryear{{Tiburzi} et~al.,}{{Tiburzi} et~al.}{2021}]{2021A&A...647A..84T}
{Tiburzi} C.,  et~al., 2021, \mn@doi [\aap] {10.1051/0004-6361/202039846}, \href {https://ui.adsabs.harvard.edu/abs/2021A&A...647A..84T} {647, A84}

\bibitem[\protect\citeauthoryear{{Vallisneri} et~al.,}{{Vallisneri} et~al.}{2020}]{2020ApJ...893..112V}
{Vallisneri} M.,  et~al., 2020, \mn@doi [\apj] {10.3847/1538-4357/ab7b67}, \href {https://ui.adsabs.harvard.edu/abs/2020ApJ...893..112V} {893, 112}

\bibitem[\protect\citeauthoryear{{Wyithe} \& {Loeb}}{{Wyithe} \& {Loeb}}{2003}]{2003ApJ...590..691W}
{Wyithe} J. S.~B.,  {Loeb} A.,  2003, \mn@doi [\apj] {10.1086/375187}, \href {https://ui.adsabs.harvard.edu/abs/2003ApJ...590..691W} {590, 691}

\bibitem[\protect\citeauthoryear{{Xu} et~al.,}{{Xu} et~al.}{2023}]{2023RAA....23g5024X}
{Xu} H.,  et~al., 2023, \mn@doi [Research in Astronomy and Astrophysics] {10.1088/1674-4527/acdfa5}, \href {https://ui.adsabs.harvard.edu/abs/2023RAA....23g5024X} {23, 075024}

\bibitem[\protect\citeauthoryear{{Zic} et~al.,}{{Zic} et~al.}{2022}]{2022MNRAS.516..410Z}
{Zic} A.,  et~al., 2022, \mn@doi [\mnras] {10.1093/mnras/stac2100}, \href {https://ui.adsabs.harvard.edu/abs/2022MNRAS.516..410Z} {516, 410}

\bibitem[\protect\citeauthoryear{{van Haasteren}}{{van Haasteren}}{2024}]{2024ApJS..273...23V}
{van Haasteren} R.,  2024, \mn@doi [\apjs] {10.3847/1538-4365/ad530f}, \href {https://ui.adsabs.harvard.edu/abs/2024ApJS..273...23V} {273, 23}

\bibitem[\protect\citeauthoryear{{van Haasteren}, {Levin}, {McDonald}  \& {Lu}}{{van Haasteren} et~al.}{2009}]{2009MNRAS.395.1005V}
{van Haasteren} R.,  {Levin} Y.,  {McDonald} P.,   {Lu} T.,  2009, \mn@doi [\mnras] {10.1111/j.1365-2966.2009.14590.x}, \href {https://ui.adsabs.harvard.edu/abs/2009MNRAS.395.1005V} {395, 1005}

\makeatother
\end{thebibliography}





\bsp	
\label{lastpage}
\end{document}